\documentclass[english,journal=jctcce,manuscript=article,email=true,etalmode=truncate,maxauthors=0]{achemso}

\usepackage[utf8]{inputenc}
  \usepackage{geometry} 
  \usepackage{stackengine}
  \usepackage{titlesec}

  \usepackage[T1]{fontenc} 
  \usepackage{amsmath}     
  \usepackage{amssymb}
  \usepackage{bm}
  \usepackage{mathrsfs}
  \usepackage{etoolbox}
  \usepackage{environ}
  \usepackage[version=4]{mhchem}
  \usepackage{booktabs}
  \usepackage{adjustbox}
  \usepackage{hhline}

  \usepackage{setspace}    
  \usepackage{float}       
  \newfloat{chart}{htbp}{loc}
  \newfloat{graph}{htbp}{loh}
  \newfloat{scheme}{htbp}{los}

  \usepackage{graphicx}    

  \usepackage[english]{babel}
  \usepackage{array}

  \usepackage{subcaption}  
  \usepackage{outlines}    
  \usepackage[normalem]{ulem} 

  \usepackage[numbers,sort&compress,super]{natbib}
  \usepackage{natmove}

  \usepackage{cancel}      
  \usepackage{pdfpages}  
  \usepackage[font=scriptsize,labelfont=bf]{caption}

  \usepackage{hyperref} 
  \usepackage{cleveref}	
  	\crefname{figure}{Figure}{Figures}
  	\crefname{table}{Table}{Tables}
  	\crefname{equation}{Eq.}{Eqs.}
  	\crefname{section}{Section}{Sections}
  	\crefname{subsection}{Section}{Sections}
  	\crefname{subsubsection}{Section}{Sections}
  	\crefname{algorithm}{Algorithm}{Algorithms}

  \usepackage{tikz}

    \usepackage[ruled,vlined]{algorithm2e}

  \usepackage{todonotes}

\usepackage{epstopdf}
\AppendGraphicsExtensions{.tiff}

\title{A CPD-enabled low-scaling environment solver in a coupled cluster based static quantum embedding theory}
\author{Karl Pierce}
\affiliation{Department of Mathematics, University of Maryland, College Park, MD 20742, USA}
\author{Muhammad Talha Aziz}
\affiliation{Department of Mathematics, Rensselaer Polytechnic Institute, Troy, NY 12180, USA}
\author{Avijit Shee}
\affiliation{Department of Chemistry, University of California, Berkeley, CA 94720, USA}
\author{Fabian M. Faulstich}
\affiliation{Department of Mathematics, Rensselaer Polytechnic Institute, Troy, NY 12180, USA}
\email{faulsf@rpi.edu}
\date{\today}

\begin{document}

\begin{abstract}
We incorporate a canonical polyadic decomposition (CPD) based low-level solver as a means to accelerate the environment-level solver for the recently developed MPCC embedding framework. Using CPD, we both factorize the three dominant order-three density-fitting two-electron integral (DF TEI) tensors and develop a novel formulation that reduces the storage complexity of the low-level solver from $\mathcal{O}(N^3)$ to $\mathcal{O}(NR)$, where $R$ is the CPD rank, and the computational scaling of the most time-consuming contractions from $\mathcal{O}(N^4)$ to $\mathcal{O}(NR^2)$. We provide benchmarks on representative chemical environments, namely water clusters $\ce{(H_2O)_n}$ with $n = 1$ to $6$ and linear alkane chains $\ce{C_nH_{2n+2}}$ with $n = 1$ to $6$. For both test sets, using the CPD-compressed DF TEI tensors reproduces the DF reference convergence behavior of the low-level solver, the subsequent high-level step, and the fully self-consistent MPCC iterations, while introducing only small, rank-controlled shifts in absolute energies. At a fixed tolerance in the absolute MPCC energy, the CP ranks required for these tensor approximations increase linearly with system size. Chemically relevant energy differences are likewise preserved, as demonstrated for water-cluster dissociation energies and in a proof-of-concept embedding calculation of methane in a four-water cluster.
\end{abstract}

\maketitle

\section{Introduction}

Coupled cluster (CC) theory has long been regarded as one of the most reliable and systematically improvable approaches for describing electron correlation in molecular systems.\cite{bartlett2007coupled,crawford2007introduction,kummel1991origins,vcivzek1991origins,bartlett2005theory,paldus2005beginnings,arponen1991independent,bishop1991overview} In practice, however, its applicability is limited by steep computational and storage requirements. The coupled cluster with single and double excitations method (CCSD), which is regarded as a method that offers a favorable balance between accuracy and cost for many systems, still carries an algorithmic cost of $\mathcal{O}({ O}^2{ V}^4)$ in the number of occupied (${O}$) and virtual (${ V}$) orbitals.\cite{purvis1982full}
Practical applications of CCSD are more often limited by  memory requirements as necessary tensor quantities carry a storage complexity of $\mathcal{O}(N^4)$ with system size, where $N = { O}+{ V}$.\cite{schutz2003linear}

These algorithmic and memory constraints have motivated a broad range of strategies aimed at extending correlated wavefunction methods to larger systems, including orbital localization\cite{VRG:pulay:1983:CPL,VRG:pulay:1984:JCP,VRG:ahlrichs:1975:JCP,VRG:neese:2009:JCPa,VRG:neese:2011:JCTC,Rolik:2011:JCP,Rolik:2013:JCP}, fragmentation\cite{VRG:kitaura:1999:CPL,Kristensen:2011:JCTC,Li:2002:JCC,Li:2006:JCP,Li:2009:JCP,VRG:guo:2018:JCP} and tensor decomposition strategies (i.e., the density fitting approximation\cite{VRG:whitten:1973:JCP,VRG:dunlap:1979:JCP,Vahtras:1993:CPL,VRG:jung:2005:PNAS,VRG:mintmire:1982:PRA} and the tensor hypercontraction decomposition\cite{VRG:hohenstein:2012:JCP,VRG:hohenstein:2012:JCPa,VRG:parrish:2012:JCP,Hohenstein:2013:JCP,VRG:parrish:2014:JCP,Shenvi:2013:JCP,Schutski:2017:JCP,Parrish:2019:JCP,Lee:2019:JCTC,VRG:hummel:2017:JCP,VRG:song::JCP,Hohenstein:2019:JCP,Hohenstein:2021:JCP,Hohenstein:2022:JCP,Jiang:2022:JCTC,Zhao:2023:JCTC,Datar:2024:JCTC,Schmitz:2017:JCP, Khoromskaia:2015:PCCP,VRG:pierce:2021:JCTC}).

One particularly appealing direction is to exploit the locality of electronic correlation by treating a chemically relevant subset of orbitals with a highly accurate method, while approximating the remaining degrees of freedom more economically. In the context of CC theory, this can be achieved by partitioning the cluster operator into {\it fragment} and {\it environment} components (formerly called {\it internal} and {\it external}, respectively), which are then treated at different levels of approximation.\cite{piecuch1993state,piecuch1994state} To faithfully represent the physical system, such embedding-type approaches require an explicit coupling between the fragment and environment degrees of freedom. Such coupling was established successfully in the hybrid MP2 and CCSD method by Nooijen\cite{nooijen1999combining} and Sherrill {\it et al}.\cite{bochevarov2005hybrid,bochevarov2006hybrid} Another approach similar in spirit is the multi-level CC (MLCC) theory of Koch and co-workers, which utilizes a localized active space and also defines a mechanism for relaxing the environment.\cite{myhre2014multi,folkestad2021multilevel} 

Recent work along these lines has introduced the MPCC method, a static coupled cluster embedding framework that is, in spirit, very similar: the fragment orbital space is treated with a high-level CC solver, and the environment orbital space is treated with a perturbative description.\cite{Shee2024}
The fragment is coupled to the environment via a self-consistent optimization of the global amplitudes via an effective, downfolded fragment Hamiltonian that is screened by the high-energy environment amplitudes, a feature that is essential for achieving accurate quantum embedding descriptions.\cite{van2021random,han2021investigation} A similar type of downfolding is used by Kowalski in the sub-system embedding sub-algebras (SES) CC approach.\cite{kowalski2023sub} An updated MPCC algorithm has developed that leverages the density fitting approximation to factor out the order-four, two-electron integral tensor, thereby reducing the method's computational cost\cite{shee2026towards}
In addition, extensions beyond CCSD have been introduced; these studies explore the use of CCSDT-level fragment solvers together with the perturbative treatments of three-particle interactions for the environment.\cite{shee2026addressing} Because the chemically relevant region is typically localized and remains comparatively small, the overall efficiency is then governed primarily by the scaling and memory footprint of the environment's treatment.

While the recent introduction of the density fitting approximation to the environment solver successfully reduced the storage scaling associated with the two-electron integrals, it also introduced a number of tensors whose storage grows as $\mathcal{O}(N^3)$ and whose repeated contraction can dominate both memory and data movement.\cite{schutz2003linear,werner2011efficient} This, therefore, motivates studies to further compress the density-fitting representation, to reduce the storage requirements and contraction cost of the low-level environment equations while maintaining a controlled accuracy. A natural strategy to address this limitation is to exploit low-rank structure in the tensors of the density-fitting approximation. Over the past decade, a variety of tensor factorization techniques have been developed to further compress electron repulsion integrals beyond standard density fitting approximation. Among these is the analytic canonical polyadic decomposition (CPD),\cite{Pierce:2025:JCTC:MP2,VRG:pierce:2021:JCTC,Pierce:2022:ETD,Pierce:2025:JCTC:CPB,VRG:benedikt:2011:JCP,Benedikt:2013:JCP,Benedikt:2013:MP,Bohm:2016:JCP,Schmitz:2017:JCP,Khoromskaia:2015:PCCP,Madsen:2018:JCP} 
a mathematical tool that factorizes arbitrary higher-order tensors into structured tensor products of matrices.
Because of the structure and flexibility associated the CPD, the representation has the potential to substantially reduce both the storage requirements and the cost of tensor contractions.
These properties, combined with a relatively straightforward method for improving the decomposition's accuracy makes the CPD an attractive tool to accelerate electronic structure methods.

In this work, we propose a method that replaces the density fitting integral tensors that appear in the MPCC low-level environment equations with their CPD approximations.
In doing so, we approximate these order-three tensors using sets of order-two factor matrices which may significantly reduce the overall memory overhead.
Furthermore, we derive a CPD-enhanced low-level solver that avoids forming order-three intermediates, reducing the memory requirements for the LL MPCC solver from $\mathcal{O}(N^3)$ to $\mathcal{O}(NR) \approx \mathcal{O}(N^2)$, where $R$ is the CP rank. Additionally, we reduce the scaling of the dominant tensor contractions in the low-level method from $\mathcal{O}(N^4)$ to $\mathcal{O}(NR^2) \approx \mathcal{O}(N^3)$ for all but one term.

\section{Theoretical Background}

\subsection{The MPCC method}

Coupled Cluster (CC) theory is a popular framework to recover dynamical electronic correlations that are disregarded in Hartree-Fock theory\cite{VRG:cizek:1966:JCP, Cizek:1969:ACP,Cizek:1971:IJQC,VRG:crawford:2000:RCC,VRG:tajti:2004:JCP,VRG:harding:2008:JCP,Thorpe:2019:JCP,VRG:purvis:1982:JCP}.
Coupled cluster methods adopt an exponential wavefunction ansatz, i.e.,
\begin{equation}
|\Psi\rangle = e^{T}|\Phi_0\rangle, \qquad 
T=\sum_{n\ge 1} \frac{1}{(n!)^2} T_n ,
\end{equation}
with
\begin{equation}
T_n
=\sum_{i_1\cdots i_n}\sum_{a_1\cdots a_n}
t_{i_1\cdots i_n}^{a_1\cdots a_n}\,
a_{a_1}^\dagger\cdots a_{a_n}^\dagger a_{i_n}\cdots a_{i_1}
=\sum_{i_1\cdots i_n}\sum_{a_1\cdots a_n}
t_{i_1\cdots i_n}^{a_1\cdots a_n}\,
X_{i_1\cdots i_n}^{a_1\cdots a_n},
\end{equation}
where $X$ are particle-hole excitation operators,\cite{Cizek1966} and $i,j,k,\ldots$ and $a,b,c,\ldots$ label the occupied orbitals ($O$) and virtual orbitals ($V$), respectively. Assuming intermediate normalization $\langle \Phi_0 | \Psi\rangle = 1$, untruncated CC can characterize solutions to the Schrödinger equation, since
\begin{equation}
H |\Psi\rangle = E |\Psi\rangle
\quad \Leftrightarrow \quad
\left\lbrace
\begin{aligned}
\langle \Phi_0 | e^{-T} H e^{T} |\Phi_0\rangle &= E,\\
\langle \Phi_\mu | e^{-T} H e^{T} |\Phi_0\rangle &= 0,
\end{aligned}
\right.
\end{equation}
giving rise to a stationary (saddle-point) problem governed by the CC Lagrangian
\begin{equation}
\mathcal{L}(t,\lambda) 
=
\langle \Phi_0 | e^{-T} H e^{T} |\Phi_0\rangle
+ \sum_\mu \lambda_\mu \langle \Phi_\mu | e^{-T} H e^{T} |\Phi_0\rangle
=
\langle \Phi_0 | (I + \Lambda)e^{-T} H e^{T} |\Phi_0\rangle.
\end{equation}
Since the untruncated CC equations rapidly become numerically intractable, truncations are commonly employed. The subject of this work is the CCSD variant, where $T$ comprises excitation operators up to a maximum excitation rank of two, i.e., 
\begin{equation}
T = 
\sum_{i,a} t_i^a X_i^a
+ \frac{1}{4} \sum_{i,j,a,b}  t_{ij}^{ab} X_{ij}^{ab}. 
\end{equation}
However, with algorithmic scaling of $O^2 V^4\approx \mathcal{O}(N^6)$, even CCSD becomes computationally demanding when naively applied to larger systems. 
The MPCC framework,~\cite{Shee:2024:JCP} mitigates this cost by using orbital localization and embedding techniques. 
With the MPCC framework, the full orbital subspace is partitioned and different levels of theory are applied to each partition, reducing the overall computational cost.
The idea is that orbitals which are strongly correlated with chemically relevant degrees of freedom are treated with a high-level theory (i.e., CCSD) and weakly correlated orbitals are treated at a lower-level theory (i.e., perturbation theory). 
We denote the orbitals treated with a higher-level theory as fragment (F) orbitals, and those treated with a lower-level theory as environment (E) orbitals.

A partition of a given (localized) orbital set $\{\phi_i\}_{i=1}^K$ is specified by disjoint index sets that assign each occupied and virtual orbital to either the fragment or the environment. For an $N$-electron system, the occupied space is written as 
\begin{equation}
O= O_{\rm E} \cup O_{\rm F} = \{\phi_i:1\le i\le N\} ~{\rm where}~
O_{\rm E}=\{\phi_i:i\in\mathcal{I}_{\rm occ}^{\rm E}\}~{\rm and}~
O_{\rm F}=\{\phi_i:i\in\mathcal{I}_{\rm occ}^{\rm F}\},
\end{equation}
with $\mathcal{I}_{\rm occ}^{\rm E}\cap\mathcal{I}_{\rm occ}^{\rm F}=\emptyset$ and
$\mathcal{I}_{\rm occ}^{\rm E}\cup\mathcal{I}_{\rm occ}^{\rm F}=[\![N]\!]$.
The virtual space is partitioned analogously, i.e., 
\begin{equation}
V 
= V_{\rm E}\cup V_{\rm F} 
= \{ \phi_i : N < i < K \}~{\rm where}~
V_{\rm E}=\{\phi_i:i\in\mathcal{I}_{\rm vir}^{\rm E}\}~{\rm and}~
V_{\rm F}=\{\phi_i:i\in\mathcal{I}_{\rm vir}^{\rm F}\},
\end{equation}
with $\mathcal{I}_{\rm vir}^{\rm E}\cap\mathcal{I}_{\rm vir}^{\rm F}=\emptyset$ and $\mathcal{I}_{\rm vir}^{\rm E}\cup\mathcal{I}_{\rm vir}^{\rm F}=[\![K]\!] \setminus [\![N]\!].$
We denote the number of fragment and environment orbitals by $N_{\rm F}=|\mathcal{I}_{\rm occ}^{\rm F}|+|\mathcal{I}_{\rm vir}^{\rm F}|$ and $N_{\rm E}=|\mathcal{I}_{\rm occ}^{\rm E}|+|\mathcal{I}_{\rm vir}^{\rm E}|$, respectively. 
We slightly abuse notation by using $O$ and $V$ to denote both the sets of occupied and virtual orbitals, respectively, and their cardinalities. The intended meaning should be clear from context.

Similar to other embedding strategies, the choice of orbitals used to define the fragment space is crucial for achieving the desired accuracy. 
One method for constructing this orbital fragmentation is via the active valence active space (AVAS) protocol,\cite{sayfutyarova2017automated} which offers an automated and chemically motivated fragmentation procedure. 
The AVAS method identifies a set of orbitals associated with a chosen valence manifold.
Starting from a compact reference description of the region of interest, AVAS determines the combination of orbitals that most strongly overlaps with this reference.
A minimal atomic orbital basis (MINAO) is often used for this purpose because it naturally reflects valence bonding/antibonding character, while the full electronic-structure calculation may employ a larger basis set. 
An additional practical advantage of AVAS is that higher angular momentum functions can be incorporated into the active space in a straightforward way by selecting a slightly enlarged reference basis. 
This provides a systematic handle to expand the fragment space and, in turn, to improve the description of dynamical correlation within the active subspace.

Given a valid orbital partition, we adopt the standard coupled-cluster ansatz by defining the fragment cluster operator and environment cluster operator as
\begin{equation}
T_n^{\rm F}=
\sum_{\substack{i_1,\ldots,i_n\in \mathcal I_{\rm occ}^{\rm F}\\
a_1,\ldots,a_n\in \mathcal I_{\rm vir}^{\rm F}}}
t_{i_1\cdots i_n}^{a_1\cdots a_n}\,
a_{a_1}^\dagger\cdots a_{a_n}^\dagger a_{i_n}\cdots a_{i_1},
\quad{\rm and}\quad
T_n^{\rm E}=T_n-T_n^{\rm F}.
\end{equation}
We denote by ${\bf t}^{\rm F}$ and ${\bf t}^{\rm E}$ the set of amplitudes tensors associated with the $T^{\rm F}$ and $T^{\rm E}$ operators, respectively.
By construction, $T^{\rm F}$ contains only those excitation operators that act entirely within the fragment orbital subspace. 
In contrast, $T^{\rm E}$ contains both (i) excitations that involve only environment orbitals and (ii) mixed fragment--environment excitations. The mixed terms encode correlation effects that connect the two subsystems, such as excitations from fragment occupied orbitals into environment virtual orbitals, or simultaneous excitations spanning both regions.

In a quantum-embedding spirit, the fragment and environment cluster operators are evaluated at different levels of theory. 
In particular, $T^{\rm F}$ is obtained from a conventional coupled-cluster treatment (the high-level, HL, component), while $T^{\rm E}$ is approximated using a perturbative (low-level, LL) description. Equivalently, the similarity-transformed Hamiltonian entering the projective equations is handled at different accuracy in the two sectors. For the fragment, we retain the full similarity transformation,
\begin{equation}
\overline{H}({\bf t}^{\rm F}; {\bf t}^{\rm E})
= e^{-T^{\rm E}-T^{\rm F}} H e^{T^{\rm F}+T^{\rm E}},
\end{equation}
whereas for the environment we employ a reduced, perturbative approximation denoted $\widetilde{H}({\bf t}^{\rm E}; {\bf t}^{\rm F})$ (vide infra).
This construction yields the coupled set of projective equations
\begin{align}
     \langle \Phi_\mu^{\rm F}| \overline{H}^{\rm F}({\bf t}^{\rm F}; {\bf t}^{\rm E})| \Phi_0 \rangle = {}& 0, \label{Eq:projF} \\
     \langle \Phi_\mu^{\rm E}| \widetilde{H}^{\rm E}({\bf t}^{\rm E}; {\bf t}^{\rm F}) | \Phi_0 \rangle = {}& 0, \label{Eq:projEB}
\end{align}
where $|\Phi_\mu^{\rm F}\rangle$ and $|\Phi_\mu^{\rm E}\rangle$ denote excited Slater determinants in the fragment and environment spaces, respectively.
\Cref{Eq:projF,Eq:projEB} result from stationarity of the MPCC Lagrangian,
\begin{equation}
\label{Eq:LagrangianGen}
\begin{aligned}
\mathcal{L}({\bf t},\boldsymbol{\lambda}) 
= \langle \Phi_0 | \overline{H} | \Phi_0 \rangle + \langle \Phi_0 | \Lambda^{\rm F} \overline{H}^{\rm F}({\bf t}^{\rm F}; {\bf t}^{\rm E}) | \Phi_0 \rangle + \langle \Phi_0 | \Lambda^{\rm E} \widetilde{H}^{\rm E}({\bf t}^{\rm E}; {\bf t}^{\rm F}) | \Phi_0 \rangle, 
\end{aligned}    
\end{equation}
where $\Lambda^{Y}=\sum_{\mu\in Y}\lambda_\mu^{Y} X_\mu^{Y}|$ for $Y\in\{{\rm F},{\rm E}\}$, and the first term is the coupled-cluster energy for the full (fragment + environment) system, hence, the unlabeled $\overline{H}$. Note that both $\overline{H}^{\rm F}$ and $\widetilde{H}^{\rm E}$ in Eqs.~\eqref{Eq:projF} and~\eqref{Eq:projEB} depend on the complete set of amplitudes, i.e., ${\bf t}^{\rm F}$ \emph{and} ${\bf t}^{\rm E}$. In practice, the fragment equations are solved while holding ${\bf t}^{\rm E}$ fixed (updating ${\bf t}^{\rm F}$), and the environment equations are solved while holding ${\bf t}^{\rm F}$ fixed (updating ${\bf t}^{\rm E}$). The two problems are therefore coupled and are converged self-consistently via a nested iteration scheme of \textit{macro}- and \textit{micro}-iterations.

During a \textit{macro}-step, we first solve the LL equations to obtain the environment cluster operator $T^{\rm E}$. We then build an effective interaction for the fragment by similarity-transforming the bare Hamiltonian with $T^{\rm E}$, i.e.,
\begin{equation}
\label{eq:WF}
    W^{\rm F} = e^{-T^{\rm E}} H e^{T^{\rm E}},
\end{equation}
and subsequently restricting all indices to the fragment subspace. Although $H$ is a two-body operator, this procedure generally generates higher particle-rank (effective many-body) terms in $W^{\rm F}$ (e.g., effective three-body contributions and beyond).
Within each \textit{micro}-cycle, we solve the fragment amplitude equations using $W^{\rm F}$. The resulting ${\bf t}^{\rm F}$ then enters the LL environment problem and updates the next \textit{macro}-iteration. The operator $W^{\rm F}$ is often referred to as a (static) downfolded Hamiltonian: it isolates an effective fragment problem with reduced degrees of freedom, which can then be treated with a higher-accuracy solver. Related static constructions have been explored by Kowalski \textit{et al.} in both unitary \cite{metcalf2020resource} and non-unitary \cite{kowalski2023sub} forms, and by Evangelista \textit{et al.} \cite{huang2023leveraging} using the driven similarity renormalization group (DSRG) framework. In contrast to our embedding approach, these studies typically solve the downfolded Hamiltonian as a standalone problem rather than enforcing self-consistency with an explicit environment update.


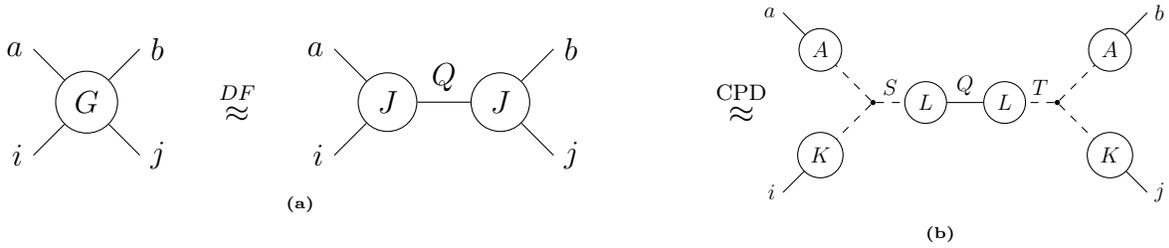
\begin{figure}[h]
\begin{tabular}{cc}
\newcommand*{\Scale}[2][4]{\scalebox{#1}{$#2$}}%

\begin{subfigure}[c]{0.49\textwidth}
\begin{tikzpicture}[scale=1, transform shape]
\node[circle, draw, minimum size=0.5cm] (g) at (0,0) {$G$};
\draw (g) -- ++(135:1) node[left] {$a$};
\draw (g) -- ++(45:1)  node[right] {$b$};
\draw (g) -- ++(-135:1) node[left] {$i$};
\draw (g) -- ++(-45:1)  node[right] {$j$};

\node at (2,0) {$\overset{DF}{\approx}$};

\node[circle, draw, minimum size=0.5cm] (J1) at (4,0) {$J$};
\node[circle, draw, minimum size=0.5cm] (J2) at (5.5,0) {$J$};

\draw (J1) -- ++(135:1) node[left] {$a$};
\draw (J1) -- ++(-135:1) node[left] {$i$};
\draw (J2) -- ++(45:1)  node[right] {$b$};
\draw (J2) -- ++(-45:1) node[right] {$j$};

\draw (J1) -- node[above] {$Q$} (J2);
\end{tikzpicture}
\caption{}
\label{fig:1a}
\end{subfigure}
&

\begin{subfigure}[c]{0.49\textwidth}
\centering
\begin{tikzpicture}[scale=0.7, transform shape]

\node at (-2,0) {\scalebox{1.5}{$\overset{\mathrm{CPD}}{\approx}$}};

\node[circle, draw, minimum size=0.5cm] (X1) at (-0.5,1) {$A$};
\node[circle, draw, minimum size=0.5cm] (X2) at (5,1) {$A$};
\node[circle, draw, minimum size=0.5cm] (Y1) at (1.5,0) {$L$};
\node[circle, draw, minimum size=0.5cm] (Y2) at (3,0) {$L$};
\node[circle, draw, minimum size=0.5cm] (X3) at (-0.5,-1) {$K$};
\node[circle, draw, minimum size=0.5cm] (X4) at (5,-1) {$K$};

\node [circle, fill=black, inner sep=1pt] (J6) at (0.5,0) {};
\node [circle, fill=black, inner sep=1pt] (J7) at (4,0) {};

\draw (X1) -- ++(135:1) node[left] {$a$};
\draw (X3) -- ++(-135:1) node[left] {$i$};
\draw (X2) -- ++(45:1)  node[right] {$b$};
\draw (X4) -- ++(-45:1) node[right] {$j$};

\draw[dashed] (J6) -- (X1);
\draw[dashed] (J6) -- (X3);
\draw[dashed] (J6) -- node[above] {$S$} (Y1);
\draw[dashed] (J7) -- (X2);
\draw[dashed] (J7) -- (X4);
\draw[dashed] (J7) -- node[above] {$T$} (Y2);
\draw (Y1) -- node[above] {$Q$} (Y2);
\end{tikzpicture}
\caption{}
\label{fig:1b}
\end{subfigure}






\end{tabular}
\caption{(a) Graphical representation of the four-index tensor $G_{ij}^{ab}$ decomposed using the DF approximation. 
(b) Representation of a CPD approximation TEI tensor where the CPD is applied to each DF TEI tensor.}
\label{fig:tensor_appx}
\end{figure}
\subsection{Tensor Decomposition Review}

In this section, we review the canonical polyadic decomposition (CPD) and outline how it is incorporated into the MPCC formalism. Broadly, tensor decompositions aim to expose (often low-rank) structure in higher-order tensors by expressing them in terms of interconnected lower-order tensors. Such factorizations can be used to reconstruct the original tensor, but more commonly, they are employed directly within computational algorithms as a surrogate representation. Rewriting an algorithm in terms of the decomposed tensors can substantially reduce both storage requirements and computational cost. In the present work, we make use of the density-fitting approximation\cite{VRG:whitten:1973:JCP,VRG:dunlap:1979:JCP,Vahtras:1993:CPL,VRG:jung:2005:PNAS,VRG:mintmire:1982:PRA} and CPD.\cite{VRG:carroll:1970:P,VRG:Harshman:1970:WPP}

\subsubsection{The Density Fitting Approximation}
The density fitting (DF) approximation of the two-electron integral (TEI) tensor is a standard tensor decomposition used in electronic structure methods.
Elements of the TEI tensor can be expressed in a basis of single particle function functions $\{ \phi_p \}_{p=1}^N$ as
\begin{align}\label{eq:g}
    G^{pq}_{st} = \iint \phi^*_{p}(r_1) \phi^*_q(r_2) g(r_1, r_2) \phi_s(r_1) \phi_t(r_2) dr_1 dr_2.
\end{align}
where $g(r_1, r_2)$ may be any positive kernel.
For this work, we restrict our scope to the Coulomb interaction kernel $g(r_1, r_2) \equiv \Vert r_1 - r_2 \Vert_2^{-1}$.
The DF approximation decomposes the order-$4$ TEI tensor into the following low-rank representation
\begin{align}\label{eq:SQG}
    G^{pq}_{st} \overset{\mathrm{DF}}{\approx} \sum_{Q=1}^X J^{Q}_{ps} J^{Q}_{qt}
\end{align}
where $X$ represents an optimized, predetermined auxiliary basis set which grows linearly with system size, i.e., typically between $2$ to $3$ times the dimension $V$.
We denote the tensor $J$ as the DF TEI tensor.
It should be noted that the DF approximation can also be constructed using the Cholesky decomposition\cite{VRG:beebe:1977:IJQC,VRG:lowdin:1965:JMP,Lowdin:2009:IJQC,Folkestad:2019:JCP} or the related chain-of-spheres (COSX) method.\cite{VRG:izsak:2011:JCP,Izsak:2012:MP,VRG:dutta:2016:JCP,VRG:izsak:2013:JCP,VRG:neese:2009:CP,Kossmann:2010:JCTC,Kossmann:2009:CPL}
A pictorial representation of the DF approximation can be found in \cref{fig:1a}.

The DF approximation formally reduces the computational storage complexity of the TEI tensor from $\mathcal{O}(N^4)$ to $\mathcal{O}(N^3)$. However, notice that the DF approximation requires that the two indices associated with a given particle are represented using the {\it same} three-center DF tensor. This structural restriction is well known to limit the ability of the DF approximation to reduce the computational complexity of many high-scaling electronic structure methods. 
However, methods which introduce the canonical polyadic decomposition (CPD) of the three-center TEI tensor using an analytic decomposition,\cite{Pierce:2025:JCTC:MP2,VRG:pierce:2021:JCTC,Pierce:2022:ETD,Pierce:2025:JCTC:CPB,VRG:benedikt:2011:JCP,Benedikt:2013:JCP,Benedikt:2013:MP,Bohm:2016:JCP,Schmitz:2017:JCP,Khoromskaia:2015:PCCP,Madsen:2018:JCP}, the pseudospectral \cite{VRG:friesner:1985:CPL,Friesner:1986:JCP,Langlois:1990:JCP,VRG:ringnalda:1990:JCP,Friesner:1991:ARPC,Martinez:1995:JCP,VRG:martinez:1992:JCP,Martinez:1994:JCP,Ko:2008:JCP,Martinez:1993:JCP} method, and the tensor hypercontraction\cite{VRG:hohenstein:2012:JCP,VRG:hohenstein:2012:JCPa,VRG:parrish:2012:JCP,Hohenstein:2013:JCP,VRG:parrish:2014:JCP,Shenvi:2013:JCP,Schutski:2017:JCP,Parrish:2019:JCP,Lee:2019:JCTC,VRG:hummel:2017:JCP,VRG:song::JCP,Hohenstein:2019:JCP,Hohenstein:2021:JCP,Hohenstein:2022:JCP,Jiang:2022:JCTC,Zhao:2023:JCTC,Datar:2024:JCTC,Schmitz:2017:JCP, Khoromskaia:2015:PCCP} (THC) approaches have shown to successfully reduce the computational complexity of many high-scaling electronic structure methods.

\subsubsection{The Canonical Polyadic Decomposition}
The CPD is a tensor decomposition which maps an order-$N$ tensor into a sum of $R$ rank-$1$ tensor where a rank-$1$ tensor is defined as the outer product of $N$ vectors.
By collecting the set of vectors that span one mode of the original tensor, we can construct the factor matrix representation of the CPD.
For example, the CPD of a DF TEI tensor can be written as
\begin{align}
    J^{Q}_{ai} = \sum_S^R A_{aS} K_{iS} L_{QS}
\end{align}
where the matrices $A$, $K$ and $L$ are the so called CP factor matrices.
For reference, we will denote the CP rank index as $S$ and $T$ and refer to the dimension of the index with $R$.
Decomposing both 3-center DF TEI tensors in \cref{eq:SQG} results in the network shown in \cref{fig:tensor_appx}b.
The effectiveness of the CPD depends explicitly on the CP rank\cite{Hastad:1990:algorithm, VRG:hillar:2013:JA} and,
unfortunately, there exists no closed form algorithm to determine this rank.
Therefore, the value is revealed by constructing multiple rank-$R$ CPD approximations and choosing the value that satisfies a problem's predetermined accuracy thresholds.

\section{The Low-Level Problem}

The flexibility associated with lower-order expansion in $\widetilde{H}^{\rm E}$ provides us with various options to gain a computational advantages, as was elaborated in an earlier work.\cite{Shee2024}
In the following, we will briefly explain the most useful approximation, the relaxed scheme, which leads to both qualitative and quantitative accuracy. 
From this scheme, it has been recognized that the orbital relaxation effect of the environment are vitally important and, according to the Thouless theorem, \cite{thouless1960stability} the $e^{T_1}$ component of the CC ansatz can capture this effect. 
Therefore, a perturbation theory was defined in terms of the $e^{T_1}$ transformed Hamiltonian, i.e.,
\begin{equation}
\label{Eq:mp_partition}
\widetilde{H} = e^{-T_1} H e^{T_1} = E_{\mathrm{cl}}\,I + \tilde{F} + \tilde{V} ,
\end{equation}
where $E_{\mathrm{cl}}$ is the scalar (zero-body) contribution. Following an MP-type partitioning, we take $\tilde{F}$ as the zeroth-order part and $\tilde{V}$ as the first-order (fluctuation) contribution.
We then define the first-order amplitude equations for the singles and doubles amplitudes as
\begin{align}
    \langle \Phi_\mu^{\rm E} | \tilde{F} + [\tilde{F}, T_2] + [\tilde{F}, T_1] | \Phi_0 \rangle &= 0,~\forall  \mu = {\binom{a}{i}}  ,\\
    \langle \Phi_\mu^{\rm E} | \tilde{V} + [\tilde{F}, T_2] | \Phi_0 \rangle  &= 0,~\forall \mu = {\binom{ab}{ij}}.
\end{align}
%
In the MPCC framework, the high-level (HL) component restricts the most accurate, and most expensive, solver to a comparatively small subsystem, leaving the majority of the system to the the low-level (LL) solver. 
Therefore, devising methods to reduce the scaling of the LL treatment without compromising accuracy is central to MPCC. Given its close resemblance to MP2 theory and the proven success of tensor decomposition strategies in that context, including DF and THC, some of the authors have recently introduced an MPCC implementation that leverages the DF approximation in the environment treatment.\cite{shee2026towards} In this work, we extend that effort by introducing an additional compression of the resulting order-three DF integral tensors using the CPD.

\subsection{A CPD enhanced Low-Level Solver} 

Incorporating DF into the MPCC LL solver described above yields an algorithm with $\mathcal{O}(N^4)$ computational cost and $\mathcal{O}(N^3)$ storage complexity.\cite{shee2026towards}
Although DF improves the scaling of the LL solver (especially compared to the cost of CCSD), computational memory requirements will become the limiting factor for large systems with extended environments. In particular, the storage of the order-3 DF integral tensors quickly becomes unmanageably large. As a step towards reducing the computational storage complexity of the modified DF-MPCC algorithm, we introduce the CPD approximation of the environment's order-3 DF TEI tensors, specifically
\begin{equation}
\label{eq:cpd_J}
J_{ij}^Q \approx \sum_S^{R_{oo}} I_{iS} I_{jS} M_{QS}, \quad
J_{ab}^Q \approx \sum_S^{R_{vv}} C_{aS} C_{bS} V_{QS}, \quad 
J_{ia}^Q \approx \sum_S^{R_{ov}} K_{iS} A_{aS} L_{QS}.
\end{equation}
We assume that each DF integral tensor is decomposed separately and, therefore, the factor matrices of the same orbital index may differ across the different approximations, i.e., $I_{is} \neq K_{is}$, $C_{as} \neq A_{as}$ and $M_{QS} \neq L_{QS}$.
Also, we allow the CP rank to be different for each approximated tensor; the dimension of each of these CP ranks will be discussed in \cref{sec:comp_det}.
In this CPD-based implementation, we reformulate the DF-LL solver to introduce no order-3 intermediate tensors. 
In the SM readers may find the complete CP-DF-LL-solver algorithm; here we focus on evaluating the computational cost and memory usage in the novel algorithm. We follow the set-enumeration convention of the table in the SM.

\paragraph{1. Compute Intermediate:}
While this procedure avoids the explicit formation of the order-3 tensor in this step, we compute the intermediate $\hat{X}^Q\approx X^Q$ for convenience via
\begin{align}
\label{intermediate}
    \hat{X}^Q = 2\sum_{S} L_{QS} \,\Big[\,\sum_{i}K_{iS} \,\Big(\,\sum_{a} A_{aS}t^a_i\,\Big)\,\Big].
\end{align}
Following the order of contractions indicated by the parentheses, the innermost
summation has a cost of
$O V R \approx \mathcal{O}(N^2 R)$.
The subsequent contraction is $\sum_i K_{iS} A_{iS}$ is general tensor product which effectively computes the dot product $\sum_i K_iA_i$ for vectors of $K$ and $A$ associated with columns of matching $S$. 
This product can be done efficiently with specialized kernels that leverage the stride option of BLAS functions with a cost of $O R \approx \mathcal{O}(NR)$.
The final contraction has a cost of
$XR \approx \mathcal{O}(NR)$. 
In \cref{sec:results}, we demonstrate that the CP ranks in Eq.~\eqref{eq:cpd_J} grows linearly with system size, therefore the computational scaling of this intermediate is now $\mathcal{O}(N^3)$.

\paragraph{2. Compute $\bar{F} \approx \tilde F$:}
Next, we reformulate the tensor $\bar{F} \approx \tilde{F}$ tensors. We begin with the occupied-occupied block
\begin{equation}
\label{eq:cpd_Fij}
\begin{aligned}
\bar{F}_{i j}
&= F_{i j} 
+ \sum_{S} I_{iS} \,\Big(\, I_{jS}\,\big(\, \sum_{Q} M_{QS} \hat{X}^{Q} \,\big)\, \Big) \\
&\quad- \sum_{S} I_{iS} 
   \Big(\,\sum_T\Big[\,
   \sum_{Q} M_{QS} L_{QT}
   \big(\, \sum_{k} I_{kS} K_{kT} \,\big)\, \Big]
    \Big[\sum_{a} A_{aT} t^a_j\Big]
   \Big).
\end{aligned}
\end{equation}
In the second term of \cref{eq:cpd_Fij} the innermost contraction is straightforward and gives the intermediate $M_S$ with a cost of $X R\approx \mathcal{O}(N^2)$.
Next we have the general tensor product $\bar{I}_{jS} = I_{jS} M_{S}$, which scales the vector $I_j$ by a value $m$ for each matching $S$ column. 
This tensor product has a cost of $OR \approx \mathcal{O}(N^2)$.
Finally we compute $\sum_S I_{iS} \bar{I}_{jS}$ which has a cost of $O^2 R \approx \mathcal{O}(N^3)$. In the third term of \cref{eq:cpd_Fij}, we first contract the three inner-most sums, resulting in the intermediates
\begin{align}
    \bar{M}_{ST} = \sum_{Q} M_{QS} L_{QT},\quad
    \bar{I}_{ST} = \sum_{k} I_{kS} K_{kT},\quad
    \bar{A}_{jT} = \sum_{a} A_{aT} t^a_j.
\end{align}
The cost of these contractions is $X R^2\approx \mathcal{O}(N^3)$, $ OR^2 \approx \mathcal{O}(N^3)$, and $O V R \approx \mathcal{O}(N^3)$, respectively. From here, we compute the Hadamard tensor product $[\bar{M} \odot \bar{I}]_{ST} = \bar{M}_{ST}\bar{I}_{ST}$ with a cost of $R^2\approx \mathcal{O}(N^2)$.
Finally we contract $\sum_{ST} I_{iS} [\bar{M} \odot \bar{I}]_{ST} \bar{A}_{jT}$
with a cost of $ O R^2 + O^2R \approx \mathcal{O}(N^3)$.
With this formulation, the construction of $\bar{F}_{ij}$ scales as $\mathcal{O}(N^3)$.

Next, we approximate the virtual-virtual block $\tilde{F}_{a b} \approx \bar{F}_{ab}$ via
\begin{equation}
\begin{aligned}
\label{eq:cpd_Fab}
\bar{F}_{a b} 
    &= F_{a b} 
      + \sum_{S} C_{aS} C_{bS} \,\big(\, \sum_Q V_{QS} \hat{X}^Q \,\big) \\ 
    &\quad - \sum_{S} A_{aS}  \Big[\sum_{T} C_{bT} 
      \Big( \sum_Q L_{QS} V_{QT}\,\Big)\,
      \Big( \sum_{k} K_{kS} \big(\sum_{c} C_{cT} t^c_k\big) \Big)
      \Big],
\end{aligned}
\end{equation}
The second term in \cref{eq:cpd_Fab} can be evaluated in the same way as the second term in \cref{eq:cpd_Fij}, resulting in a $\mathcal{O}(N^3)$ computational scaling.
For the third term in \cref{eq:cpd_Fab}, we first contract over the $c$ index creating the intermediate $C_{kT} = \sum_{c} C_{cT} t^c_k$ with a cost of $V O R \approx \mathcal{O}(N^3)$.
Next, we contract over the $k$ index as $[KC]_{ST} = \sum_{k} K_{kS} C_{kT}$ with a cost of $ O R^2 \approx \mathcal{O}(N^3)$. 
With this intermediate, we compute the Hadamard product $L_{ST} = [NV]_{ST} [KC]_{ST}$
with a cost of $R^2 \approx \mathcal{O}(N^2)$ and where $[NV]_{ST} = \sum_{Q} L_{QS} V_{QT}$ with a cost of $X R^2 \approx \mathcal{O}(N^3)$.
Finally, we are left with the contraction
$\sum_{ST} A_{aS} L_{ST} C_{bT}$
which, similar to $\bar{F}_{ij}$, can be evaluated with a cost of $VR^2 + V^2 R \approx \mathcal{O}(N^3)$.
Using this approximation, computing $\bar{F}_{ab}$ scales as $\mathcal{O}(N^3)$ with linear scaling CP rank.

Lastly we approximate the occupied-virtual block $\tilde{F}_{jb} \approx \bar{F}_{jb}$ via
\begin{equation}
\label{eq:cpd_Fjb}
\begin{aligned}
\bar{F}_{j b}
    &=
    F_{jb} + \sum_{S} A_{bS} \Big(K_{jS} \big( \sum_{Q} L_{QS} \big) \hat{X}^Q\Big) \\ 
    &\quad - \sum_{T}A_{bT} \Big[\, \sum_{S} K_{jS} 
        \Big( \sum_Q L_{QS} L_{QT}\Big) 
        \Big( \sum_i K_{iT} \big( \sum_{a} A_{aS} t^a_i\big) \Big)\Big]
\end{aligned}
\end{equation}
These contractions can be evaluated in a very similar manner to those in \cref{eq:cpd_Fab}.
Therefore, the cost of evaluating the second term in \cref{eq:cpd_Fjb} is $X R + O R + V O R \approx \mathcal{O}(N^3)$ and the cost of evaluating the third term is $2 V OR + 2 O R^2 + X R^2 + R^2 \approx \mathcal{O}(N^3)$, by following the order laid out by the parenthesis.

\paragraph{3. Compute $\Omega$:} Next, we approximate the $T_1$ surrogate $\Omega_{ai}  \approx \bar\Omega_{ai} $
\begin{equation}
\label{cpd_om}
\begin{aligned}
\bar \Omega_{ai} 
    &= F_{ai} 
        + \sum_{S} A_{aS} \Big(\,I_{iS} \big(\,\sum_{Q} L_{QS} \hat{X}^Q\, \big)\,\Big) \\ 
        &\quad - \sum_{S} C_{aS} \Big[\, \sum_{T} I_{iT} \Big(\, \sum_Q V_{QS} M_{QT}\, \Big) \Big(\, \sum_jI_{jT} \big(\,\sum_{b} C_{bS} t^b_j\big)\,\Big)\Big]
\end{aligned}
\end{equation}
Interestingly, the evaluation of the terms in \cref{cpd_om} can be evaluated in the same way as the terms in \cref{eq:cpd_Fab,eq:cpd_Fjb}. Therefore, the scaling of this term is, also, $\mathcal{O}(N^3)$.

\paragraph{6. Computation of $T_2$-factors:}
The computation of $Y^{Q\alpha}_{bj}$, unfortunately, can not be directly reduced below $\mathcal{O}(N^4)$ because of the specific structure of the contractions. 
However, it is possible to devise a strategy which generates intermediates which require less than $\mathcal{O}(N^3)$ storage using either the DF or CPD approximation. 
Because this step must only be computed once at the conclusion of the LL optimization procedure and does not explicitly require the utilization of CPD approximated integral tensors, we derive the reduced-scaling algorithm in the SM.

\section{Computational details}\label{sec:comp_det}
Calculations were performed using the PySCF package~\cite{sun2015libcint,sun2018pyscf,sun2020recent} (version 2.9.0) on a MacBook Air equipped with an Apple M3 processor, comprising 4 performance and 4 efficiency cores.
To assess the impact of the CPD approximation within this workflow, we emulate its use by replacing the exact DF-TEI tensors with tensors reconstructed from their CPD representation. We defer a fully optimized implementation to future work.
Accordingly, the aim of this study is twofold: (1) to present a reduced-scaling DF-LL framework, and (2) to benchmark the accuracy of the CPD approximation both in the resulting DF-LL formulation and within the full MPCC optimization loop.

We primarily investigate two classes of molecular systems that represent common local environments in solvated systems and hydrocarbon-rich compounds: (i) water clusters containing 1–6 water molecules in TIP4P-optimized geometries,\cite{jorgensen:1983:JCP,Wales:1998:CPL} and (ii) linear alkane chains containing 1–6 carbon atoms. These test sets are widely used, probe distinct interaction regimes, and are sufficiently small to enable systematic benchmarks of both accuracy and computational scaling. 
As a proof of principle for the simulation of solvent systems, a class of systems which are typically targeted by embedding calculations, we consider a system of a methane molecule embedded in a cluster of 4 water molecules ($\ce{CH_4} \cdots \ce{H_2O}$).
For all numerical experiments, we employ the cc-pVTZ (TZ) orbital basis sets (OBS) with the corresponding cc-pVTZ-RI (TZ-RI) density-fitting basis set (DFBS).~\cite{Dunning1989,Kendall1992} 
In \cref{sec:results}, we primarily present results for the $\ce{(H2O)6}$ water cluster using the TZ/TZ-RI basis. Corresponding results obtained with the cc-pVDZ (DZ) basis and its associated cc-pVDZ-RI (DZ-RI) density-fitting set,\cite{Dunning:1989:JCP,VRG:weigend:2002:JCP,Hattig:2005:PCCP} as well as results for $\ce{C6H14}$ computed in the TZ/TZ-RI and DZ/DZ-RI bases, are provided in the Supporting Materials (SM).

We optimize the CPD approximation of each three-center integral tensor
separately via a standard alternating least squares (ALS) algorithm.\cite{VRG:kroonenberg:1980:P,VRG:beylkin:2002:PNAS} 
Factor matrices are initialized following the procedure outlined in Ref.~\citenum{VRG:pierce:2021:JCTC} with column vectors drawn independently from a uniform distribution on $[-1, 1]$. As a means to simplify the analysis of presented results, unless otherwise noted, we fix the CP rank for two of the DF-TEI tensors. For the three-center integral decompositions, we choose CP ranks proportional to the auxiliary basis dimension $X$: for $J_{ij}^Q$ we set $R_{oo}=X$; for $J_{ai}^Q$ we set 
$R_{ov}=2X$ in the TZ/TZ-RI calculations; and for $J_{ab}^Q$, which is typically more challenging to compress with CPD,~\cite{Pierce:2022:ETD} we vary $R_{vv}$ over the range $1.5X$ to $3.5X$. 

We construct the fragment space using the AVAS procedure. The considered fragments include all atoms in the molecular system, so that the embedding partition primarily targets to separate a chemically motivated valence manifold from the remaining orbital space rather than selecting a spatial subset of atoms. The AVAS reference set is defined elementwise from the valence shell of each unique atomic species present in the molecule and is constructed using a minimal atomic orbital basis. Specifically, for first row elements (H and He), the reference includes the 1s manifold, for second row main group elements (Li through Ne), the 2s and 2p manifolds, and for third row main group elements (Na through Ar), the 3s and 3p manifolds. In all cases, the minimal atomic orbital basis is taken to be STO-3G, which serves as the MINAO reference for the AVAS procedure.

\section{Results and Discussion}\label{sec:results}
In this section, we investigate the impact of the CPD approximated DF TEI integral tensors on the convergence and accuracy of the low-level (LL) solver.
Because the tensor quantities computed in the LL solver impact the optimization of the high-level (HL) equation through the Lagrangian framework, we also provide an analysis for the convergence of the HL method and the MPCC procedure as a whole.

\subsection{The CP-DF-LL Solver Convergence}
\label{sec:ll_conv}

First, we study the convergence behavior of the CP-DF-LL method compared to the DF-LL method. Since the LL solver requires an initialization from a second-order correction to the HF state (a result of Brillouin's theorem), the results reported here are taken starting from the second macro-iteration of the MPCC optimization. 

\begin{figure}[htbp]
    \centering
    \begin{subfigure}{0.49\textwidth}
        \centering
        \includegraphics[width=\linewidth]{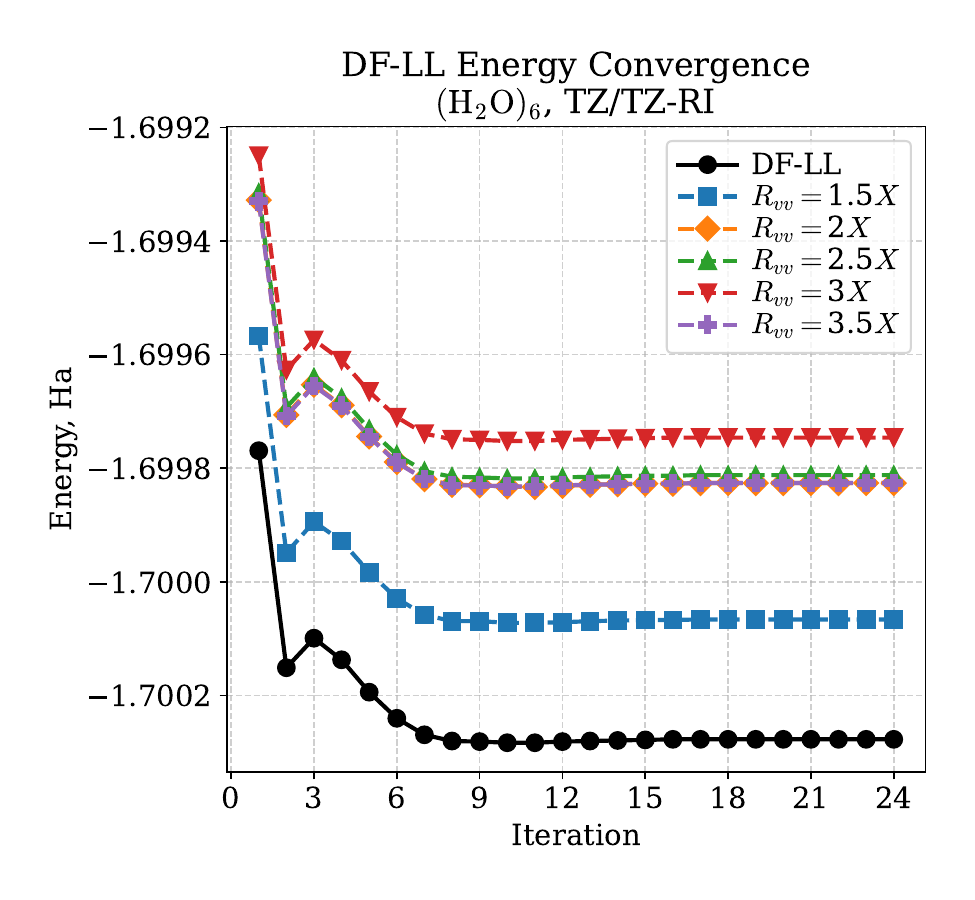}
        \caption{}
        \label{fig:cc2_cgs_a}
    \end{subfigure}
    \begin{subfigure}{0.49\textwidth}
        \centering
        \includegraphics[width=\linewidth]{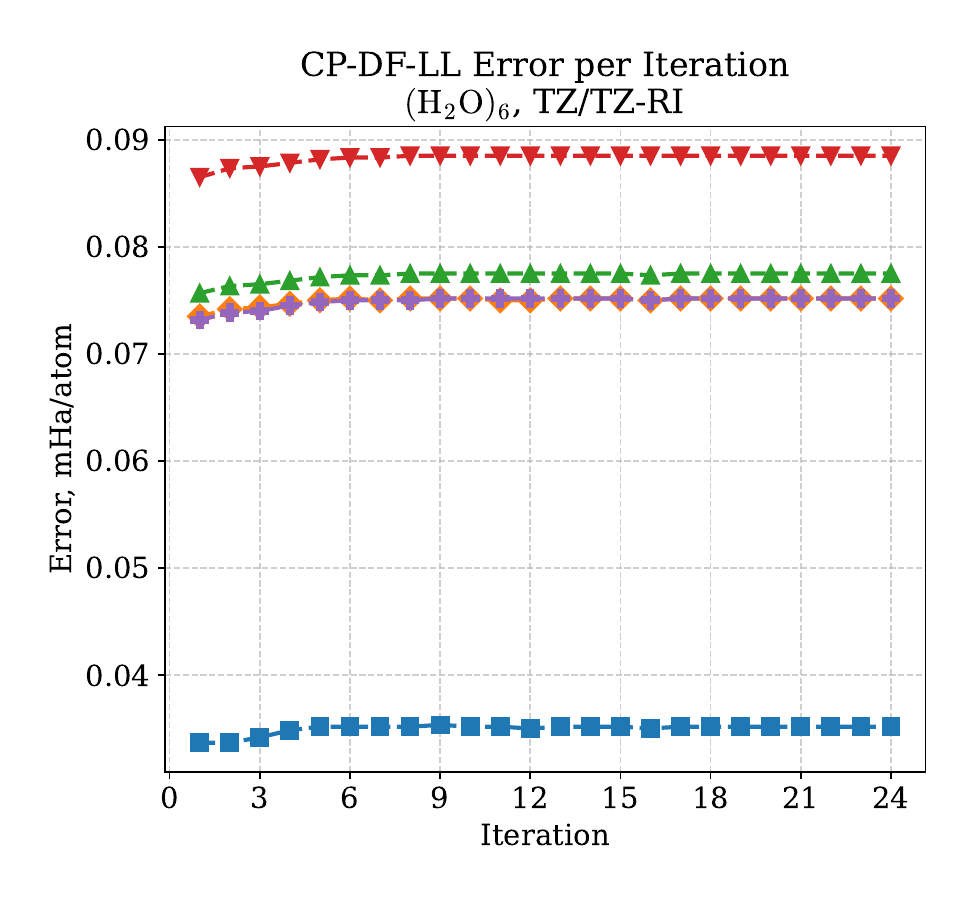}
        \caption{}
        \label{fig:cc2_cgs_b}
    \end{subfigure}
    \caption{(a) LL energy and (b) LL energy error per non-hydrogen atom, both reported as a function of LL iteration, for a 6-water cluster in the TZ/TZ-RI basis.}
    \label{fig:cc2_cgs}
\end{figure}

\cref{fig:cc2_cgs_a} considers the LL convergence behavior for a $\ce{({H}_2{O})_6}$ cluster in the TZ/TZ-RI basis. Note that the LL energy is used only as a proxy for the CPD error and is not computed explicitly in practical MPCC calculations. \Cref{fig:cc2_cgs_a} shows that replacing the DF-TEI integrals with their CPD approximation preserves the qualitative convergence behavior of the LL solver, while introducing an energy shift with a maximum deviation of $\delta E \approx 5\times10^{-4}$~Ha across the CP ranks tested. This trend is consistent across all systems and basis sets considered, see the SM for additional results and numerical details.
\Cref{fig:cc2_cgs_b} shows the difference between the CP-DF-LL energy and the DF-LL energy per LL-micro iteration. The discrepancy remains well-behaved throughout the optimization process, further indicating that the LL solver is robust with respect to CPD approximations in the DF-TEI tensors.

\begin{figure}[ht!]
    \begin{subfigure}{0.49\textwidth}
        \centering
        \includegraphics[width=\linewidth]{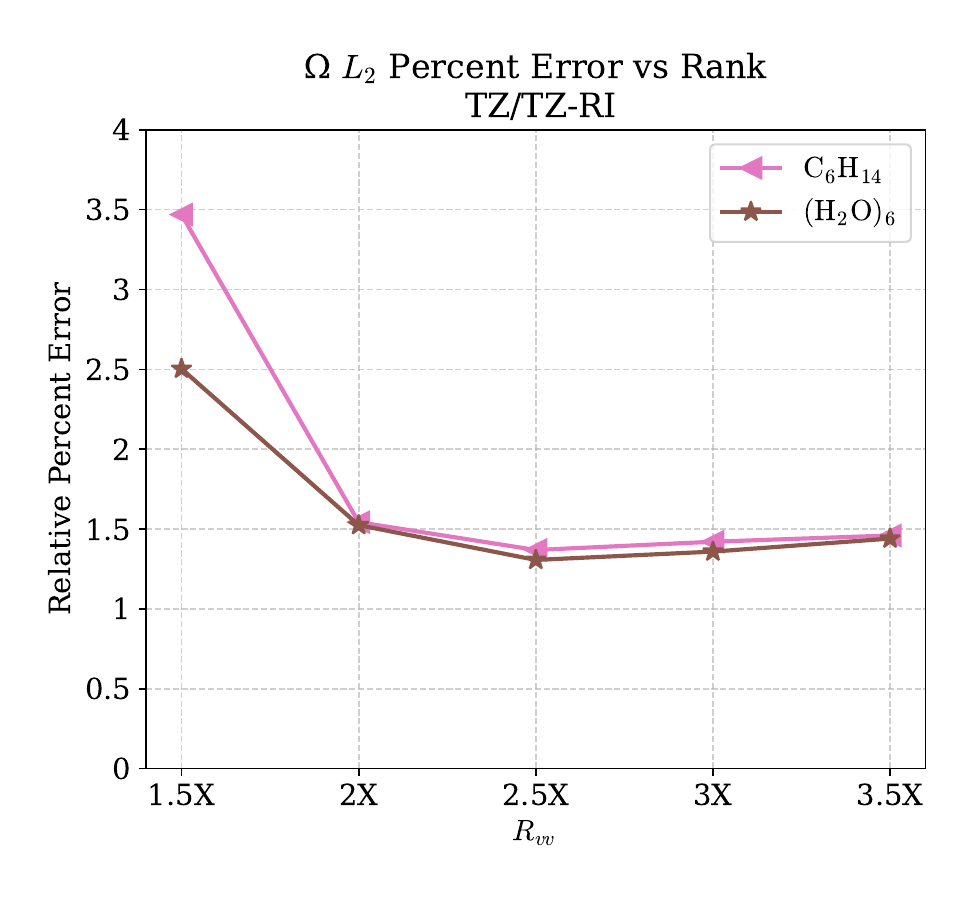}
        \caption{}
        \label{fig:omega_subplots_a}
    \end{subfigure}
    \begin{subfigure}{0.49\textwidth}
        \centering
        \includegraphics[width=\linewidth]{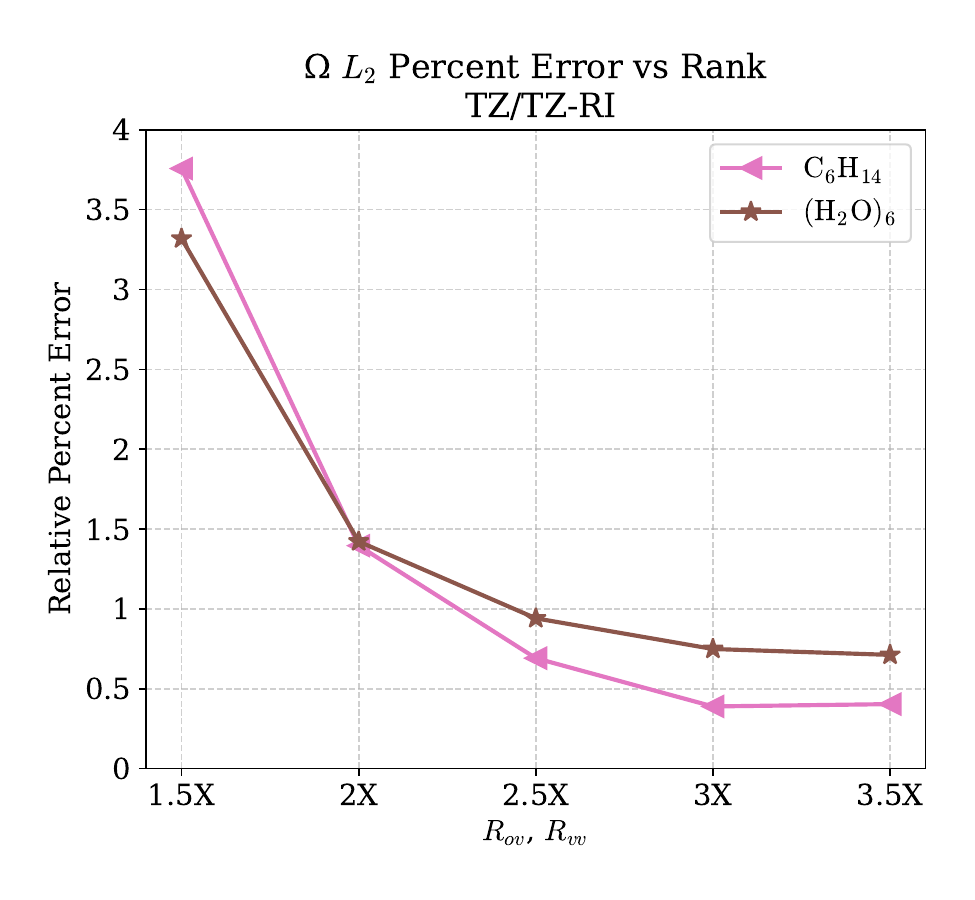}
        \caption{}
        \label{fig:omega_subplots_b}
    \end{subfigure}

    \caption{$L_2$ relative percent error in $\Omega$ for a 6-water cluster and hexane molecule in the TZ/TZ-RI basis.
    In (a) only the rank of the CPD approximation of $J^Q_{ab}$ is modified and in (b) the ranks of the CPD approximation of both $J^Q_{ab}$ and $J^Q_{ai}$ are modified simultaneously.}
    \label{fig:Omega_subplots}
\end{figure}

Moreover, \Cref{fig:cc2_cgs} reveals an unexpected trend: Increasing the CP rank of $J_{ab}^Q$ (i.e., $R_{vv}$) can reduce the accuracy of the CP-DF-LL energies. We attribute this to differing degrees of error cancellation among the CPD-approximated DF integral tensors. For small $R_{vv}$, the CPD error in $J^{Q}_{ab}$ is relatively large, but its contribution to the LL energy is partially offset by compensating errors from the CPD approximations of the other DF tensors. As $R_{vv}$ increases, the error in $J^{Q}_{ab}$ decreases, which weakens this fortuitous cancellation and makes the remaining errors associated with the other approximated tensors more apparent. Consequently, improving the accuracy of the CPD approximations for the remaining DF-TEI tensors, particularly $J^{Q}_{ai}$, should improve the overall accuracy of the CP-DF-LL method. This interpretation is consistent with the trends observed in \cref{fig:Omega_subplots}, discussed below. 
Since the overall error associated with the CPD is relatively small compared to the error introduced by the MPCC method, as demonstrate in \cref{fig:mpcc_ccsd}, we leave the accurate initialization and optimization of the CPD for future study.

Finally, in \cref{fig:Omega_subplots} we consider the $L_2$ relative percent error in the $\Omega$ tensor computed in the LL solver method.
In \cref{fig:omega_subplots_a}, we report the $L_2$ percent error of $\Omega$ for $\ce{(H_2O)_6}$ and $\ce{C_6H_14}$ in the TZ/TZ-RI basis scanning over different values of $R_{vv}$.
We note that for small values of $R_{vv}$, the CP approximation only introduces an approximately $3$ percent error into $\Omega$, which quickly converges to about $1.5$ percent with increasing $R_{vv}$.
In \cref{fig:omega_subplots_b}, we extend this analysis by scanning over ranks $R_{vv}$ and $R_{ov}$ simultaneously. We observe that increasing the rank of both approximations decreases the reconstruction error of $\Omega$ beyond 1.5 percent. 
The remaining error in the $\Omega$ reconstruction at large rank values may be associated with the fixed convergence precision in the analytic optimization of the CPD approximation.\cite{VRG:pierce:2021:JCTC}
The results for the DZ/DZ-RI basis may be found in the SM.

\subsection{Impact of the CP-DF-LL method on the HL Optimization}
\begin{figure}[ht!]
    \begin{subfigure}{0.49\textwidth}
        \centering
        \includegraphics[width=\linewidth]{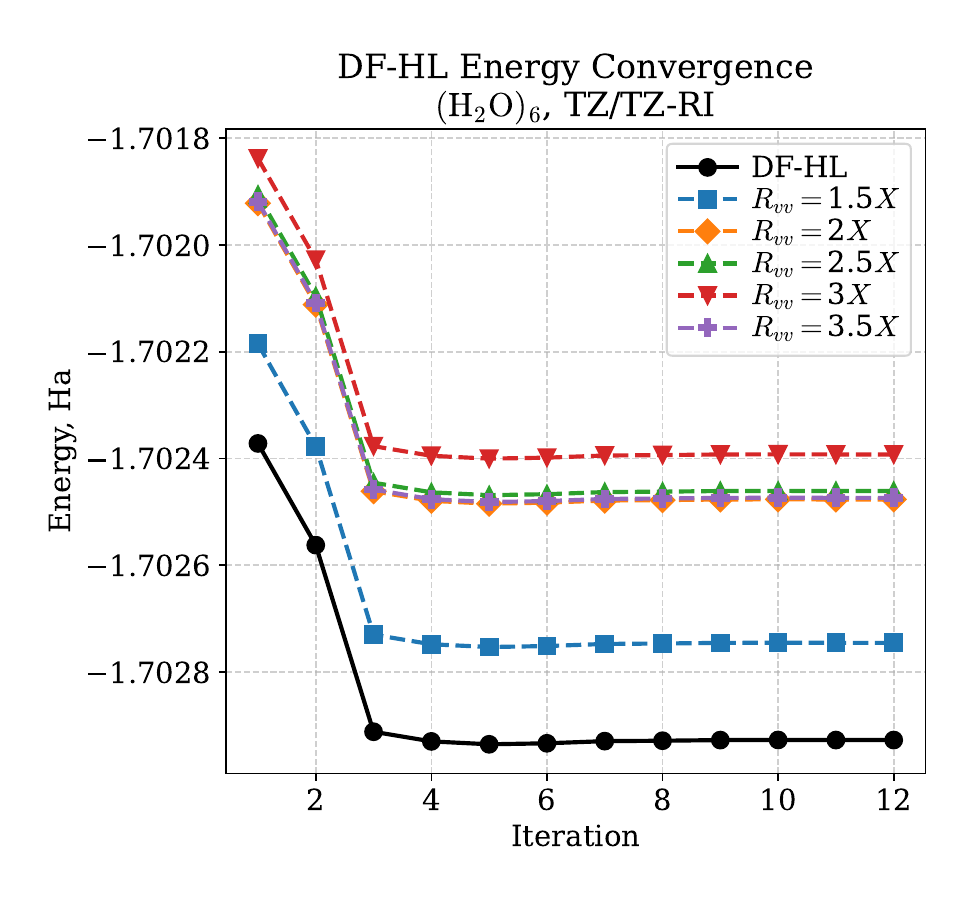}
        \caption{}
        \label{fig:ccsd_subplots_a}
    \end{subfigure}
    \begin{subfigure}{0.49\textwidth}
        \centering
        \includegraphics[width=\linewidth]{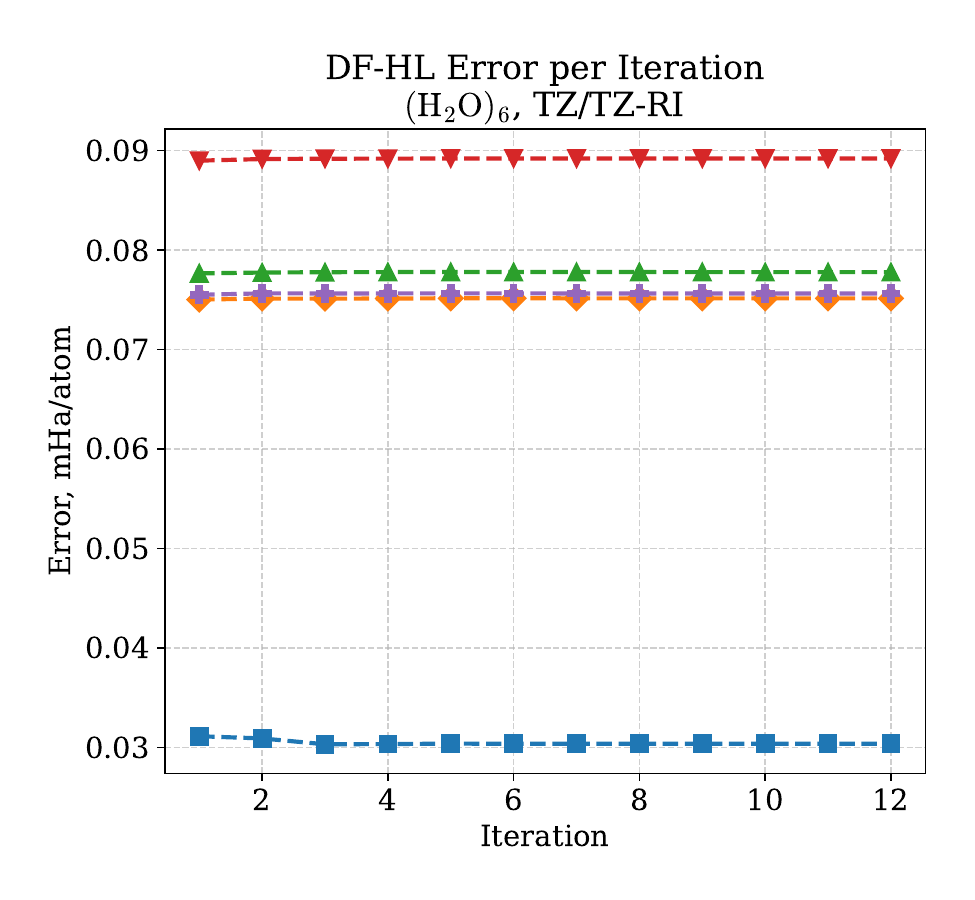}
        \caption{}
        \label{fig:ccsd_subplots_b}
    \end{subfigure}

    \caption{(a) HL energy and (b) HL energy error per non-hydrogen atom at each HL iteration during the second macro-iteration of the MPCC procedure for a 6-water cluster in the TZ/TZ-RI basis.}
    \label{fig:ccsd_subplots}
\end{figure}

In this section, we investigate the impact that a CPD approximation in the LL solver has on the HL solver.
In analogy to \cref{fig:cc2_cgs}, \cref{fig:ccsd_subplots} shows the convergence of the HL solver for a 6-water cluster in the TZ/TZ-RI basis in the second macro-iteration of an MPCC optimization, computed immediately after the LL solver.

Similar to the data in \cref{sec:ll_conv}, we see that the HL optimization is not significantly altered by the perturbations introduced by the CPD approximations made in the LL method. 
Furthermore, we see the same, positive correlation between increasing values of $R_{vv}$ and the CCSD energy.
In fact, the error in the HL energy introduced by the CPD approximation of LL, is approximately equal to the error in the LL energy.
This is interesting as it suggests that there is little propagation of error associated with approximated LL tensors in the HL optimization problem.

\subsection{Impact of the CP-DF-LL solver on the MPCC Optimization}
\begin{figure}[ht!]
    \begin{subfigure}{0.49\textwidth}
        \centering
        \includegraphics[width=\linewidth]{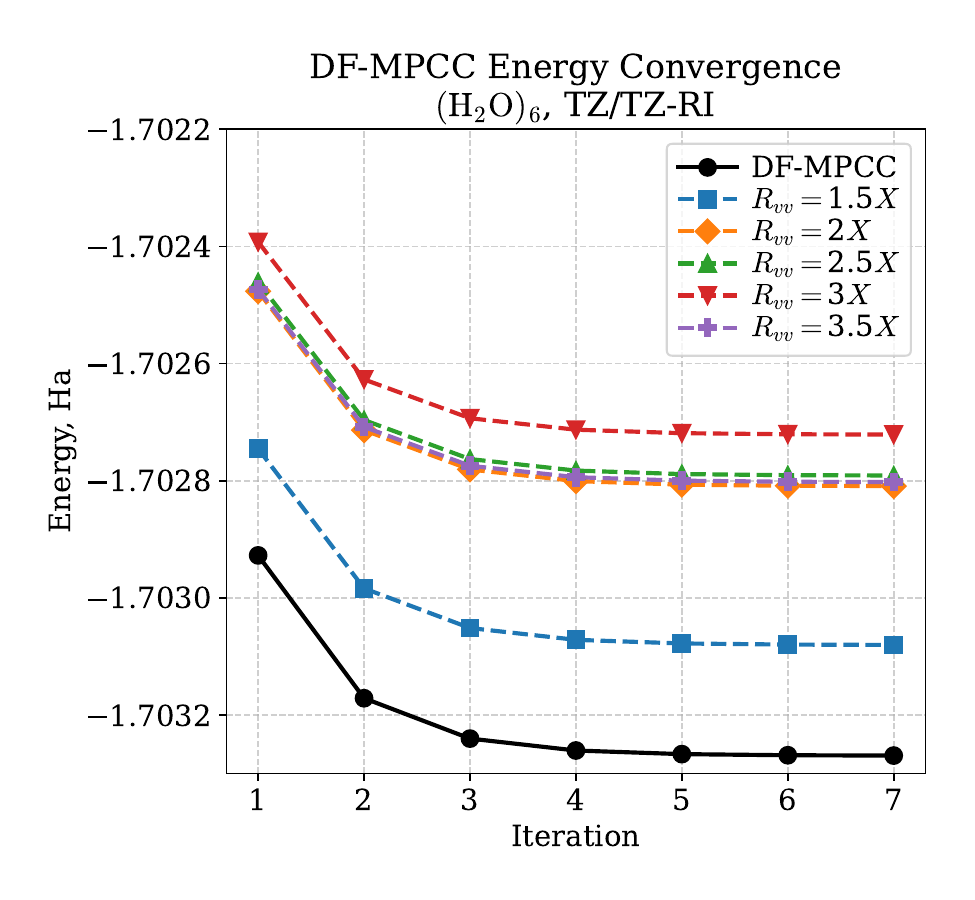}
        \caption{}
    \end{subfigure}
    \begin{subfigure}{0.49\textwidth}
        \centering
        \includegraphics[width=\linewidth]{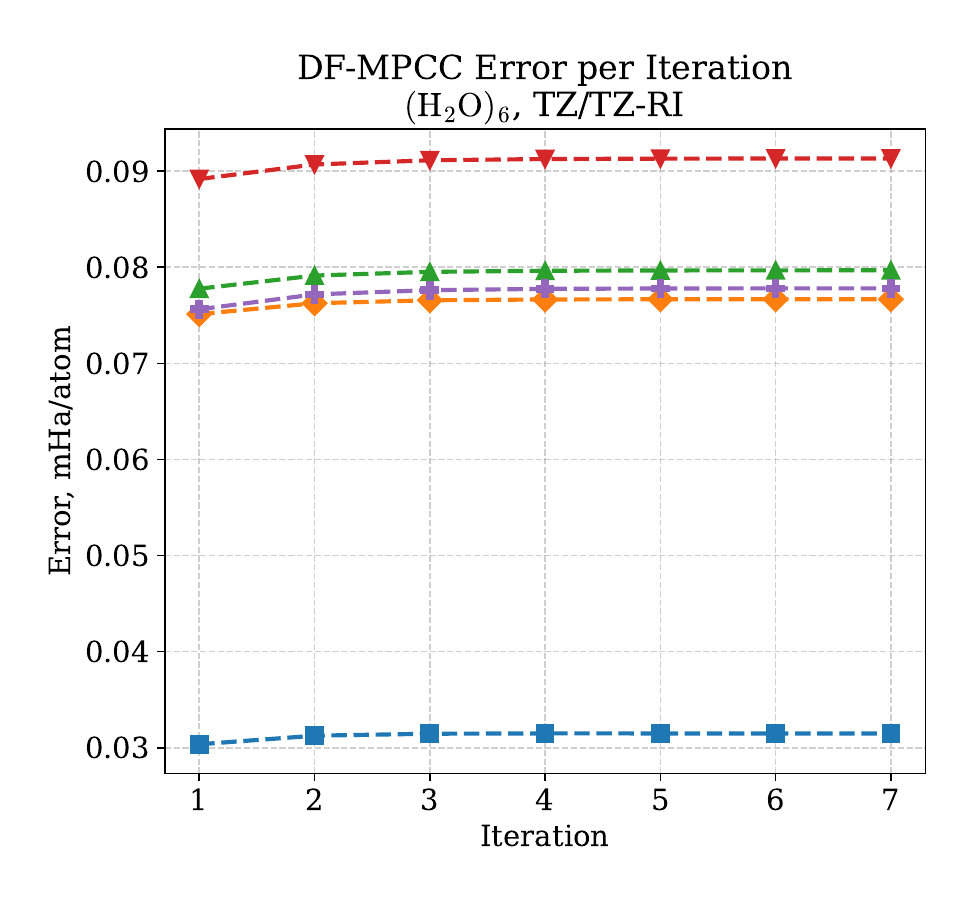}
        \caption{}
    \end{subfigure}

    \caption{(a) MPCC energy and (b) MPCC energy error per non-hydrogen atom at each MPCC iteration for a 6-water cluster in the TZ/TZ-RI basis.}
    \label{fig:mpcc_conv}
\end{figure}

Because MPCC is an alternating optimization of tensor quantities in the HL and LL procedures, we must consider how perturbations to this iterative optimization compound and affect the overall accuracy of the method.
Similar to \cref{fig:cc2_cgs,fig:ccsd_subplots}, in \cref{fig:mpcc_conv} we show the convergence of the MPCC optimization of a 6-water cluster in the TZ/TZ-RI basis.

\begin{figure}[ht!]
        \includegraphics[width=0.55\linewidth]{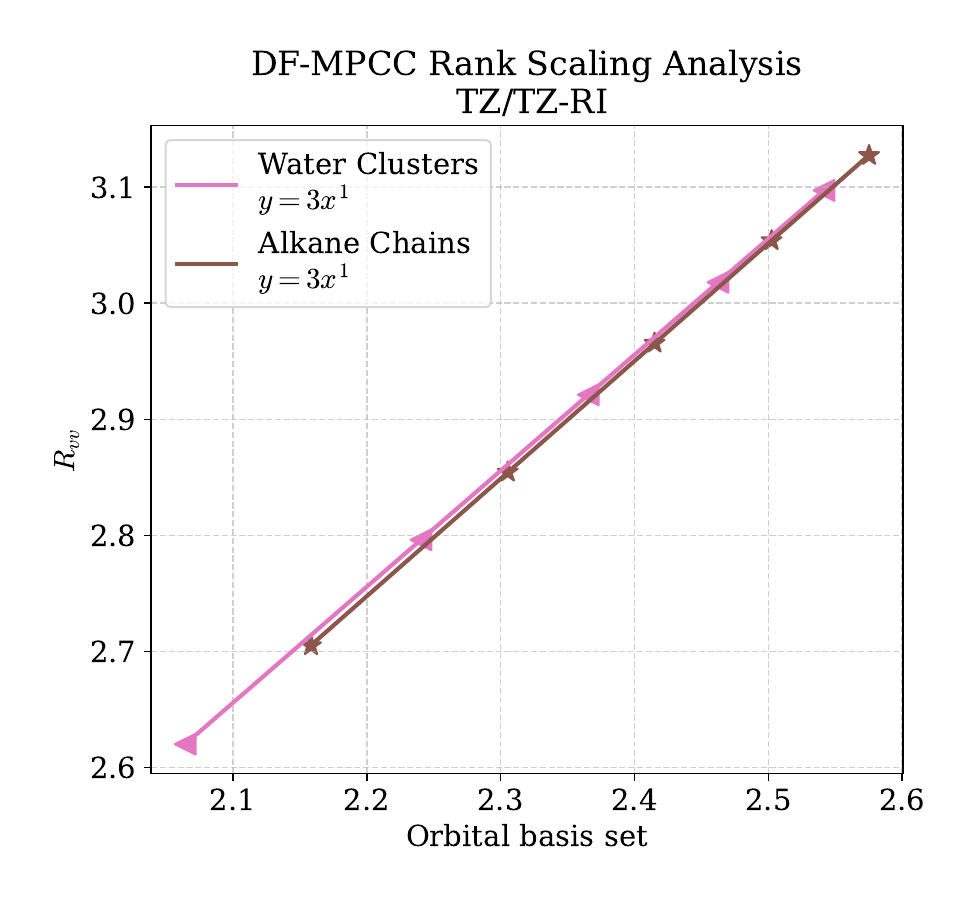}
\caption{Modeling the growth of the CP rank with system size for water molecule clusters and alkane chains in the TZ/TZ-RI basis using a threshold of 0.5mH per non-hydrogen atom.}
\label{fig:mpcc_scaling_curve}
\end{figure}

Notably, this plot indicates that, consistent with the fixed micro-iteration LL and HL results, adding the CPD has no significant impact on the qualitative convergence of the MPCC procedure.
Furthermore, we see the same positive correlation between $R_{vv}$ and the converged MPCC energy.
Moreover, we recognize that the total error in the MPCC energy is relatively small and that the error in the fixed micro-iteration LL convergence does not appear to significantly accumulate across the MPCC macro-iteration process.
This could imply that the error in the MPCC energy is more strongly correlated with the accuracy of the fragment HL optimization which, as we have demonstrated, is not significantly effected by approximations to the low-level problem. 
Furthermore, the overall accuracy of the low-level procedure can be reliably controlled through the choice of the CP rank, which allows for systematic improvements in the MPCC optimization.
\cref{fig:mpcc_scaling_curve} shows how relationship between the OBS and $R_{vv}$ at a fixed absolute error tolerance of 0.5 mH per non-hydrogen atom for both the water clusters and alkane chains in the TZ/TZ-RI basis.

We only consider the value of $R_{vv}$ because the values $R_{ov}$ and $R_{oo}$ are fixed to a value proportional to the DF auxiliary basis set.
These results demonstrate a linear scaling between the OBS and CP rank.

\begin{figure}[t]
        \centering
        \includegraphics[width=0.55\linewidth]{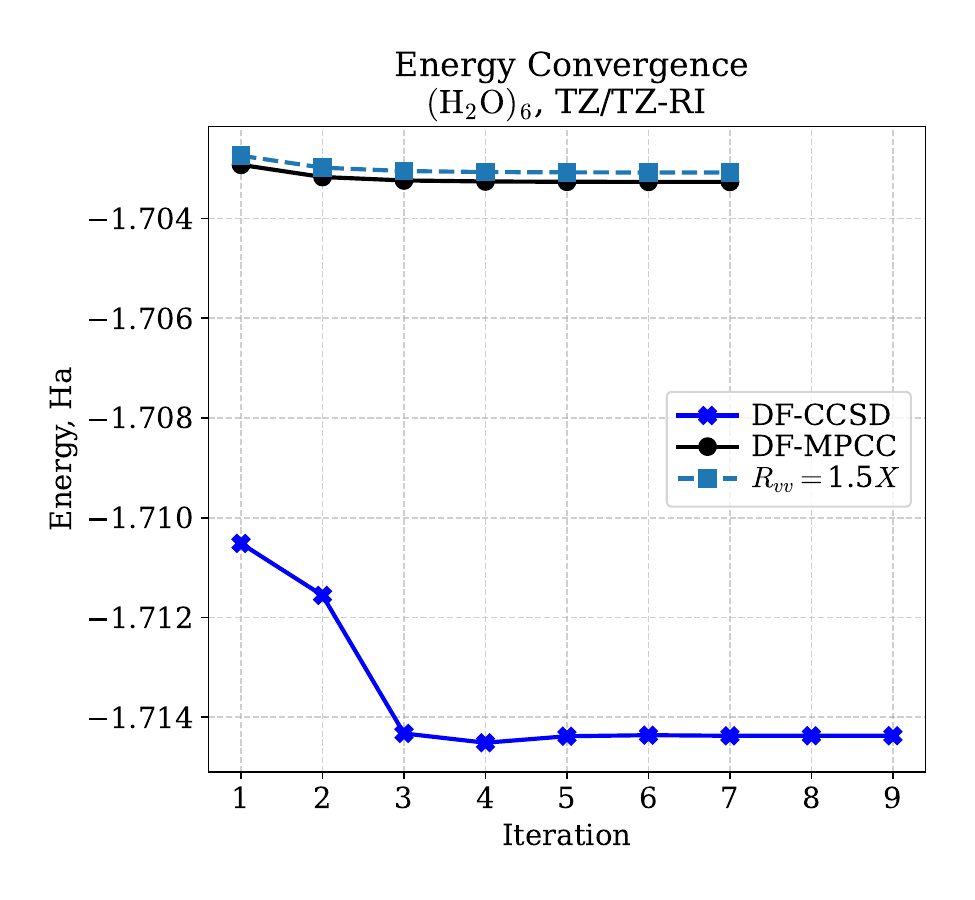}
\caption{Energy convergence curves for the canonical DF-CCSD and the MPCC optimization procedure for a 6-water cluster in the TZ/TZ-RI basis.}
\label{fig:mpcc_ccsd}
\end{figure}

\Cref{fig:mpcc_ccsd} shows the energy convergence curves of the conventional and approximated MPCC procedure compared to the canonical DF-CCSD method. 

With this figure, one can recognize that the deviations in the MPCC energy introduced by the CPD are significantly smaller than the overall error of the MPCC method compared to canonical DF-CCSD. 

\begin{figure}[t!]
    \centering
    \begin{subfigure}{0.49\textwidth}
        \centering
        \includegraphics[width=\linewidth]{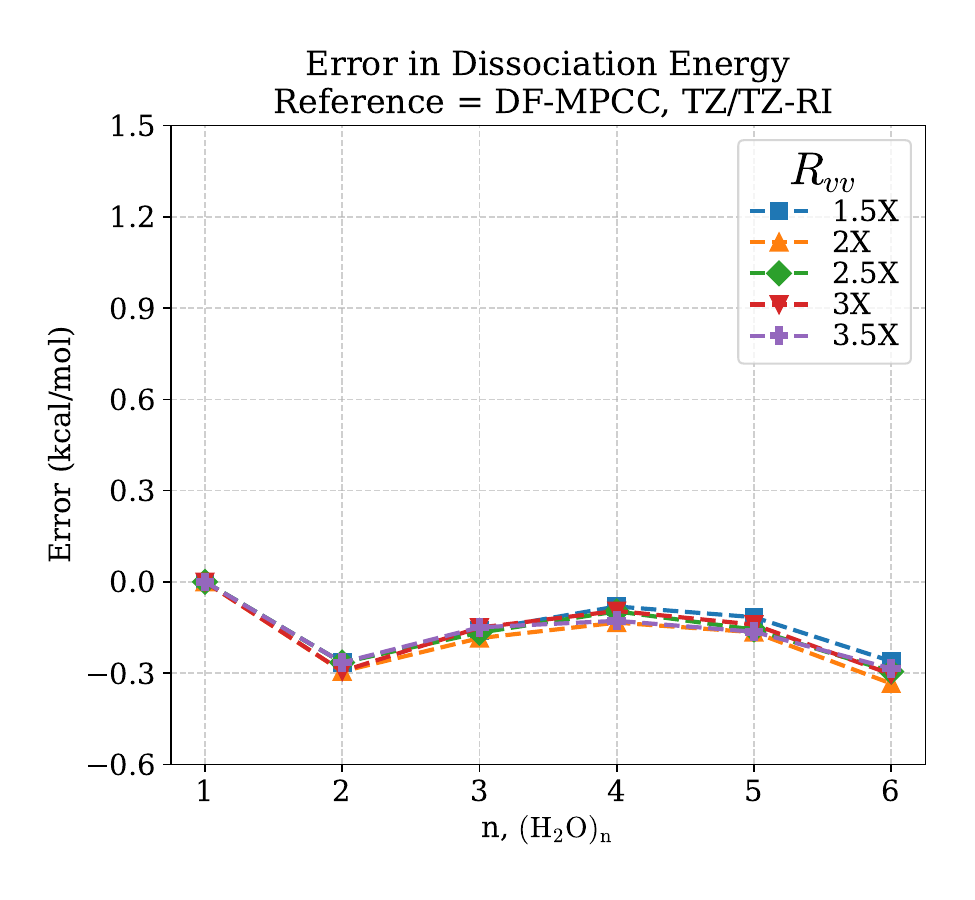}
        \caption{}
        \label{fig:mpcc_dis_err_a}
    \end{subfigure}
    \begin{subfigure}{0.49\textwidth}
        \centering
        \includegraphics[width=\linewidth]{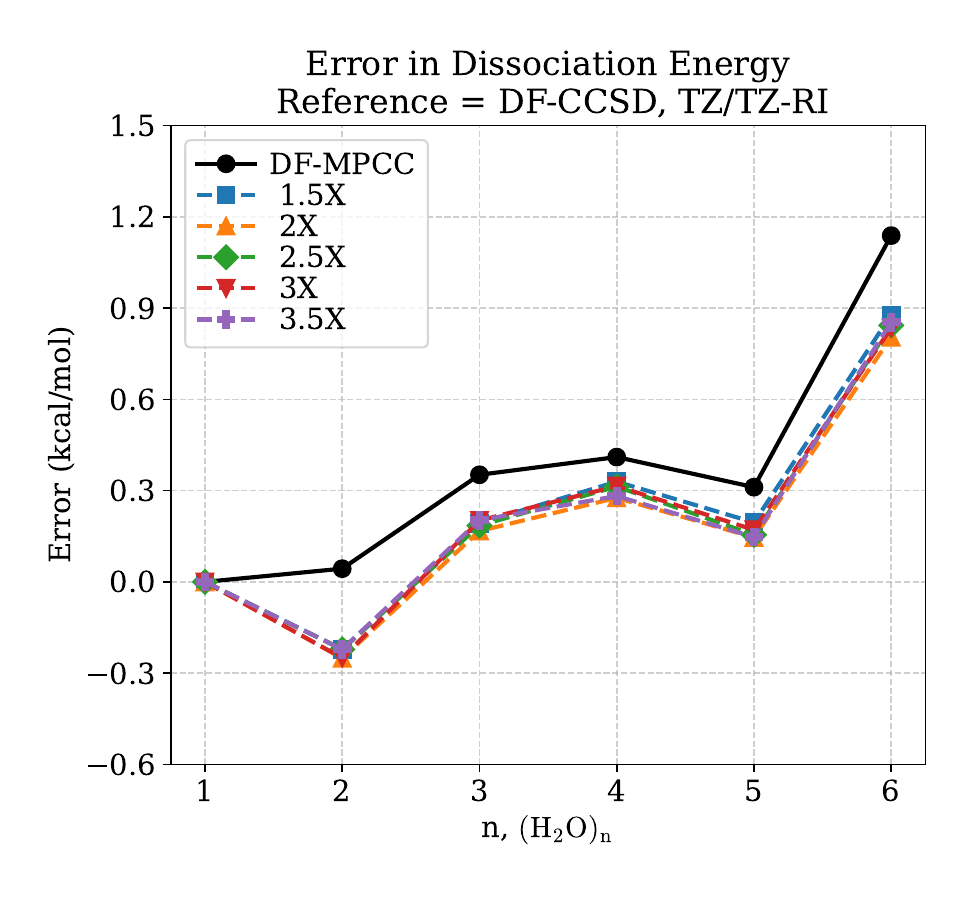}
        \caption{}
        \label{fig:mpcc_dis_err_b}
    \end{subfigure}
\caption{Dissociation energy error of water clusters with between 1 and 6 water molecules with respect to (a) MPCC and (b) DF-CCSD in the TZ/TZ-RI basis.}
    \label{fig:mpcc_dis_err}
\end{figure}

Though the absolute MPCC energy may be relatively far from the absolute DF-CCSD energy, the MPCC procedure is efficient at approximating relevant chemical energy differences.
To demonstrate this, we consider the error in water cluster disassociation energies in the TZ/TZ-RI basis in \cref{fig:mpcc_dis_err}.
In \cref{fig:mpcc_dis_err_a}, we consider the difference in the MPCC disassociation energy using the CP-DF-LL method compared to the DF-LL method.
In this plot, we recognize that the error in the MPCC disassociation energy is well within the bounds of chemical accuracy, $1$ kcal/mol, tightly clustered, and negative in sign.
Furthermore, the error in the dissociation energy in \cref{fig:mpcc_dis_err_a} does not seem to be correlated with increasing molecular system size.
\begin{figure}[b!]
    \begin{subfigure}{0.49\textwidth}
        \centering
        \includegraphics[width=\linewidth]{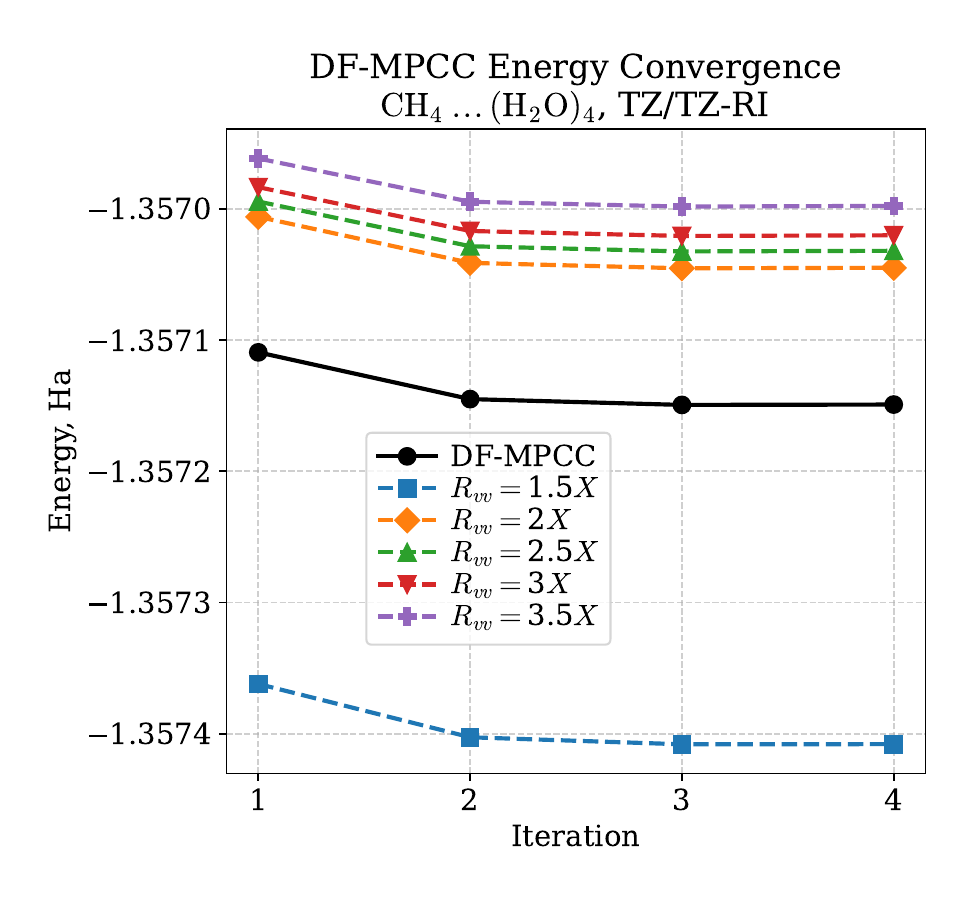}
        \caption{}
    \end{subfigure}
    \begin{subfigure}{0.49\textwidth}
        \centering
        \includegraphics[width=\linewidth]{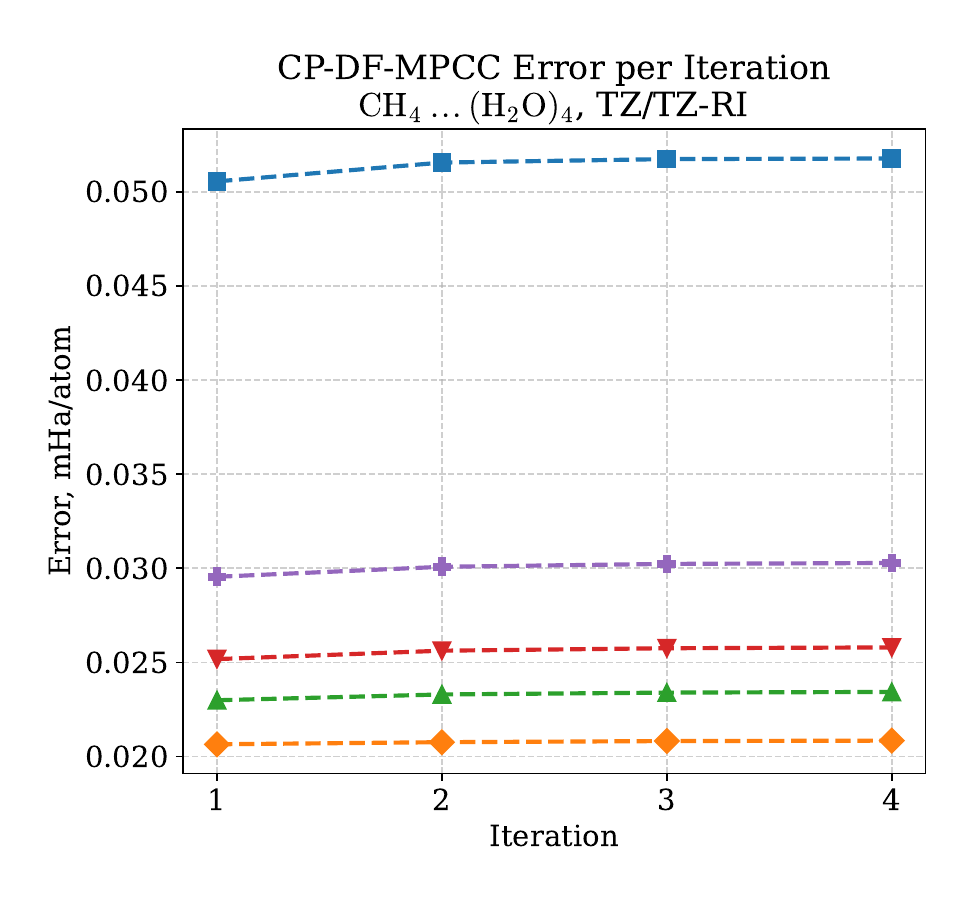}
        \caption{}
    \end{subfigure}

    \caption{(a) MPCC energy and (b) MPCC energy error per non-hydrogen atom, both reported as a function of MPCC iteration, for a $\ce{CH_4}\dots\ce{(H_2O)_4}$ cluster in the TZ/TZ-RI basis.}
    \label{fig:mpcc_m4w_conv}
\end{figure}

To put the error associated with the CPD approximation in perspective with the accuracy of the MPCC method, in \cref{fig:mpcc_dis_err_b} we plot the difference between the DF-LL and CP-DF-LL MPCC, with respect to the DF-CCSD dissociation energies.
In this plot we recognize that, in general, the error in the dissociation between MPCC and DF-CCSD is larger than the error introduced by the CPD approximation.
Also, in this case, because the error introduced by the CPD is negatively signed we see a fortuitous cancellation of errors associated with the CPD in the LL solver.
\begin{figure}[t!]
        \begin{subfigure}{0.49\textwidth}
            \includegraphics[width=\linewidth]{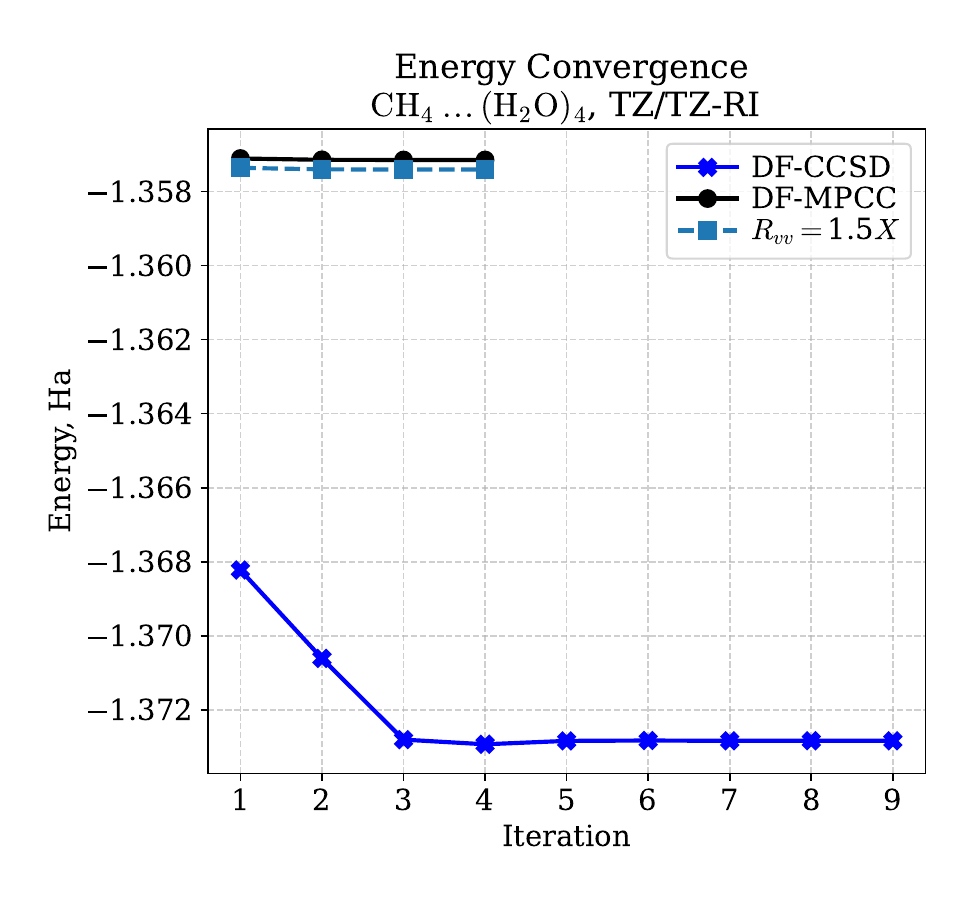}
            \caption{}
            \label{fig:solv_ccsd}
        \end{subfigure}\hfill
        \begin{subfigure}{0.49\textwidth}
            \includegraphics[width=\linewidth]{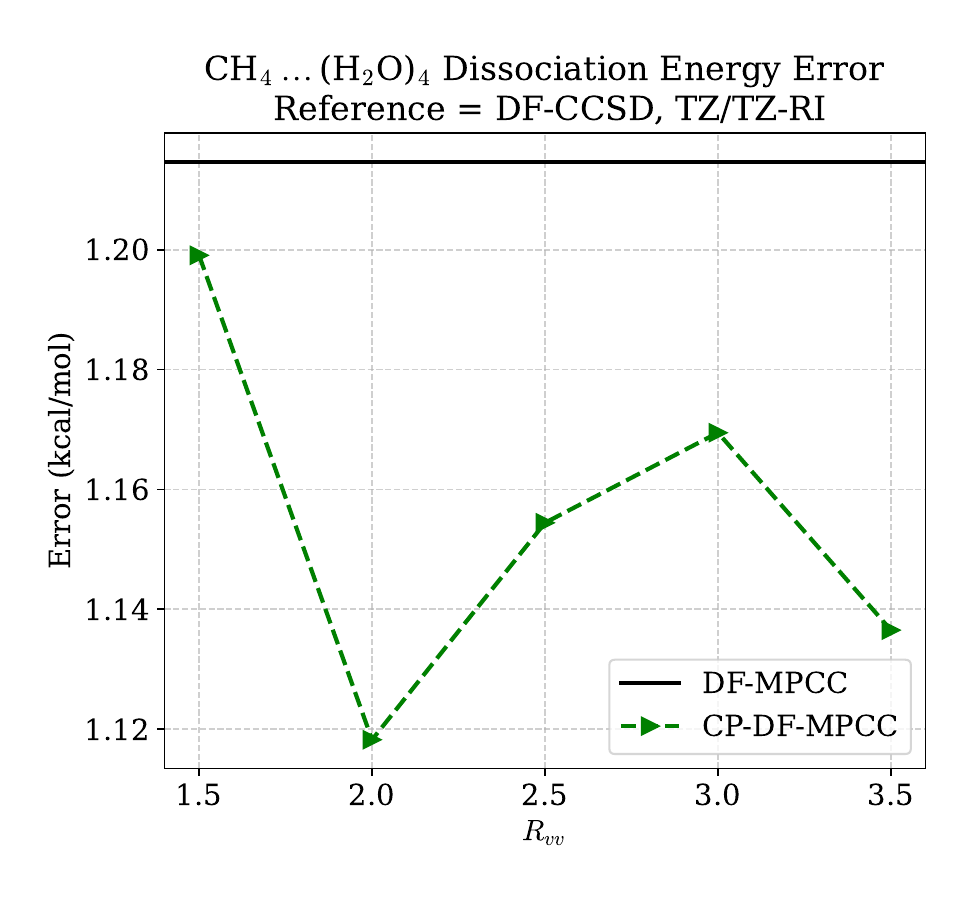}
            \caption{}
            \label{fig:solv_diss}
        \end{subfigure}
\caption{(a)Energy reported as function of DF-CCSD and MPCC iteration for a $\ce{CH_4}\dots\ce{(H_2O)_4}$ cluster in the TZ/TZ-RI basis.
(b) Error in MPCC dissociation energy compared to DF-CCSD for a $\ce{CH_4}\dots\ce{(H_2O)_4}$ cluster in the TZ/TZ-RI basis.}
\label{fig:mpcc_m4w_ccsd}
\end{figure}
\subsection{Proof-of-principle Solvated System}
Finally, in this section, we analyze the ability of the CP-approximated MPCC procedure to accurately model a small organic system (methane) solvated in a small water cluster.
In \cref{fig:mpcc_m4w_conv} we show the convergence and per iteration energy error curves for the MPCC optimization of the $\ce{CH_4}\dots\ce{(H_2O)_4}$ cluster.
Similar to the previous system, we see that the MPCC method is robust to perturbations introduced by the CP approximation and we see a similar error in the energy per non-hydrogen atom.
Finally in \cref{fig:mpcc_m4w_ccsd} we attempt to put the CP-DF-LL MPCC's performance in perspective by comparing absolute and relative energy values computed with MPCC to those computed with canonical DF-CCSD.
In \cref{fig:solv_ccsd}, we show the convergence of the MPCC methods compared to DF-CCSD.
Similar to the previous results, we see that the CP approximation introduces a relatively small error in MPCC compared to DF-CCSD.
In \cref{fig:solv_diss} we compute the error in the MPCC-based dissociation energy for the $\ce{CH_4}\dots\ce{(H_2O)_4}$ cluster.
In this figure we see that the CP approximation, again, introduces significantly less error than the MPCC approach and, again, finds a fortuitous cancellation of error.

\section{Conclusions}
The MPCC method is an emerging method to accurately model large chemical systems using QM-in-QM embedding techniques. 
The method utilizes orbital localization techniques to divide a molecule's orbital basis into two non-interacting basis sets, the fragment and environment.
Similar to other embedding methods, the number of orbitals in the environment is expected to grow rapidly with system size and, though these orbitals are treated at a low-level theory, the computational time and storage necessary for the low-level problem still quickly becomes intractable.
As a means to reduce the computational overhead of the MPCC low-level solver, we introduce the CPD approximation of three-center DF TEI tensors.
In related works, the CPD has been used to both reduce the computational storage complexity and scaling/cost of accurate electronic structure methods.
By introducing the CPD approximation of all three DF TEI tensors ($J^Q_{ab}, J^Q_{ai}\text{, and} J^{Q}_{ij}$), we are able to reduce the computational complexity of the MPCC low-level problem from $\mathcal{O}(N^4)$ to $\mathcal{O}(NR^2) \approx \mathcal{O}(N^3)$ and the storage scaling from $\mathcal{O}(N^3)$ to $\mathcal{O}(NR) \approx \mathcal{O}(N^2)$.
We demonstrate that the CPD introduces relatively small errors into the LL, HL and MPCC optimizations and that the CP rank for each DF TEI tensor scales linearly with chemical system size.
We also show that the CPD approximation has little effect on MPCC's ability to accurately predict chemically relevant energy differences, such as dissociation energy.

\section*{Acknowledgment}

This material is based upon work supported by the U.S. Department of Energy, Office of Science, Office of Advanced Scientific Computing Research and Office of Basic Energy Sciences, Scientific Discovery through Advanced Computing (SciDAC) program under Award Number DE‐SC0022198 (A.S.)

\newpage
\bibliography{references, kmp5refs}

\newpage
\includepdf[pages=-]{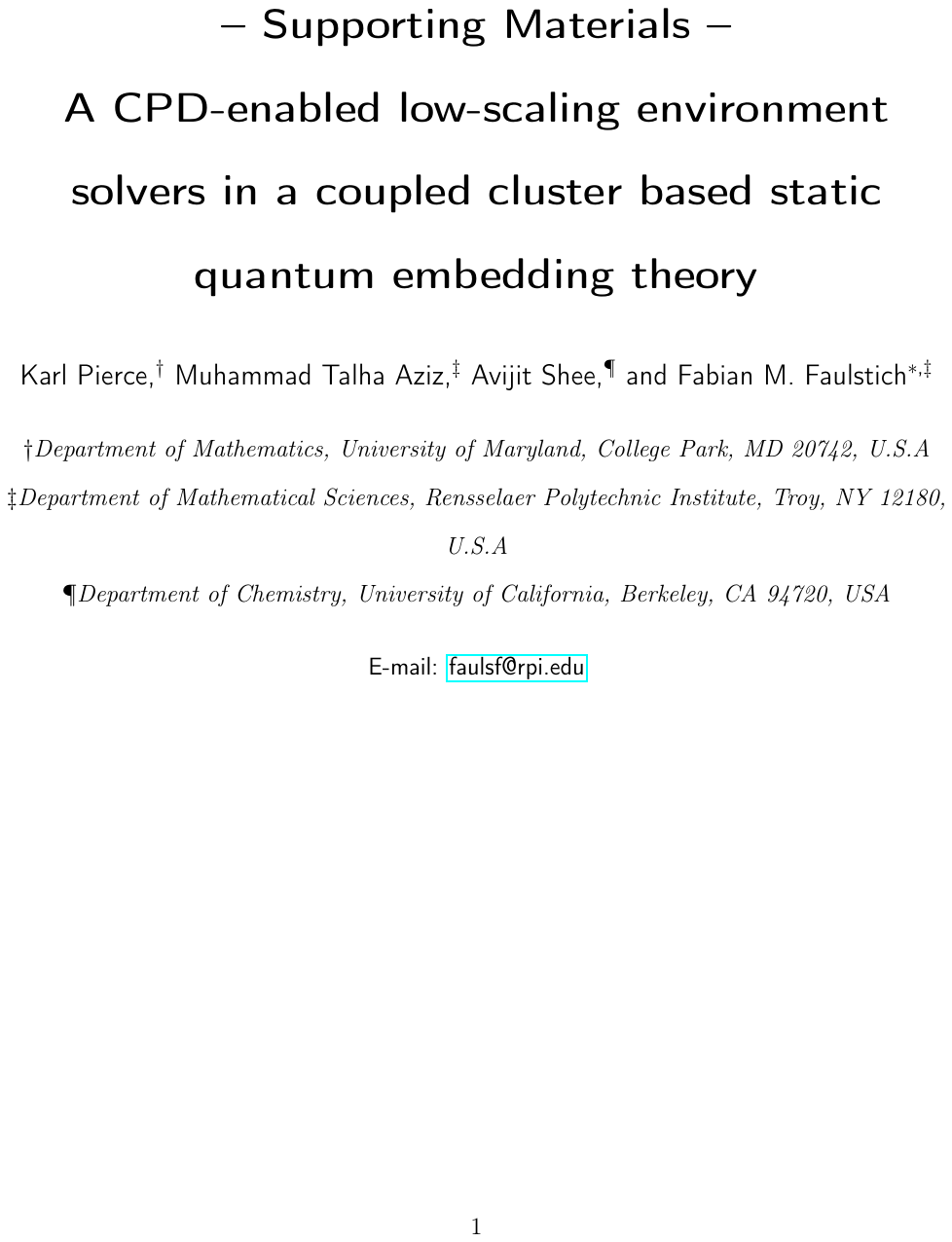}

\end{document}


\newpage
\section{A CPD-Enhanced Low-Level Solver}
\begin{table}[!hb]
    \resizebox{1\columnwidth}{!}{%
    \begin{tabular}{c|ll}
        \toprule
        \toprule
        \textbf{Step} & \textbf{Description} &\textbf{Scaling} \\
        \midrule
        \midrule
        1. & 
        Compute intermediate  & $ \mathcal{O}(N^3)$\\
        \hline
        \vspace{-3.5mm}\\
            & $\hat{X}^Q = 2\sum_{S} L_{QS} \,\Big[\,\sum_{i}K_{iS} \,\Big(\,\sum_{a} A_{aS}t^a_i\,\Big)\,\Big]$  \quad  & 
        \vspace{1mm}\\
        2. & 
        Compute $\bar F$ (i.e. $\bar F_{\rm oo}$, $\bar F_{\rm ov}$ and $\bar F_{\rm vv}$ blocks)  & $\mathcal{O}(N^{3})$\\
        \hline
        \vspace{-3mm}\\
            &  $\bar{F}_{i j} 
                = F_{i j} 
                    + \sum_{S} I_{iS} \,\Big(\, I_{jS}\,\big(\, \sum_{Q} M_{QS} \hat{X}^{Q} \,\big)\, \Big)$
                &   \vspace{1mm}\\
            &   
                    $\qquad - \sum_{S} I_{iS} 
                    \Big(\,\sum_T\Big[\,
                    \sum_{Q} M_{QS} L_{QT}
                    \big(\, \sum_{k} I_{kS} K_{kT} \,\big)\, \Big]
                    \Big[\sum_{a} A_{aT} t^a_j\Big]\Big)$ 
            \vspace{1.5mm}\\ 
            &  $\bar{F}_{a b}
                = F_{a b} 
                    + \sum_{S} C_{aS} C_{bS} \,\big(\, \sum_Q V_{QS} \hat{X}^Q \,\big)$  
                &   \vspace{1mm}\\ 
            &
                    $\qquad - \sum_{S} A_{aS}  \Big[\sum_{T} C_{bT} 
                    \Big( \sum_Q L_{QS} V_{QT}\,\Big)\,
                    \Big( \sum_{k} K_{kS} \big(\sum_{c} C_{cT} t^c_k\big) \Big)
                    \Big]$
            \vspace{1.5mm}\\ 
            &   $\bar{F}_{j b}
                = F_{jb} 
                    + \sum_{S} A_{bS} \Big(K_{jS} \big( \sum_{Q} L_{QS} \big) \hat{X}^Q\Big)$
                &   \vspace{1mm}\\
            &
                    $\qquad - \sum_{T}A_{bT} \Big[\, \sum_{S} K_{jS} 
                    \Big( \sum_Q L_{QS} L_{QT}\Big) 
                    \Big( \sum_i K_{iT} \big( \sum_{a} A_{aS} t^a_i\big) \Big)\Big]$ 
            \vspace{1mm}\\
       3. & 
       Compute $\Omega$  & $\mathcal{O}(N^{3})$\\
        \hline
        \vspace{-3mm}\\
            &   $\bar \Omega_{ai} 
                = F_{ai} 
                    + \sum_{S} A_{aS} \Big(\,I_{iS} \big(\,\sum_{Q} L_{QS} \hat{X}^Q\, \big)\,\Big)$  
                & \vspace{1mm}\\ 
            &
                $\qquad - \sum_{S} C_{aS} \Big[\, \sum_{T} I_{iT} \Big(\, \sum_Q V_{QS} M_{QT}\, \Big) \Big(\, \sum_jI_{jT} \big(\,\sum_{b} C_{bS} t^b_j\big)\,\Big)\Big]$
         \vspace{1mm}\\
        4. & 
        Update $\bar{F}$ & $\mathcal{O}(N^3)$ \\
        \hline
        \vspace{-3mm}\\
            & $\hat{F}_{b c}=\bar{F}_{b c}-\sum_{k} \bar{F}_{k c} t_{k}^{b}$ 
                &  \\
            & $\hat{F}_{k i}=\bar{F}_{k i}-\sum_{c} \bar{F}_{k c} t_{i}^{c}$ 
                &   
         \vspace{1mm}\\
        5. & 
        Diagonalize $\hat{F}$ and compute $D_{i a}^{\alpha}$ & $\mathcal{O}(N^3)$ \\
        \hline
        \vspace{-3mm}\\
            & $\left(U_{i j}, \varepsilon_{i}\right) \leftarrow \hat{F}_{b c}$ & \\
            & $\left(U_{a b}, \varepsilon_{a}\right) \leftarrow \hat{F}_{k i}$ & \\
            & $D_{i a}^{\alpha} \leftarrow\left\{\varepsilon_{r}\right\}, \alpha$ 
                & 
         \vspace{1mm}\\
        6. & Update $Y_{i a}^{Q, \alpha}$ & $\mathcal{O}(N^4)$ \\
        \hline
        \vspace{-3mm}\\
        & $\bar{J}_{bk}^Q = J_{bk}^Q + \sum_{c} J_{bc}^Q t_k^c - \sum_{j}t^{b}_j ([(\sum_{c} J^Q_{ck} t^c_j) + J^Q_{kj})$
            &  \\
        & $\hat{J}_{a i}^{Q}=\sum_{bk} U_{\mathrm{ab}} \bar{J}_{b k}^{Q} U_{\mathrm{ik}}$ 
            &  \\
        & $\bar{Y}_{ai}^{Q \alpha} \leftarrow \hat{J}_{a i}^{Q} D_{a i}^{\alpha}$ 
            &  \\
        & $Y_{b j}^{Q \alpha} \leftarrow \sum_{ai} U_{a b} \bar{Y}_{ai}^{Q, \alpha} U_{j i}$ 
            & \\
        \hline
        \bottomrule
    \end{tabular}
    }
    \caption{Working equations and their scaling for the low-level method.}
\end{table}
\paragraph{Step 6. Computation of $T_2$-factors:}
As pointed out in the manuscript, the computation of $Y^{Q\alpha}_{bj}$ cannot be reduced from $\mathcal{O}(N^4)$ even with the CPD. 
However, it is possible to devise a strategy which generates intermediates which require strictly less than $\mathcal{O}(N^3)$ using either the DF or CPD approximation.
Because the DF-based algorithm is rather complicated, we choose to omit this algorithm.
However, the strategy is much more straightforward with the CPD and can be done in the following manner.
First, we expand the contraction to form $\hat{J}^Q_{ai}$
\begin{equation}
\begin{aligned}\label{eq:Y_expanded}
    \hat{J}_{a i}^{Q}&= \sum_{bk}U_{\mathrm{ab}} \bar{J}_{b k}^{Q} U_{\mathrm{ik}} \\ 
    &= \sum_{bk} U_{ab} [J^Q_{bk} + X^Q_{bk} - \sum_{j} \bar{J}^Q_{kj} t^{b}_{j}] U_{ik}\\
    &= \sum_{bk} U_{ab}[J_{bk}^Q + \sum_{c} J_{bc}^Q t_k^c - \sum_{j} (X^Q_{kj} + J^Q_{kj})t^b_j]U_{ik} \\ 
    &= \sum_{bk} U_{ab}[J_{bk}^Q + \sum_{c} J_{bc}^Q t_k^c - \sum_{j}t^{b}_j [(\sum_{c} J^Q_{ck} t^c_j) + J^Q_{kj}]]U_{ik}
\end{aligned}
\end{equation}
From here, we can recognize that the index $Q$ is not summed over in any component of Step 6.
Therefore, we are free split Step 6 into $Q$ independent contractions by considering the matrix component of $\bar{J}$ for every different value of $Q$, i.e. $[\bar{J}^Q]_{ai}$.
\begin{align}\label{eq:jblockQ}
    [\hat{J}^{Q}]_{ai}&=
    \sum_{bk} U_{ab}[[J^Q]_{bk} + \sum_{c} [J^Q]_{bc} t_k^c - \sum_{j}t^{a}_j [(\sum_{c} [J^Q]_{ck} t^c_j) + [J^Q]_{kj}]]U_{ik}
\end{align}
where $[J^Q]_{pq}$ is the matrix of dimension $p \times q$ associated with the $Q$th index.
Next we must consider how to efficiently form sub-matrices of $J$ while avoiding the explicit storage of order-3 tensors.
By introducing the CPD approximation, we can efficiently determine sub-matrix elements of $J$ via
\begin{align}
    [J^Q]_{bk} \approx \sum_{S} (l^Q_{S} K_{kS}) B_{bS} 
\end{align}
where $l^Q_{S}$ is a vector of length $S$ associated with the $Q$th column of the factor matrix $L$.
Regardless of which method one chooses, the computational complexity of forming the $Q$th sub-matrix of $J$ is $\mathcal{O}(N^3)$.
After forming forming the $Q$th sub-matrix of each $J$ tensor in \label{eq:jblockQ} we may now compute the $Q$th sub-block of $\hat{J}$ with a scaling of $\mathcal{O}(N^3)$.
After forming $[\hat{J}^Q]$ we may immediately contract it with the remaining terms to find the $Q$th component of the $Y$ tensor, i.e.
\begin{align}
    [Y^Q]^{\alpha}_{bj} 
    = U_{ab} [\bar{Y}^Q]^\alpha_{ai} U_{ji}
    = U_{ab} \big(\, [\hat{J}^Q]_{ai} D^\alpha_{ai}\,\big) U_{ji}
\end{align}
with a computational complexity of $\mathcal{O}(N^3)$. 
Though each component now has a complexity of $\mathcal{O}(N^3)$, this contraction sequence must be performed for every $Q$ component.
Therefore, the computational complexity of this step remains as $\mathcal{O}(N^4)$.
We point out to readers that because each Q is independent from all other values, the described contraction sequence is trivially parallelizable.
Taking advantage of this degree of freedom may significantly reduce the computational cost of this step.

\newpage

\section{The CP-DF-LL Solver Convergence}

\begin{figure}[htbp]
    \begin{subfigure}{0.49\textwidth}
        \centering
        \includegraphics[width=\linewidth]{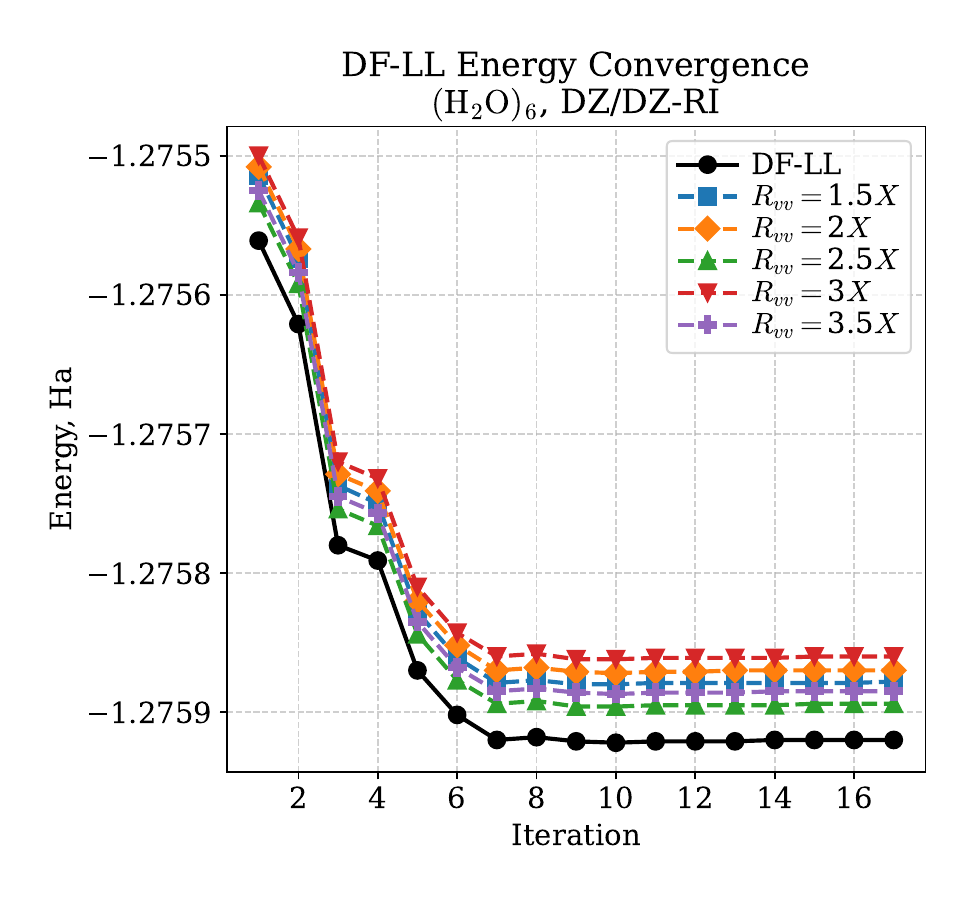}
        \caption{}
    \end{subfigure}
    \begin{subfigure}{0.49\textwidth}
        \centering
        \includegraphics[width=\linewidth]{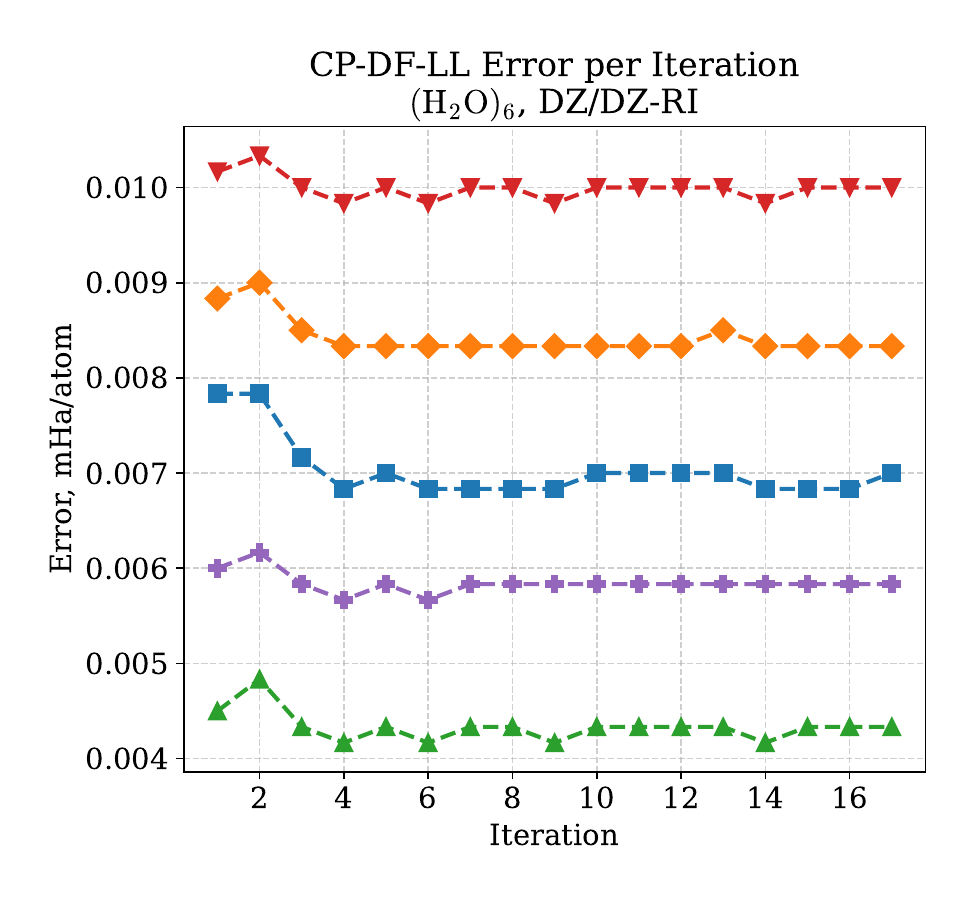}
        \caption{}
        \label{fig:ll_err_h2o_dz_err}
    \end{subfigure}
    \caption{(a) LL energy and (b) LL energy error per non-hydrogen atom, both reported as a function of LL iteration, for a 6-water cluster in the DZ/DZ-RI basis.}
    \label{fig:ll_err_h2o}
\end{figure}
\begin{figure}[ht!]
    \begin{subfigure}{0.49\textwidth}
        \centering
        \includegraphics[width=\linewidth]{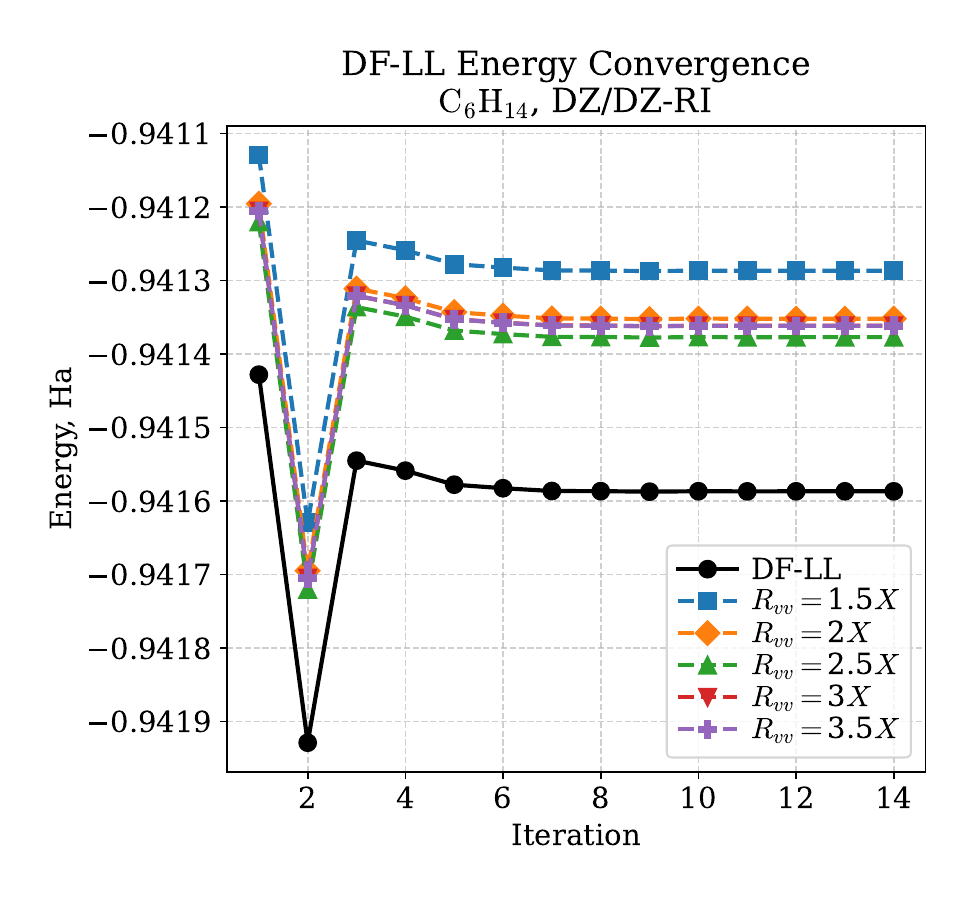}
        \caption{}
    \end{subfigure}
    \begin{subfigure}{0.49\textwidth}
        \centering
        \includegraphics[width=\linewidth]{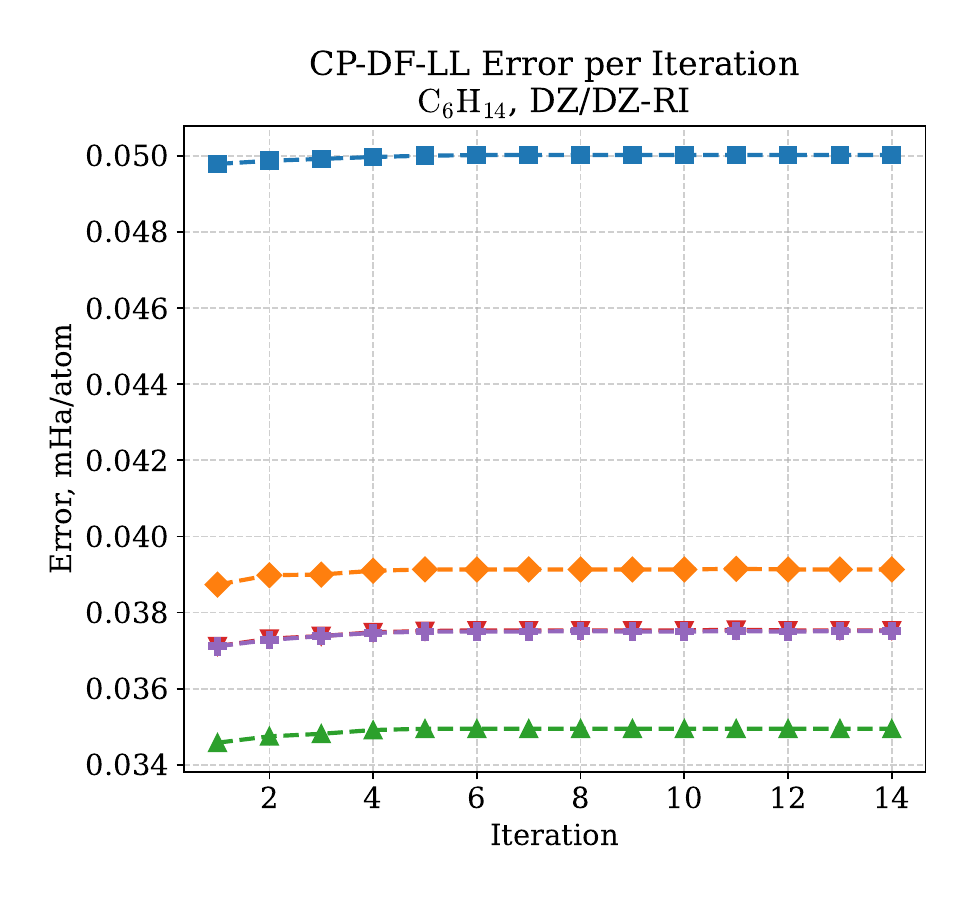}
        \caption{}
        \label{fig:ll_err_alk_dz_err}
    \end{subfigure}
    \begin{subfigure}{0.49\textwidth}
        \centering
        \includegraphics[width=\linewidth]{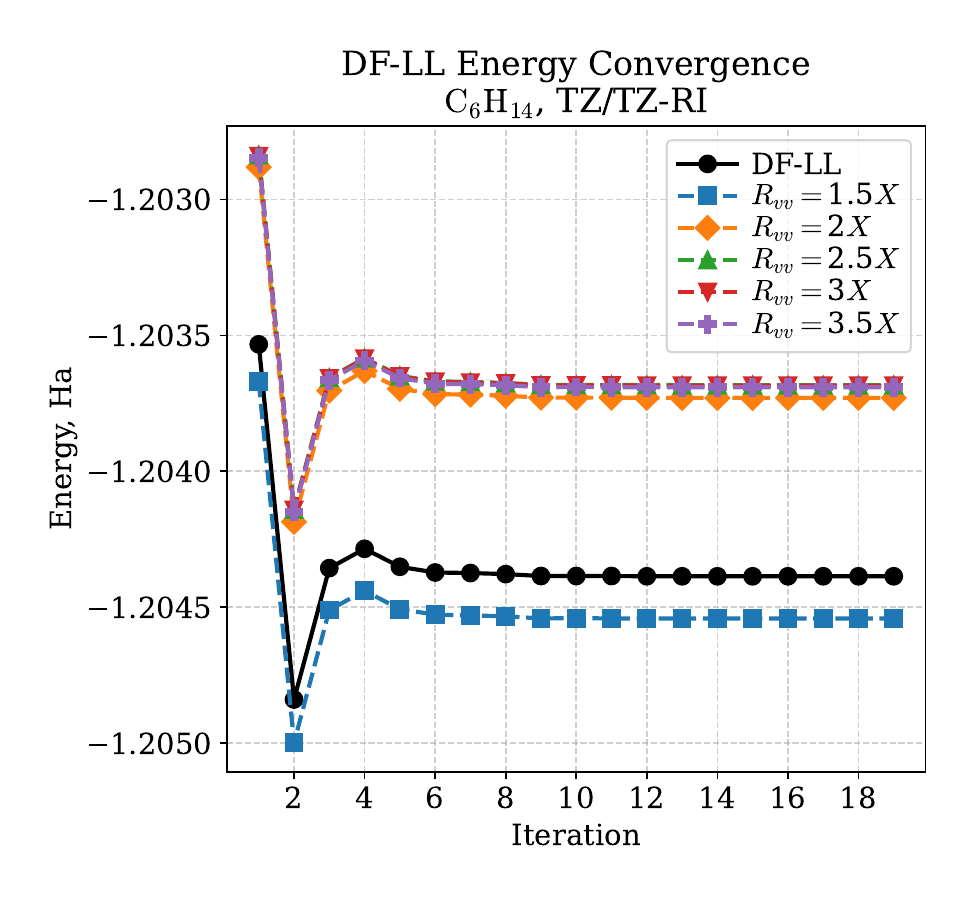}
        \caption{}
    \end{subfigure}
    \begin{subfigure}{0.49\textwidth}
        \centering
        \includegraphics[width=\linewidth]{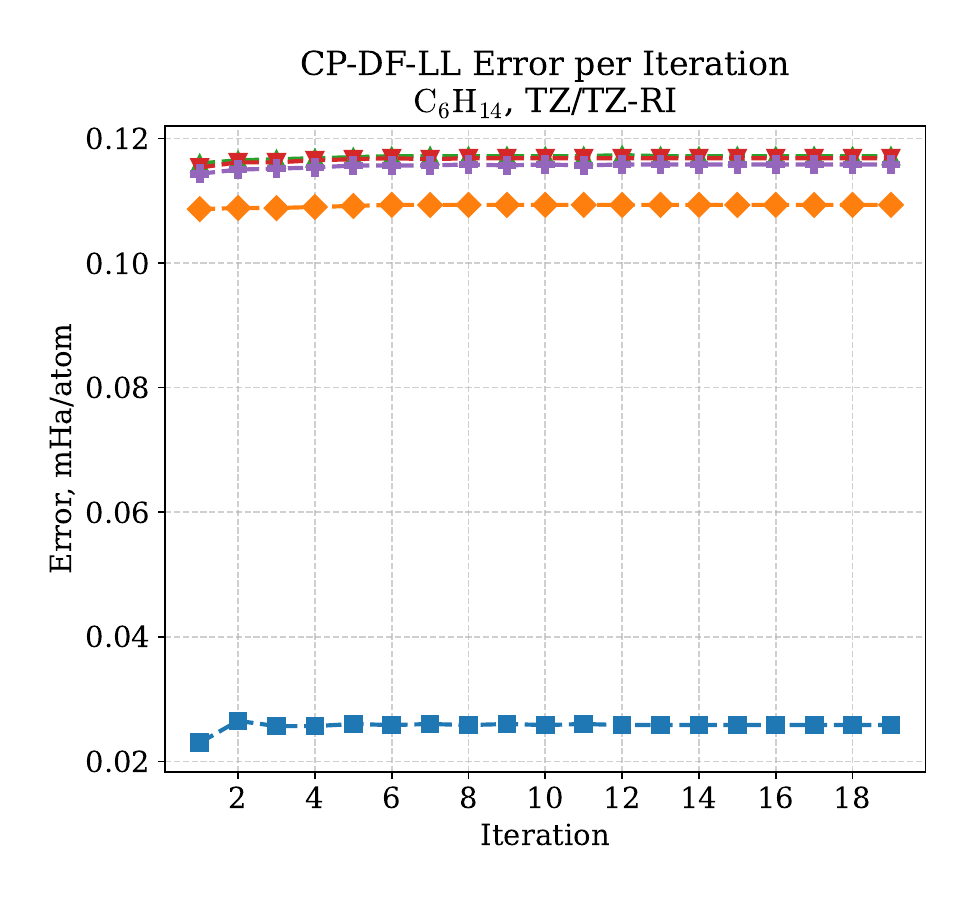}
        \caption{}
        \label{fig:ll_err_alk_tz_err}
    \end{subfigure}
    \caption{LL energy reported as a function of LL iteration for a hexane molecule in the (a) DZ/DZ-RI and (c) TZ/TZ-RI basis.
    LL energy error per non-hydrogen atom reported as a function of LL iteration for a hexane molecule in the (b) DZ/DZ-RI and (d) TZ/TZ-RI basis.}
    \label{fig:ll_err_alk}
\end{figure}
\cref{fig:ll_err_h2o} considers the LL convergence behavior for a $\ce{({H}_2{O})_6}$ cluster in the DZ/DZ-RI basis and
\cref{fig:ll_err_alk} considers the LL convergence behavior for a $\ce{C_6H_14}$ molecule in the DZ/DZ-RI and TZ/TZ-RI basis.
These results demonstrate that the CP approximation of the DF TEI tensors does not significantly impact the convergence behavior or accuracy of the LL solver.
In the results for the DZ/DZ-RI basis, \cref{fig:ll_err_h2o_dz_err,fig:ll_err_alk_dz_err}, we notice that increasing $R_{vv}$ does not monotonically improve the accuracy of the LL optimization.
This is most likely correlated with the convergence of the CP-ALS problem and can be improved by decreasing the tolerance of this optimization.\cite{VRG:pierce:2021:JCTC}
The results in \cref{fig:ll_err_alk_tz_err} demonstrate the same positive correlation between $R_{vv}$ and LL energy error as we find for the $\ce{(H_2O)_6}$ cluster.
\begin{figure}[htbp]
    \begin{subfigure}{0.49\textwidth}
        \centering
        \includegraphics[width=\linewidth]{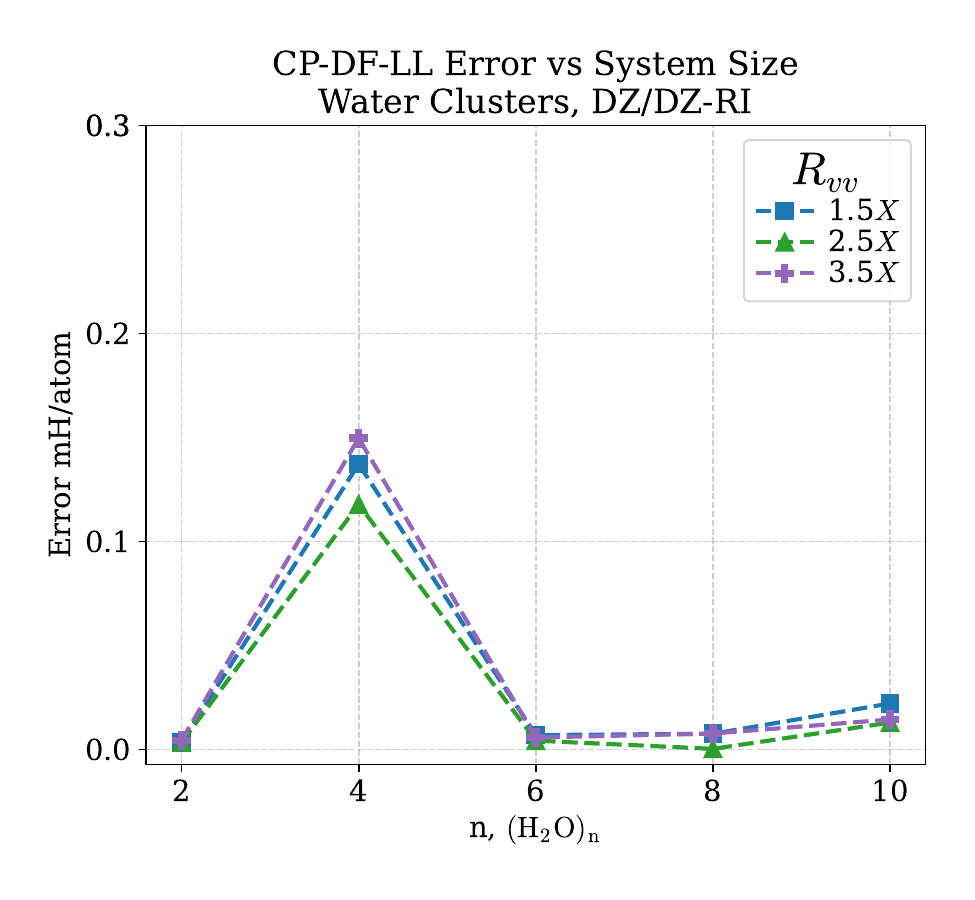}
        \caption{}
    \end{subfigure}
    \hfill
    \begin{subfigure}{0.49\textwidth}
        \centering
        \includegraphics[width=\linewidth]{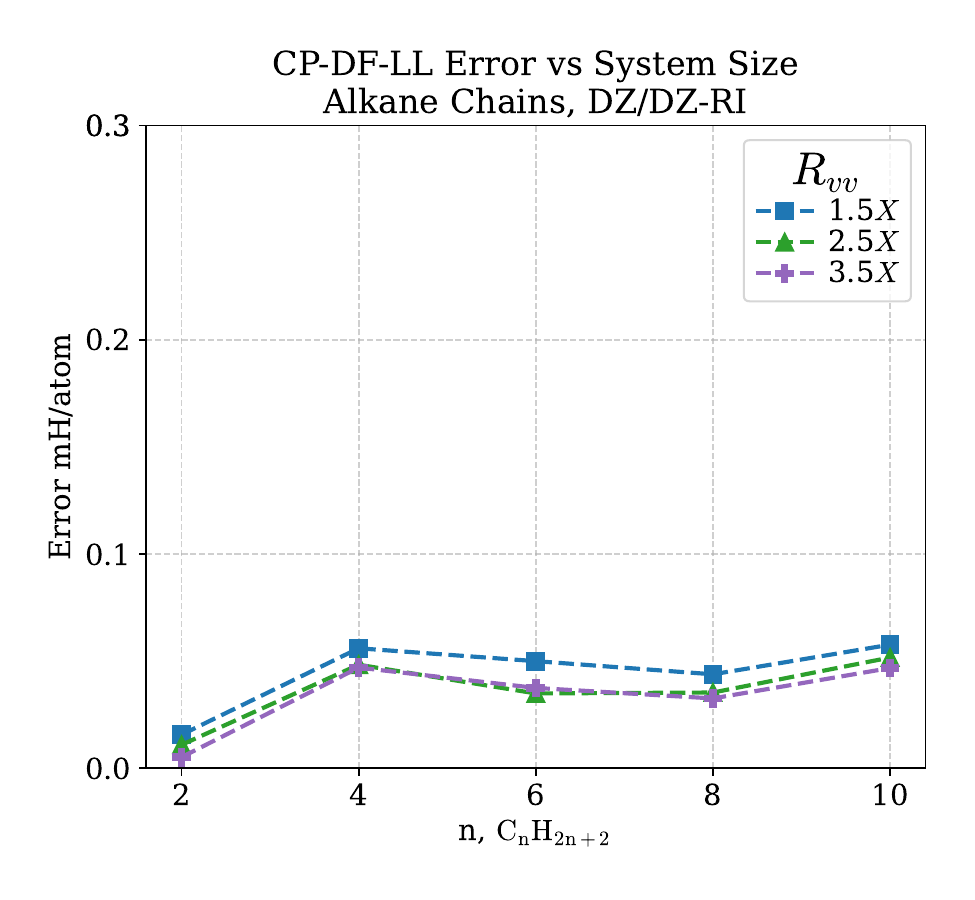}
        \caption{}
    \end{subfigure}
    \begin{subfigure}{0.49\textwidth}
        \centering
        \includegraphics[width=\linewidth]{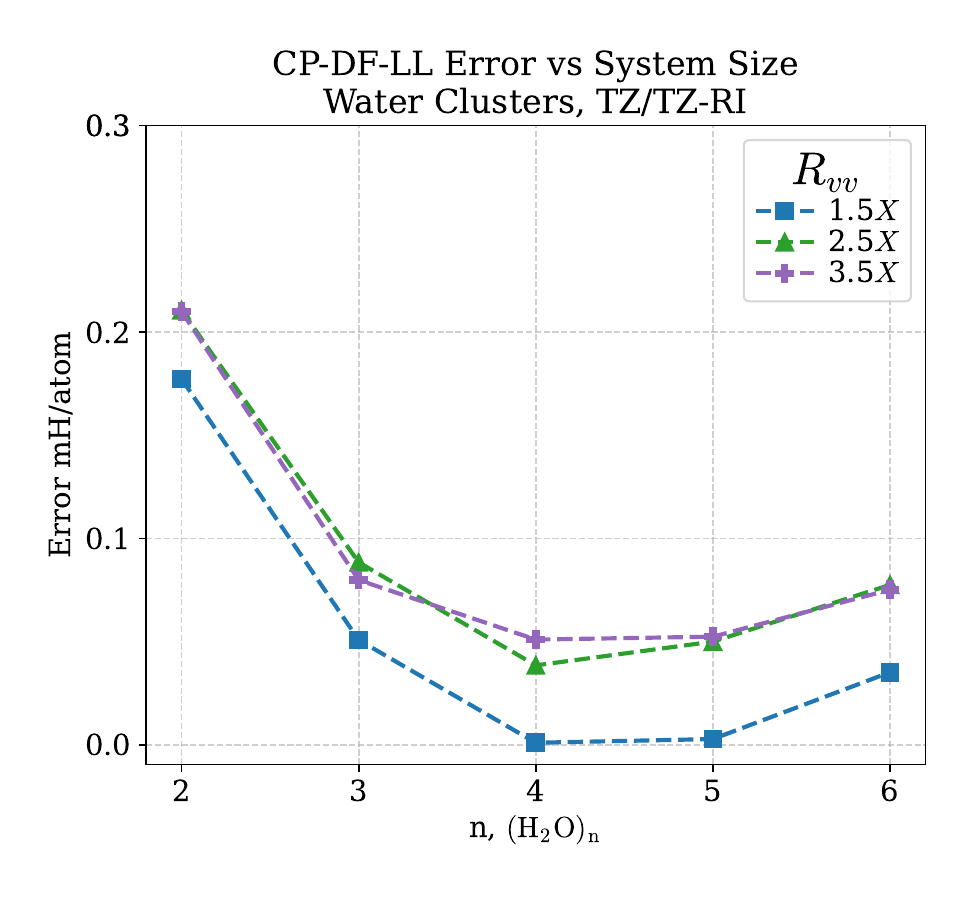}
        \caption{}
    \end{subfigure}
    \begin{subfigure}{0.49\textwidth}
        \centering
        \includegraphics[width=\linewidth]{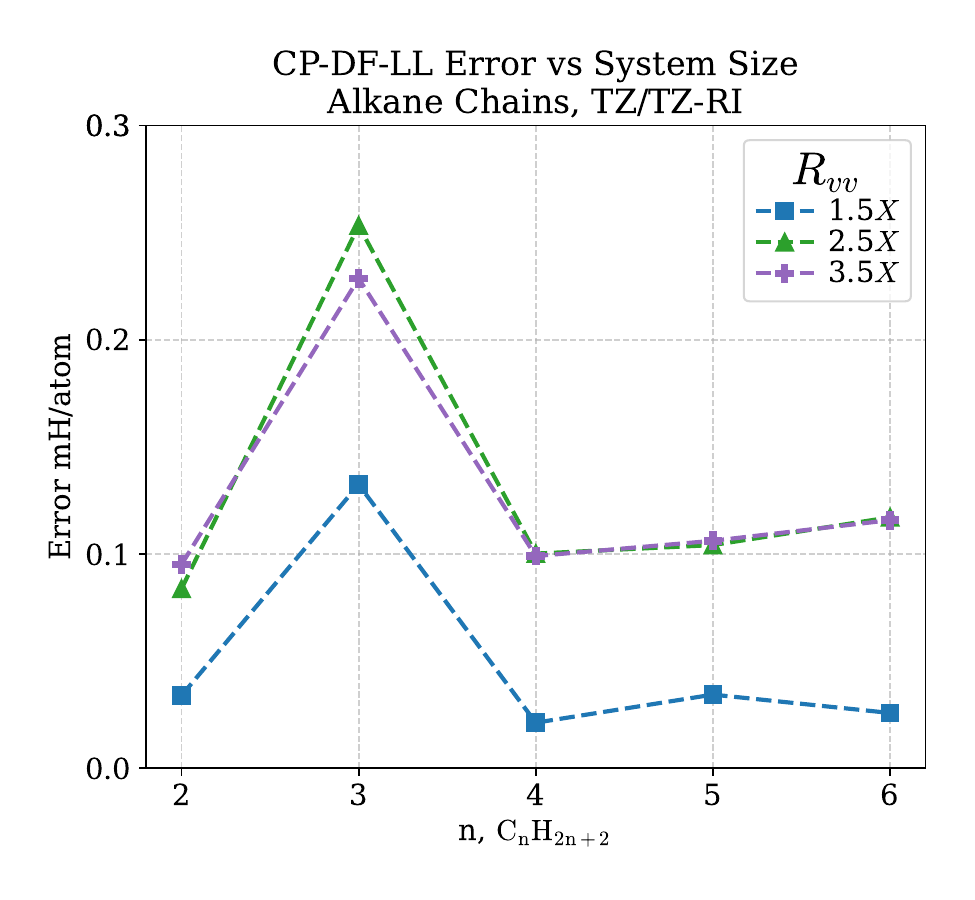}
        \caption{}
    \end{subfigure}
    \caption{Absolute LL energy error for water clusters with (a) 2-10 water molecules in the DZ/DZ-RI basis and (c) 2-6 water molecules in the TZ/TZ-RI basis and alkane chains with (b) 2-10 carbon atoms in the DZ/DZ-RI and (d) 2-6 carbon atoms in the TZ/TZ-RI basis.}
    \label{fig:cc2_err_scan}
\end{figure}
In \cref{fig:cc2_err_scan} we consider the error in the converged LL energy at the second macro iteration of the MPCC procedure for water clusters and alkane chains of increasing sizes in the DZ/DZ-RI and TZ/TZ-RI basis.
In the DZ/DZ-RI basis, we see relatively little difference in the converged LL energy with increasing $R_{vv}$.
And in the TZ/TZ-RI basis, we see that a positive correlation between LL energy error and $R_{vv}$ across both datasets and molecular size.
These results demonstrate that the error associated with the approximation of $J^Q_{ab}$ in the LL method converges relatively quickly with $R_{vv}$.
Overall, these results indicate that the LL energy error is systematic and remains well controlled across molecular size.

\begin{figure}[htbp]
    \centering

    \begin{subfigure}{0.49\textwidth}
        \centering
        \includegraphics[width=\linewidth]{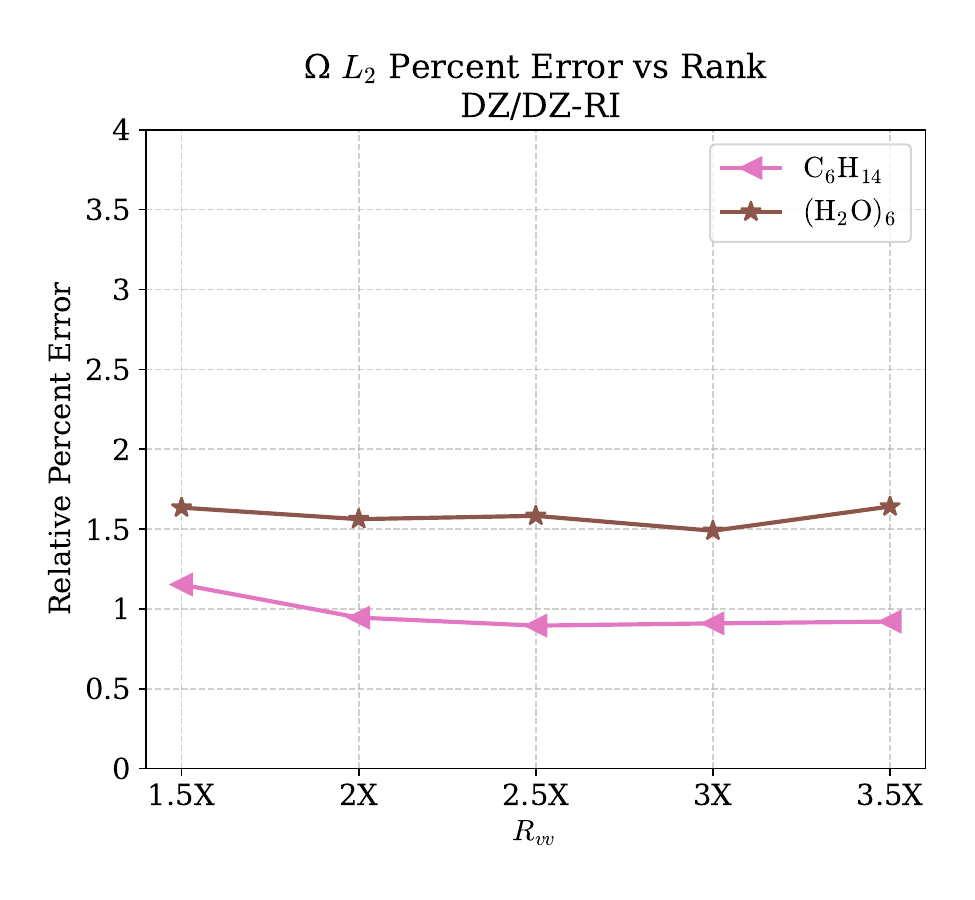}
        \caption{}
        \label{fig:OmegaDZa}
    \end{subfigure}
    \hfill
    \begin{subfigure}{0.49\textwidth}
        \centering
        \includegraphics[width=\linewidth]{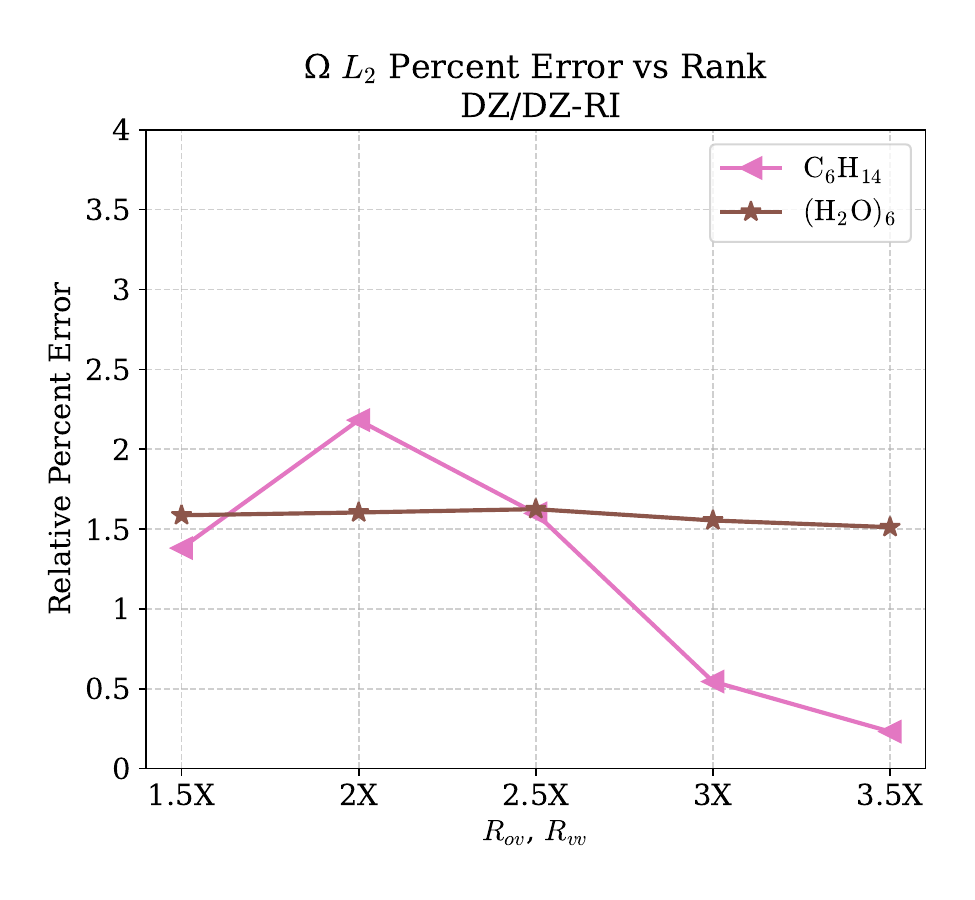}
        \caption{}
        \label{fig:OmegaDZb}
    \end{subfigure}

    \caption{$L_2$ relative percent error in $\Omega$ for a 6-water cluster and hexane molecule in the TZ/TZ-RI basis.
    In (a) only the rank of the CPD approximation of $J^Q_{ab}$ is modified and in (b) the ranks of the CPD approximation of both $J^Q_{ab}$ and $J^Q_{ai}$ are modified simultaneously.}
    \label{fig:OmegaDZ}
\end{figure}

In \cref{fig:OmegaDZa}, we report the $L_2$ percent error for $\ce{(H_2O)_6}$ and $\ce{C_6H_14}$ in the DZ/DZ-RI basis scanning over different values of $R_{vv}$.
We note that the reconstruction error of $\Omega$ does not significantly change with $R_{vv}$ and can be considered converged.
In \cref{fig:OmegaDZb}, we extend this analysis by scanning over ranks $R_{vv}$ and $R_{ov}$ simultaneously. 
We observe that increasing the rank of both approximations decreases the reconstruction error of $\Omega$ for $\ce{C_6H_14}$ but does not significantly improve $\ce{(H_2O)_6}$.
We find that remaining error associated with both of these molecules is, again, associated with the fixed convergence precision in the analytic optimization of the CPD approximation.
\newpage

\section{Impact of the CP-DF-LL solver on the MPCC Optimization}

\begin{figure}[htbp]
    \begin{subfigure}{0.49\textwidth}
        \centering
        \includegraphics[width=\linewidth]{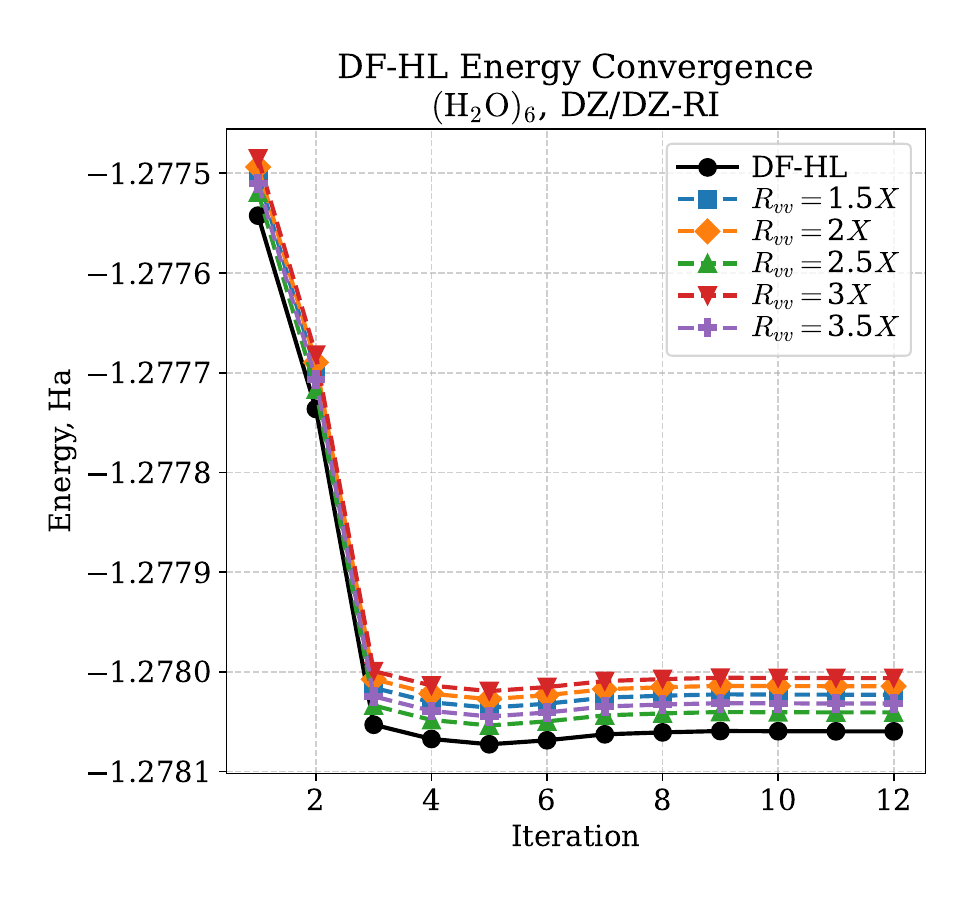}
        \caption{}
    \end{subfigure}
    \begin{subfigure}{0.49\textwidth}
        \centering
        \includegraphics[width=\linewidth]{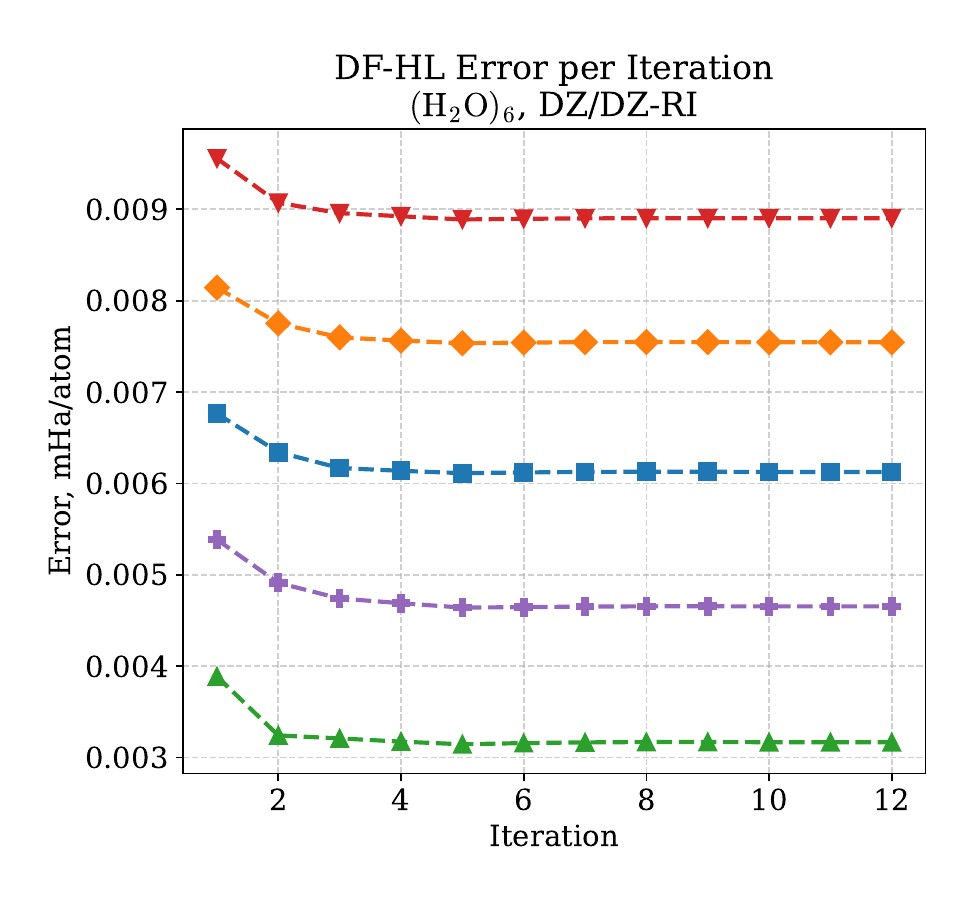}
        \caption{}
    \end{subfigure}
    \caption{(a) HL energy and (b) HL energy error per non-hydrogen atom, both reported as a function of HL iteration, for a 6-water cluster in the DZ/DZ-RI basis.}
    \label{fig:hl_err_h2o}
\end{figure}
\begin{figure}[htbp]
\begin{subfigure}{0.49\textwidth}
        \centering
        \includegraphics[width=\linewidth]{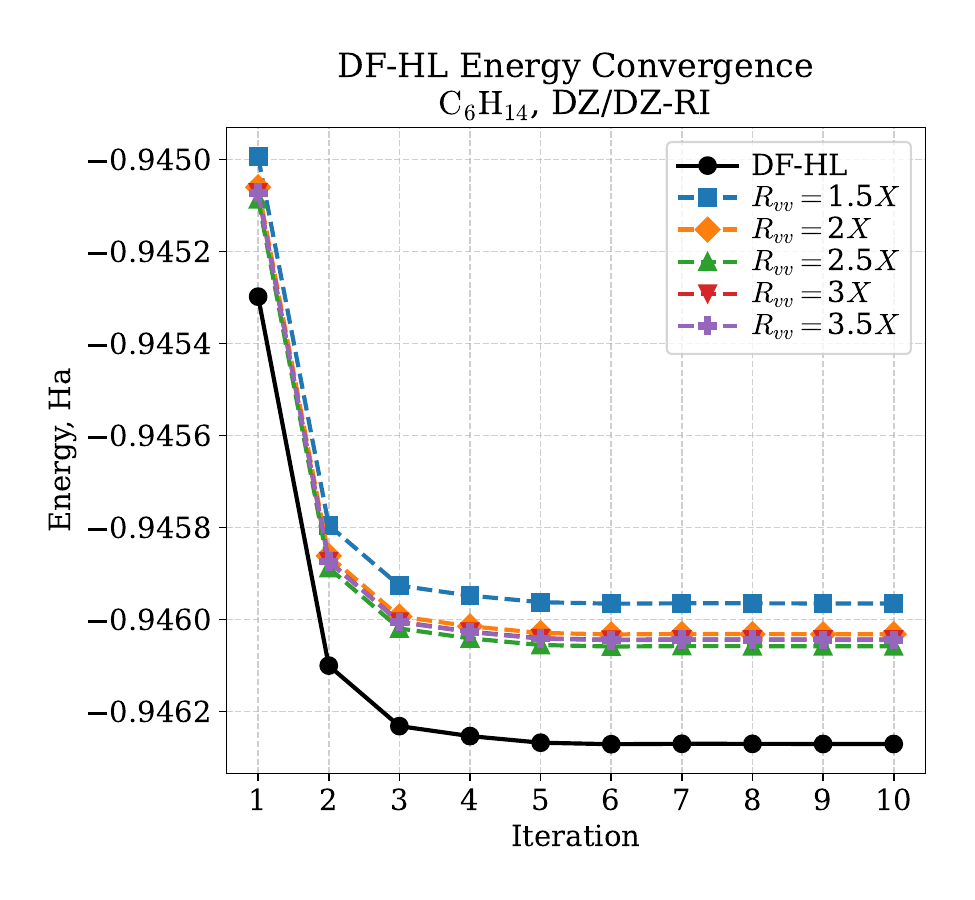}
        \caption{}
    \end{subfigure}
    \begin{subfigure}{0.49\textwidth}
        \centering
        \includegraphics[width=\linewidth]{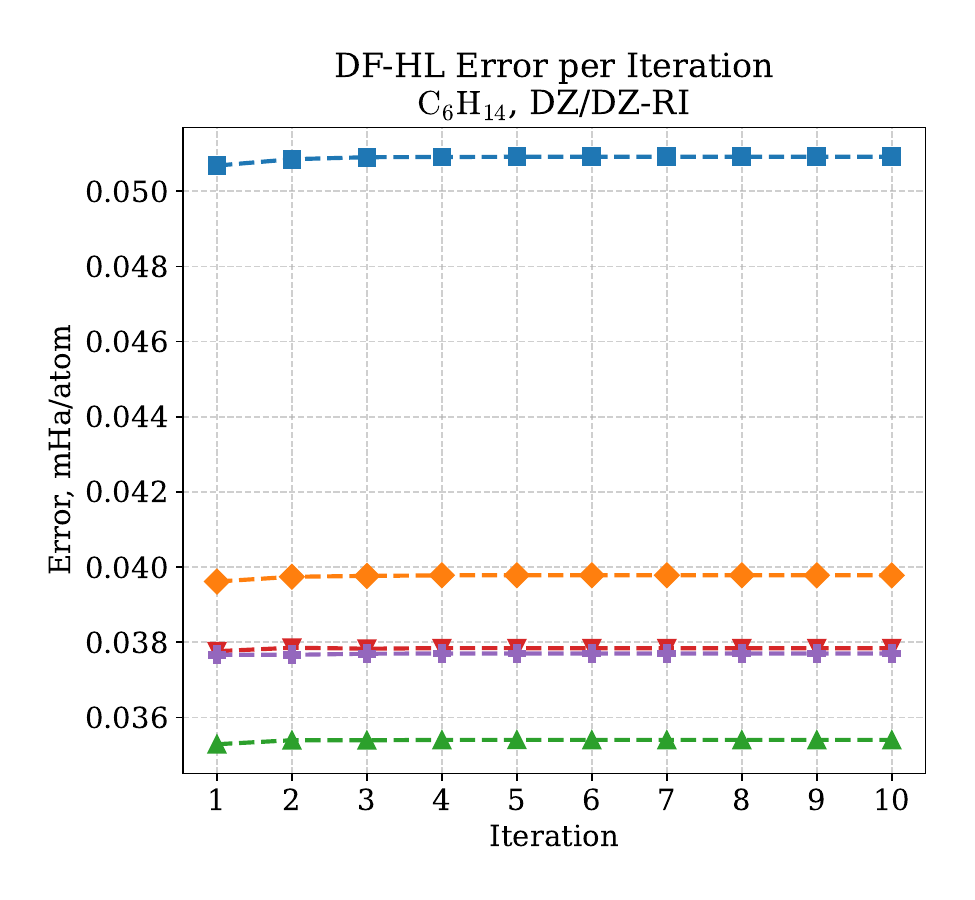}
        \caption{}
    \end{subfigure}
    \begin{subfigure}{0.49\textwidth}
        \centering
        \includegraphics[width=\linewidth]{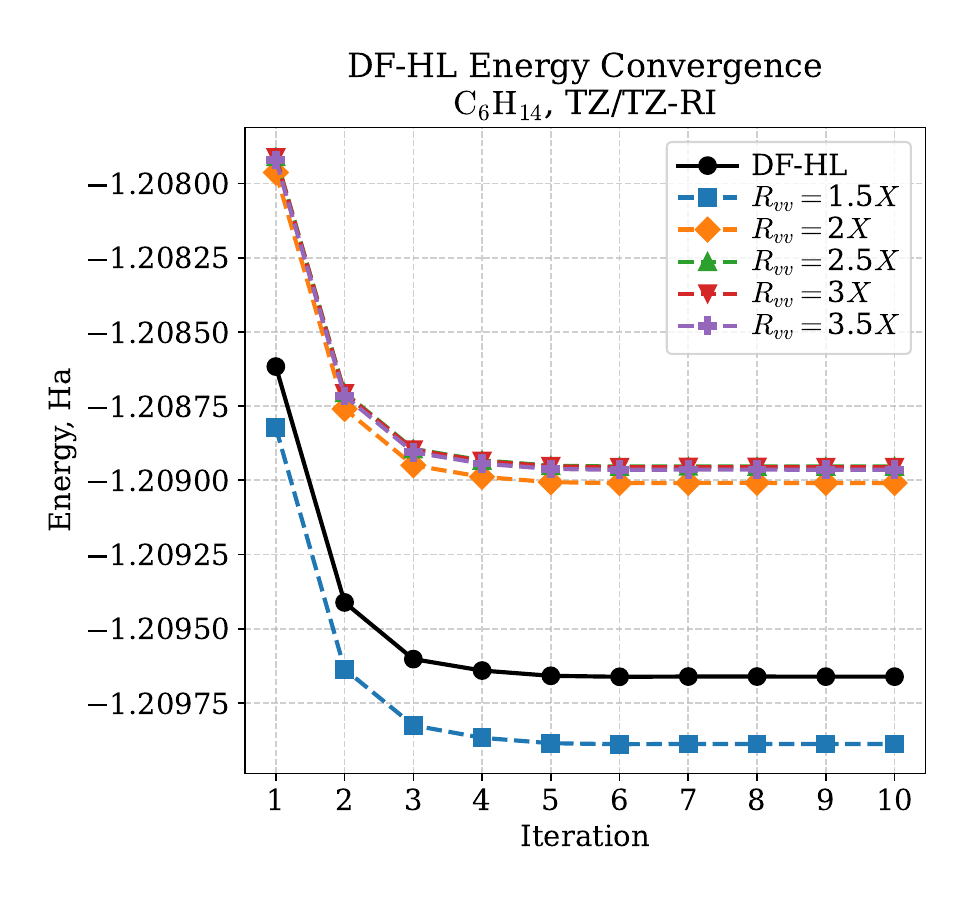}
        \caption{}
    \end{subfigure}
    \begin{subfigure}{0.49\textwidth}
        \centering
        \includegraphics[width=\linewidth]{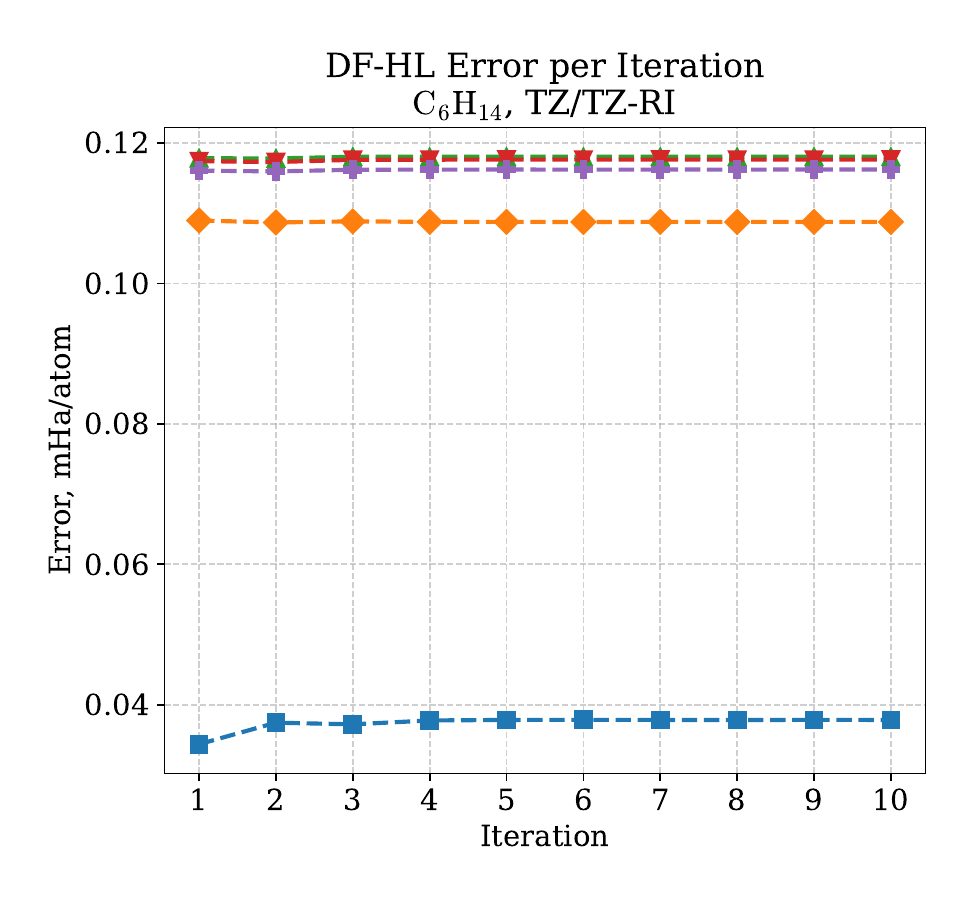}
        \caption{}
    \end{subfigure}
    \caption{HL energy reported as a function of HL iteration for a hexane molecule in the (a) DZ/DZ-RI and (c) TZ/TZ-RI basis.
    HL energy error per non-hydrogen atom reported as a function of HL iteration for a hexane molecule in the (b) DZ/DZ-RI and (d) TZ/TZ-RI basis.}
    \label{fig:hl_err_alk}
\end{figure}

\cref{fig:hl_err_h2o} considers the HL convergence behavior for a $\ce{({H}_2{O})_6}$ cluster in the DZ/DZ-RI basis and
\cref{fig:hl_err_alk} considers the HL convergence behavior for a $\ce{C_6H_14}$ molecule in the DZ/DZ-RI and TZ/TZ-RI basis.
These results closely match those found in \cref{fig:ll_err_h2o,fig:ll_err_alk} and demonstrate that introduction of the CPD does not significantly influence the convergence of the HL solver.
Furthermore, the per-iteration error in the HL energy is nearly identical to the per-iteration error in the LL energy.
\begin{figure}[htbp]
    \begin{subfigure}{0.49\textwidth}
        \centering
        \includegraphics[width=\linewidth]{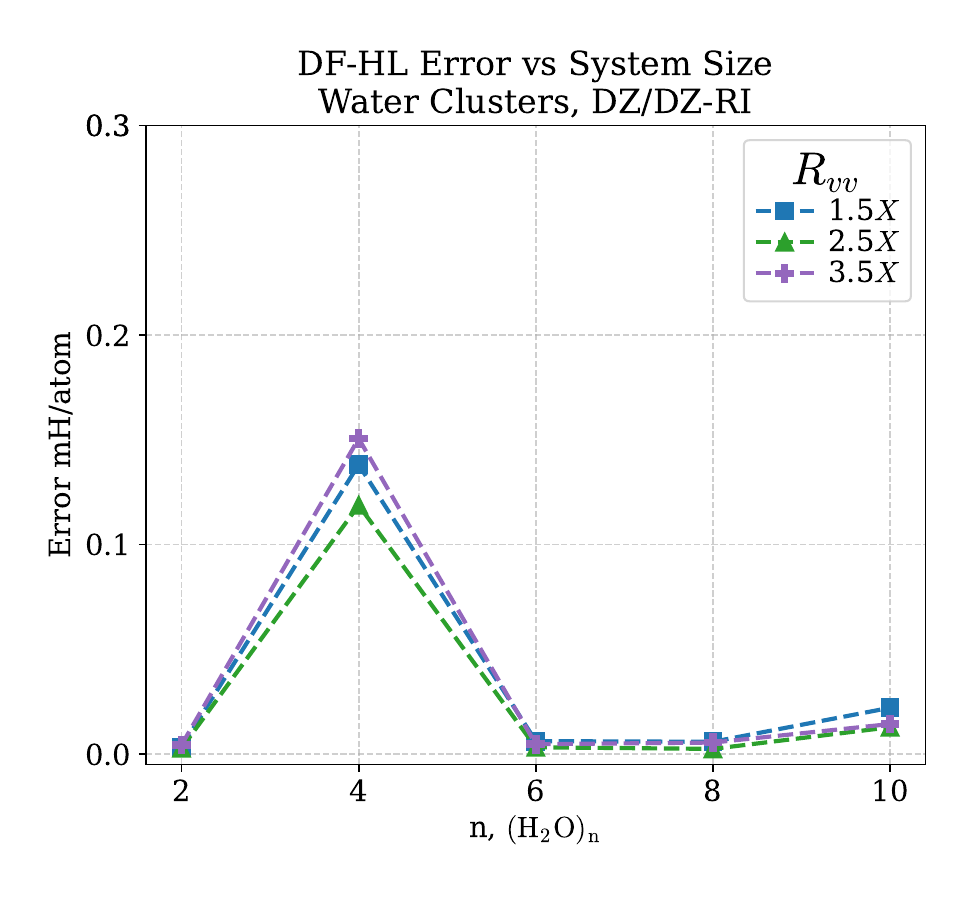}
        \caption{}
    \end{subfigure}
    \hfill
    \begin{subfigure}{0.49\textwidth}
        \centering
        \includegraphics[width=\linewidth]{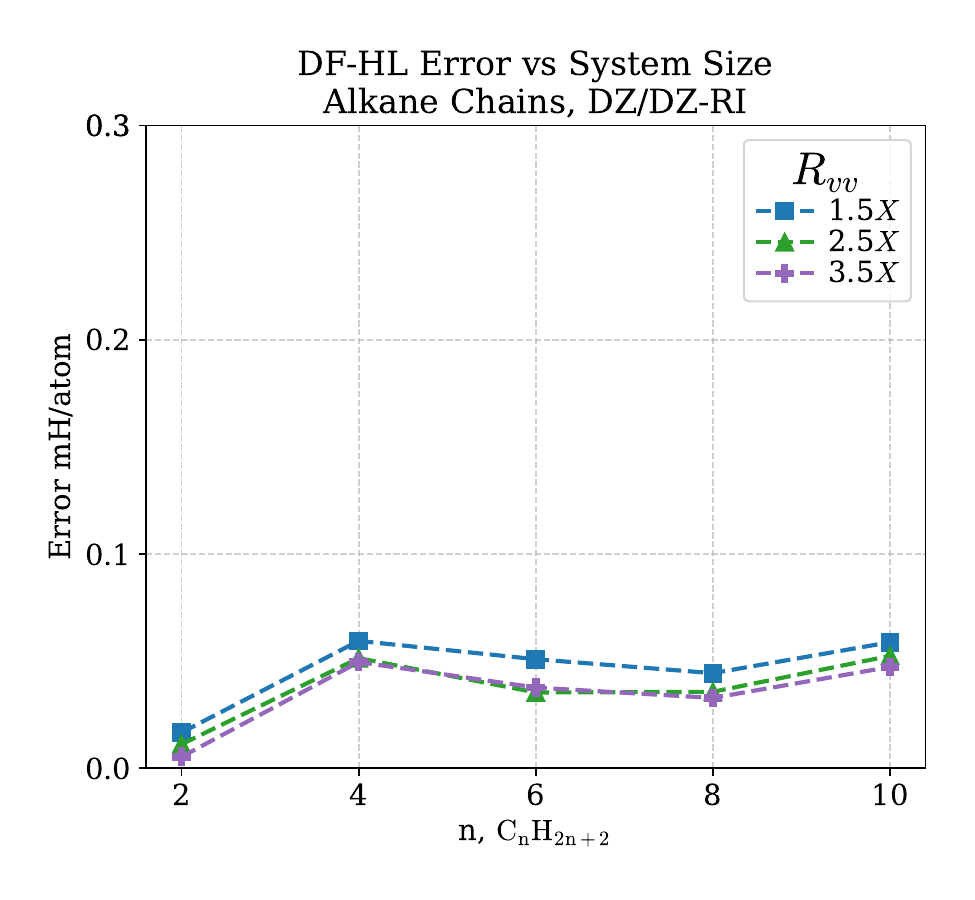}
        \caption{}
    \end{subfigure}
    \begin{subfigure}{0.49\textwidth}
        \centering
        \includegraphics[width=\linewidth]{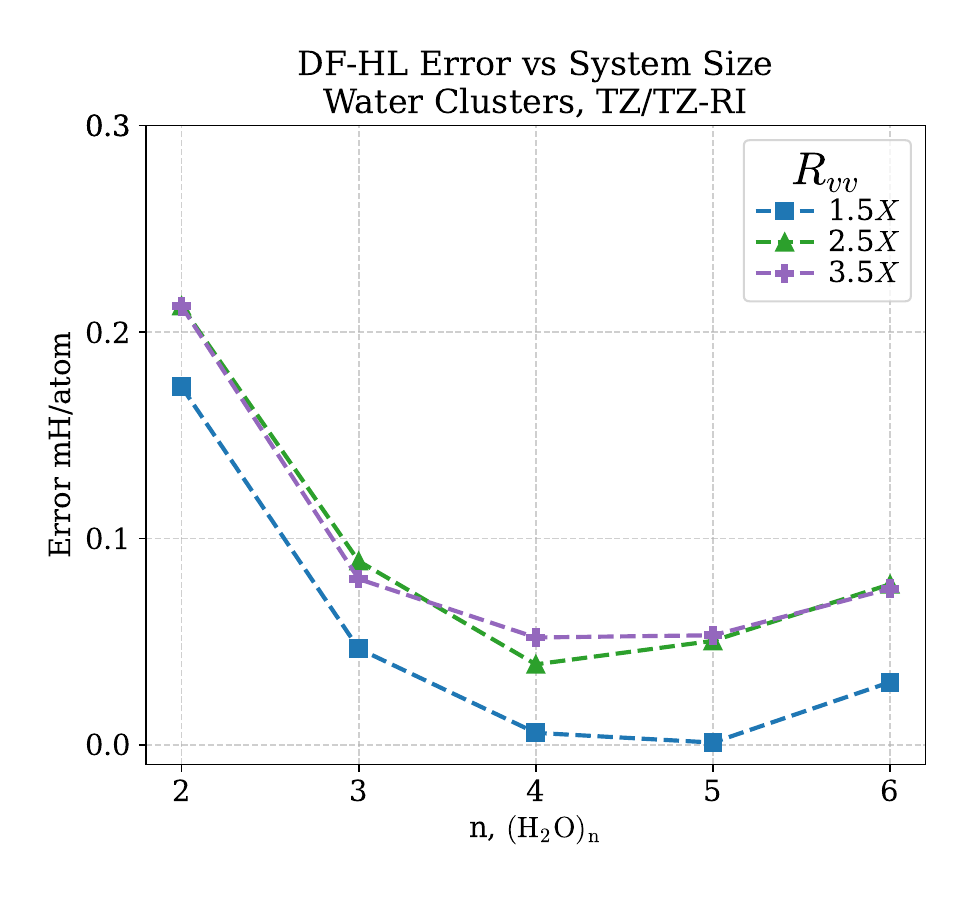}
        \caption{}
    \end{subfigure}
    \begin{subfigure}{0.49\textwidth}
        \centering
        \includegraphics[width=\linewidth]{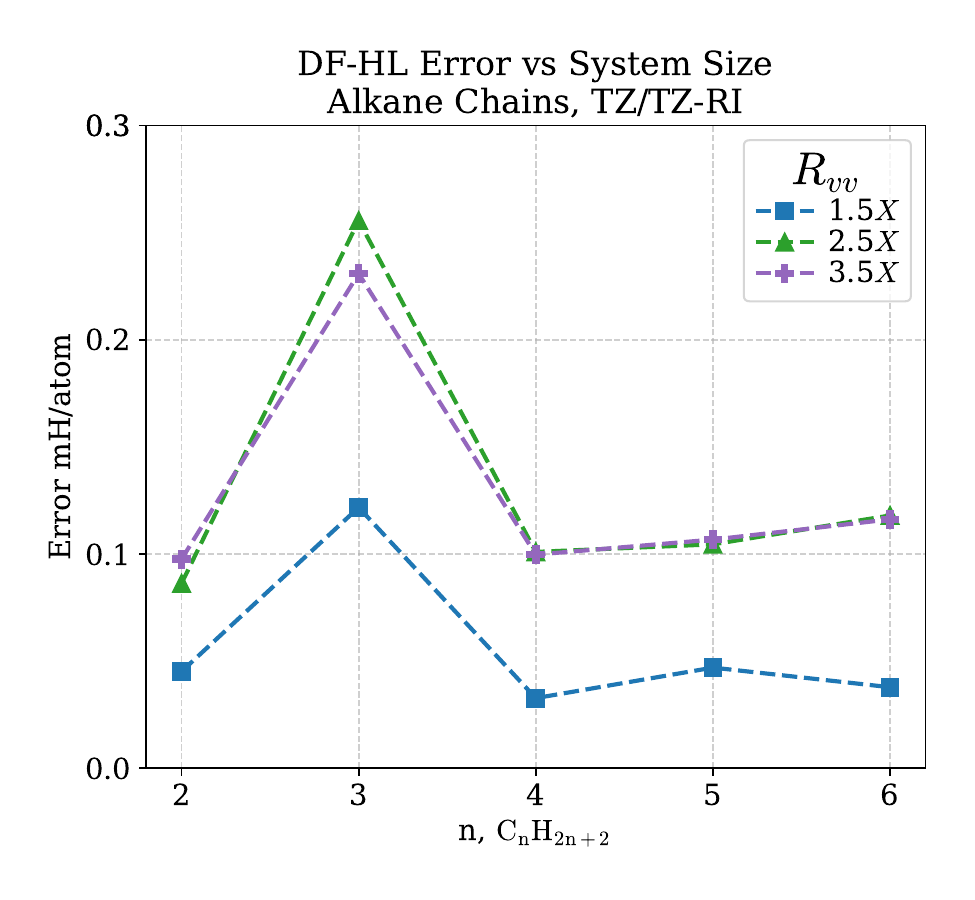}
        \caption{}
    \end{subfigure}
    \caption{Absolute HL energy error for water clusters with (a) 2-10 water molecules in the DZ/DZ-RI basis and (c) 2-6 water molecules in the TZ/TZ-RI basis and alkane chains with (b) 2-10 carbon atoms in the DZ/DZ-RI and (d) 2-6 carbon atoms in the TZ/TZ-RI basis.}
    \label{fig:ccsd_err_scan}
\end{figure}
In \cref{fig:ccsd_err_scan}, we consider the error in the converged HL energies in second macro-iteration of the MPCC procedure for water clusters and alkane chains of increasing size.
Again, these results mirror those found for the LL problem.

\newpage

\section{Impact of the CP-DF-LL solver on the MPCC Optimization}
\begin{figure}[htbp]
    \begin{subfigure}{0.49\textwidth}
        \centering
        \includegraphics[width=\linewidth]{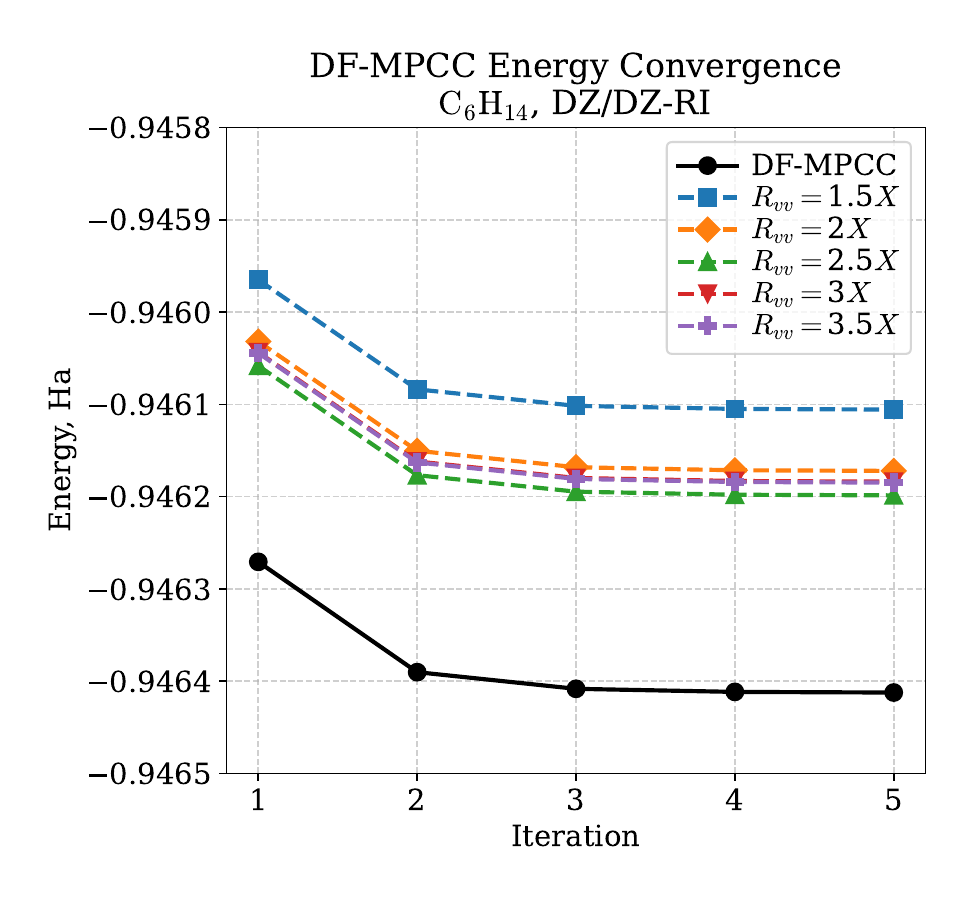}
        \caption{}
    \end{subfigure}
    \begin{subfigure}{0.49\textwidth}
        \centering
        \includegraphics[width=\linewidth]{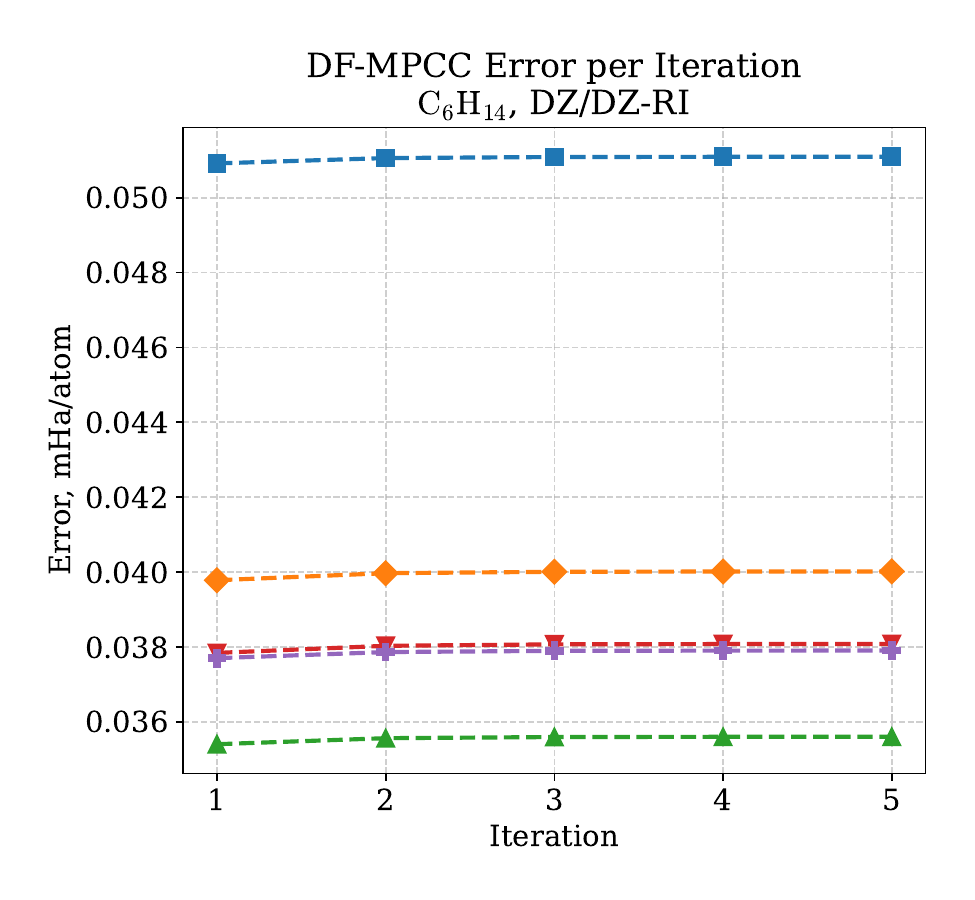}
        \caption{}
    \end{subfigure}
        \begin{subfigure}{0.49\textwidth}
        \centering
        \includegraphics[width=\linewidth]{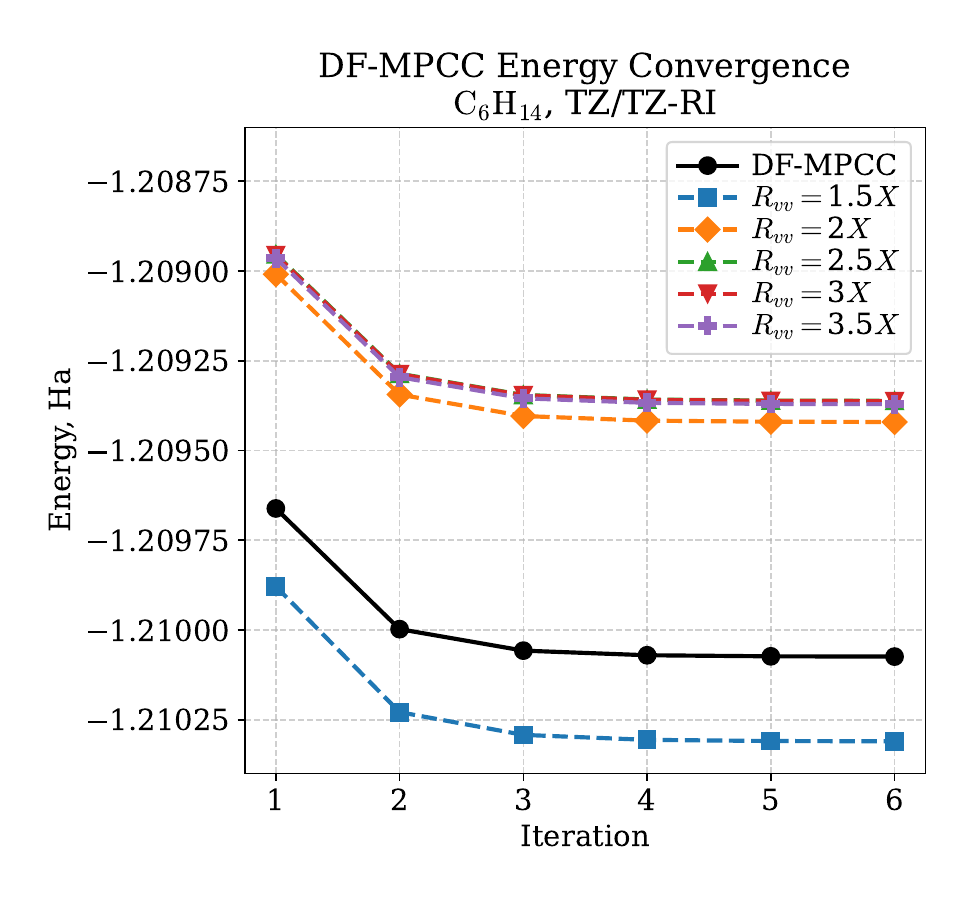}
        \caption{}
    \end{subfigure}
    \begin{subfigure}{0.49\textwidth}
        \centering
        \includegraphics[width=\linewidth]{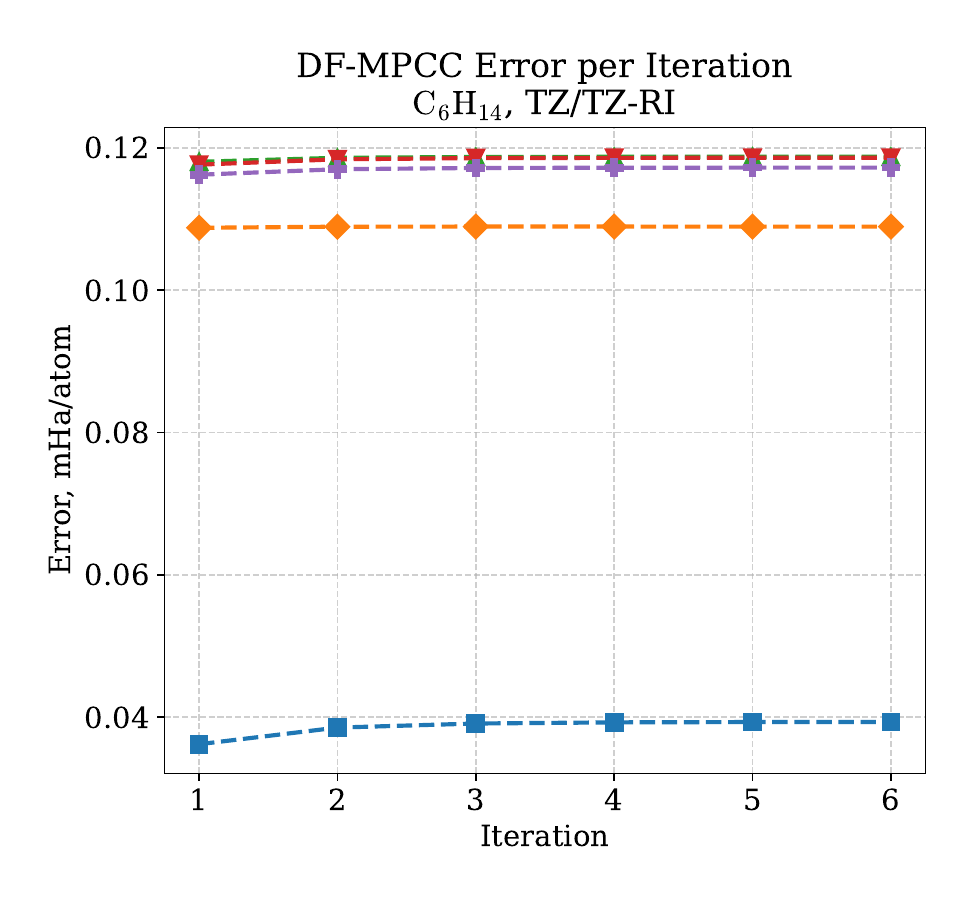}
        \caption{}
    \end{subfigure}
    \caption{
    MPCC energy reported as a function of MPCC iteration for a hexane molecule in the (a) DZ/DZ-RI and (c) TZ/TZ-RI basis.
    MPCC energy error per non-hydrogen atom reported as a function of MPCC iteration, for a hexane molecule in the (b) DZ/DZ-RI and (d) TZ/TZ-RI basis.}
    \label{fig:MPCC_err_alk}
\end{figure}
\begin{figure}[htbp]
        \begin{subfigure}{0.49\textwidth}
        \centering
        \includegraphics[width=\linewidth]{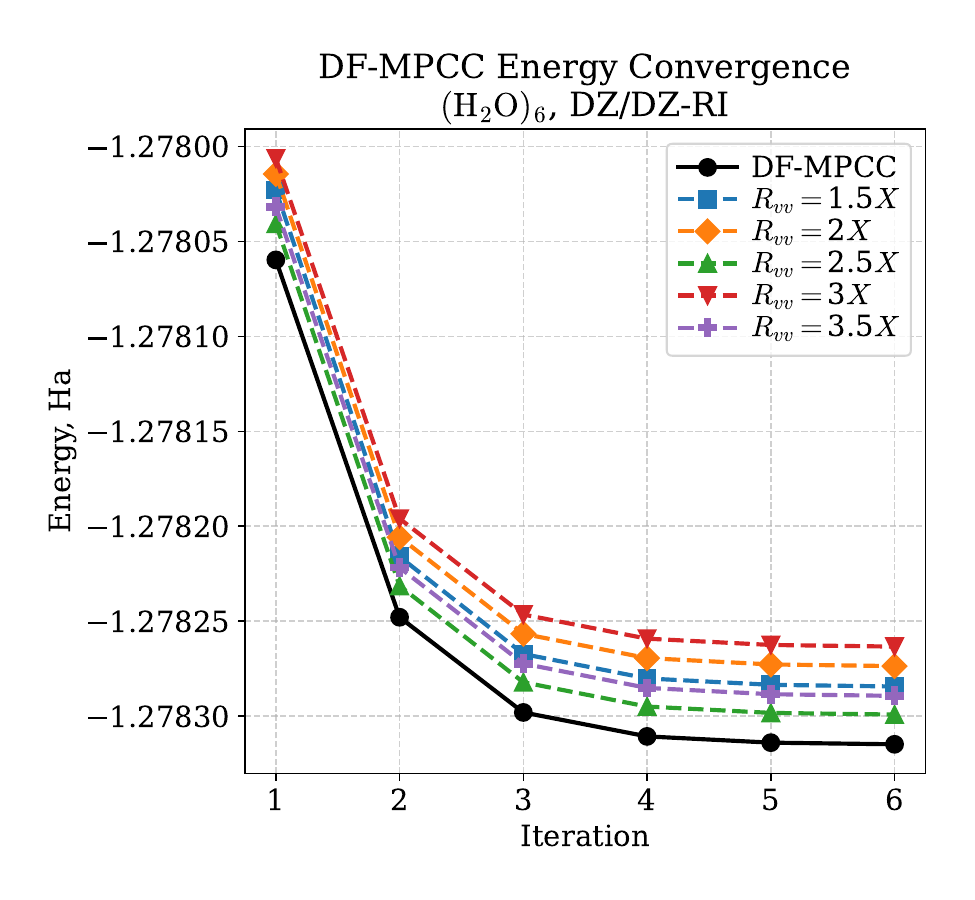}
        \caption{}
    \end{subfigure}
    \begin{subfigure}{0.49\textwidth}
        \centering
        \includegraphics[width=\linewidth]{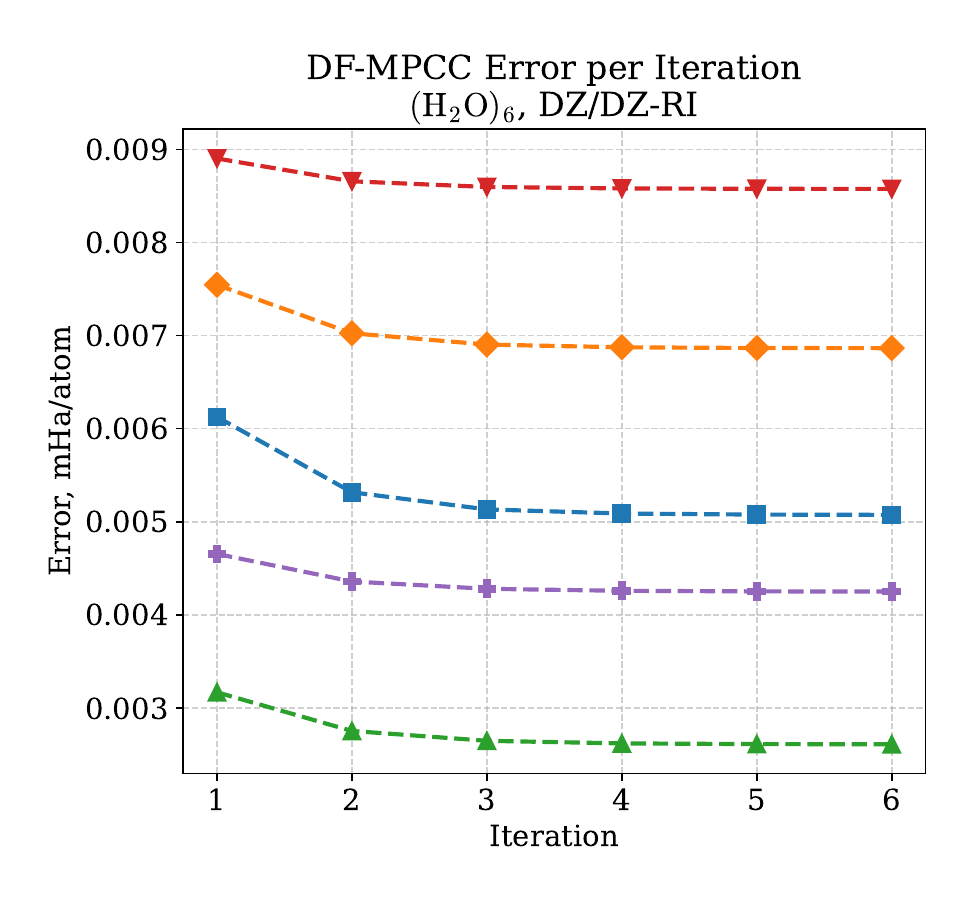}
        \caption{}
    \end{subfigure}
    \caption{(a) MPCC energy and (b) MPCC energy error per non-hydrogen atom, both reported as a function of MPCC iteration, for a 6-water cluster in the DZ/DZ-RI basis.}
    \label{fig:MPCC_err_h2o}
\end{figure}
\begin{figure}[htbp]
    \begin{subfigure}{0.49\textwidth}
        \centering
        \includegraphics[width=\linewidth]{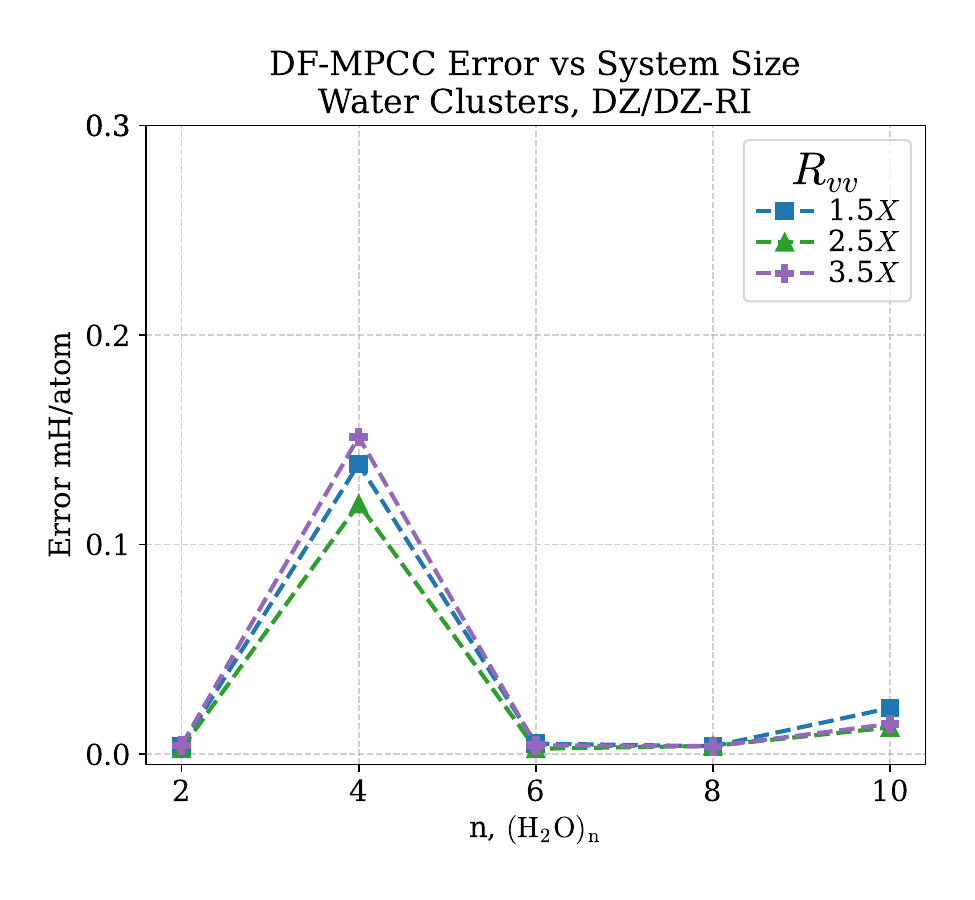}
        \caption{}
    \end{subfigure}
    \hfill
    \begin{subfigure}{0.49\textwidth}
        \centering
        \includegraphics[width=\linewidth]{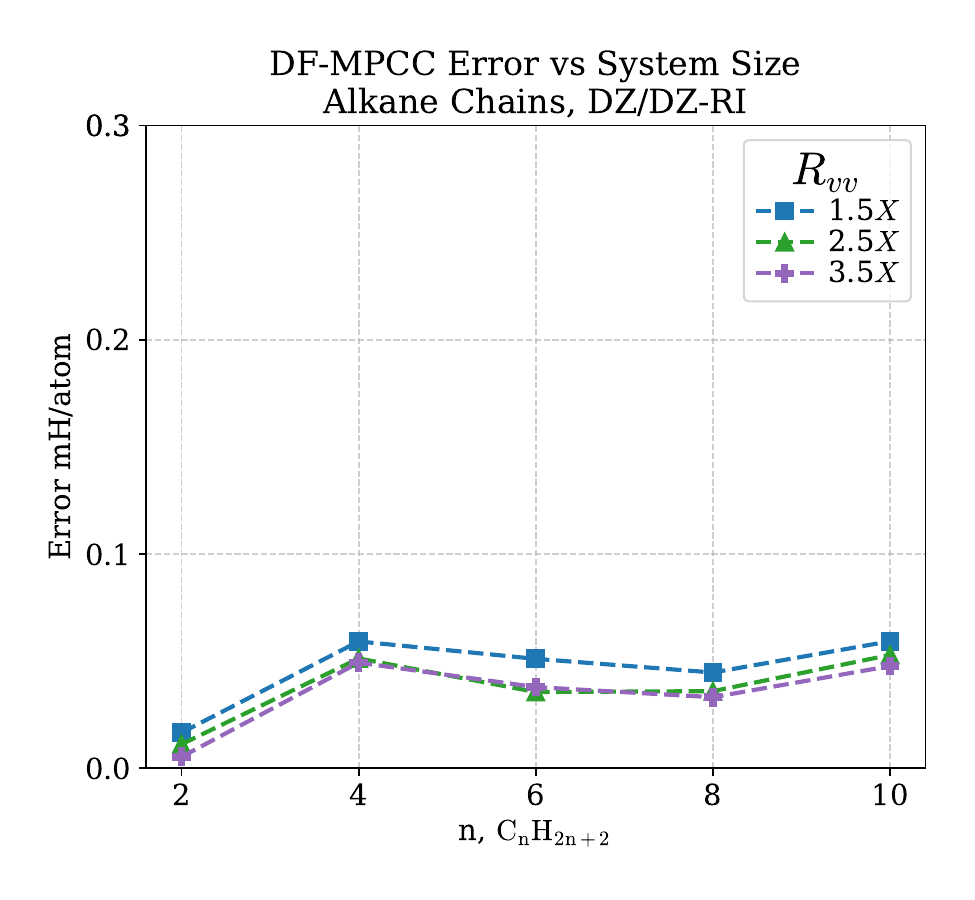}
        \caption{}
    \end{subfigure}
    \begin{subfigure}{0.49\textwidth}
        \centering
        \includegraphics[width=\linewidth]{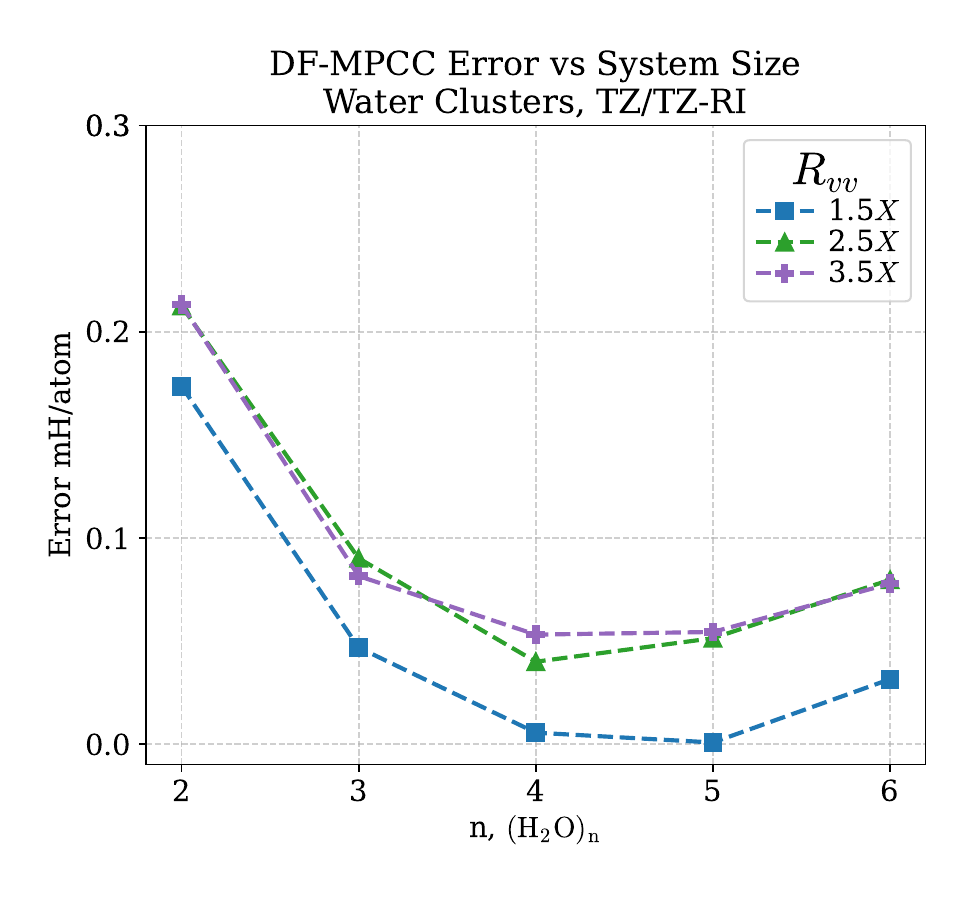}
        \caption{}
    \end{subfigure}
    \begin{subfigure}{0.49\textwidth}
        \centering
        \includegraphics[width=\linewidth]{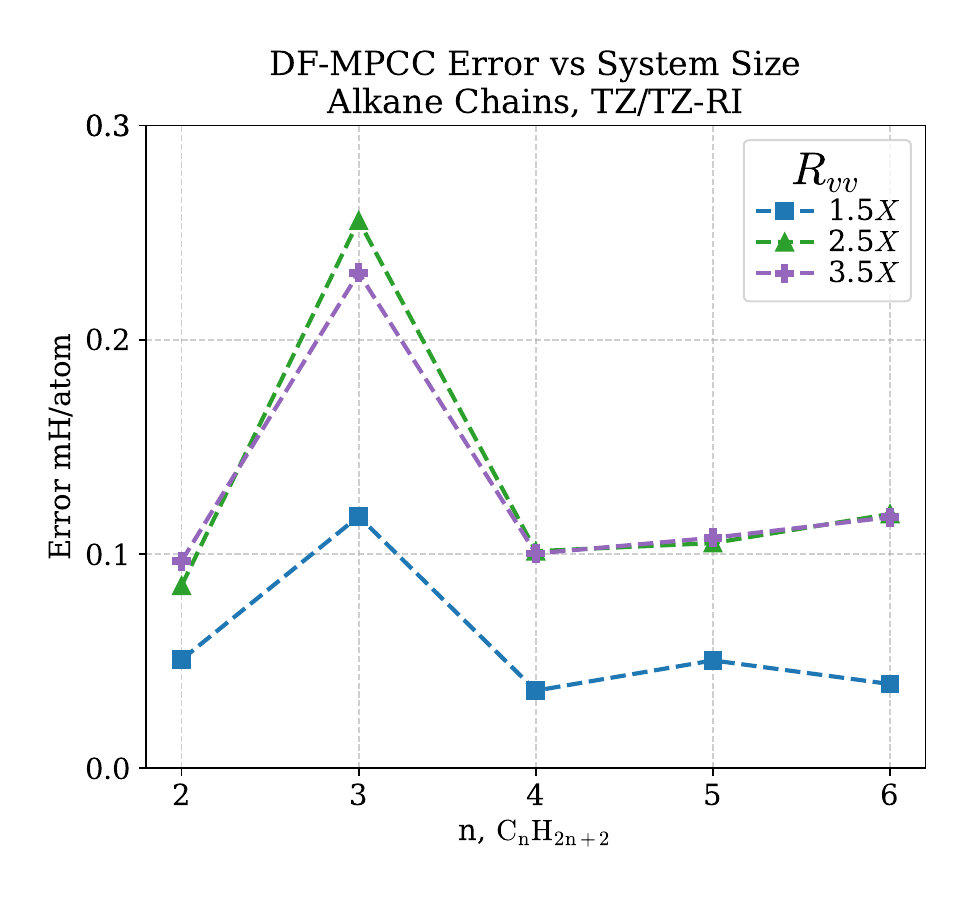}
        \caption{}
    \end{subfigure}
    \caption{Absolute MPCC energy error for water clusters with (a) water clusters with 2-10 water molecules in the DZ/DZ-RI basis, (b) alkane chains with 2-10 carbon atoms in the DZ/DZ-RI basis, (c) water clusters with 2-6 water molecules in the TZ/TZ-RI basis and (d) alkane chains with 2-6 carbon atoms in the TZ/TZ-basis.}
    \label{fig:mpcc_mols}
\end{figure}
\begin{figure}[htbp]
        \centering
        \includegraphics[width=0.7\linewidth]{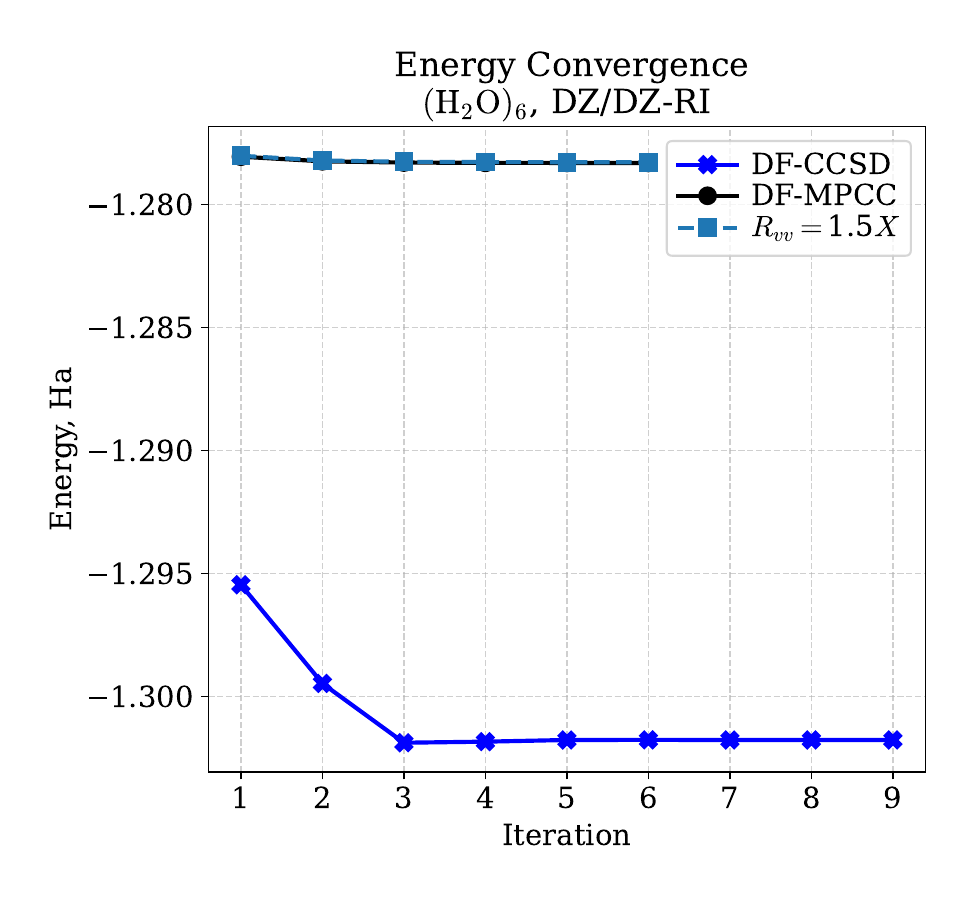}
\caption{Energy reported as function of DF-CCSD and MPCC iteration for a 6-water cluster in the DZ/DZ-RI basis.}
\label{fig:mpcc_dz_ccsd}
\end{figure}
\begin{figure}[htbp]
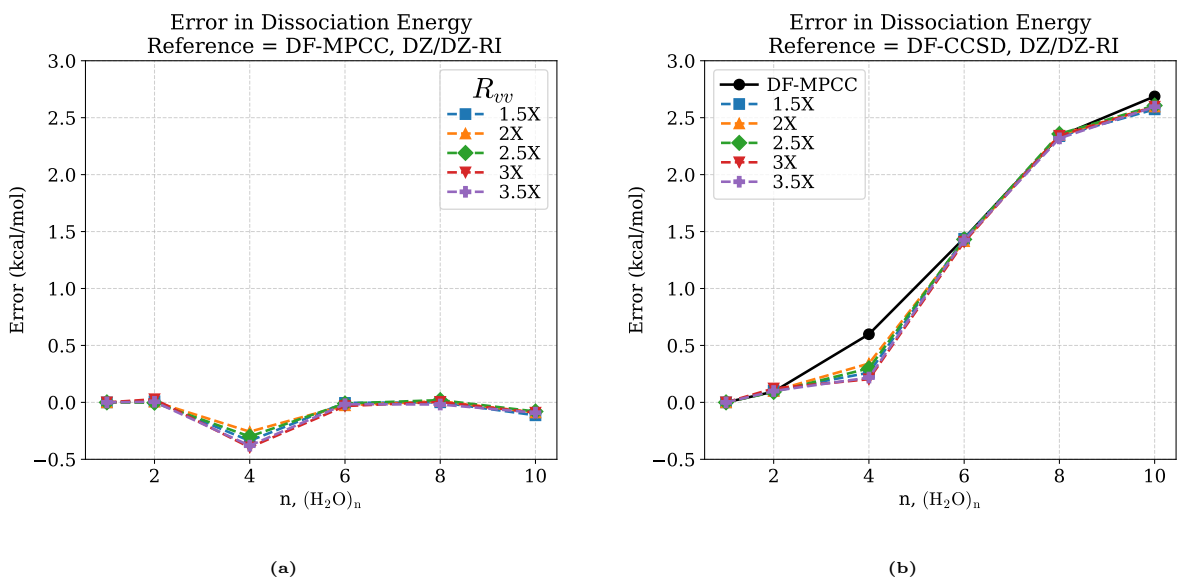

    \centering

    \begin{subfigure}{0.49\textwidth}
        \centering
        \includegraphics[width=\linewidth]{plots_for_articles/dissociation_error_MPCC_DF_all_ranks_cc-pvdz_Lvv_fix_Loo_1_signed.pdf}
        \caption{}
    \end{subfigure}
    \begin{subfigure}{0.49\textwidth}
        \centering
        \includegraphics[width=\linewidth]{plots_for_articles/dissociation_error_MPCC_DF_all_ranks_cc-pvdz_Lvv_fix_Loo_1_signed_ccsd.pdf}
        \caption{}
    \end{subfigure}
\caption{Dissociation energy error for water clusters with between 1 and 10 water molecules with respect to the DF-LL MPCC in (a) with DZ/DZ-RI and (b) in TZ/TZ-RI basis.}
    \label{fig:mpcc_dissociation_subplots}
\end{figure}
\begin{figure}[htbp]
        \centering
        \includegraphics[width=0.7\linewidth]{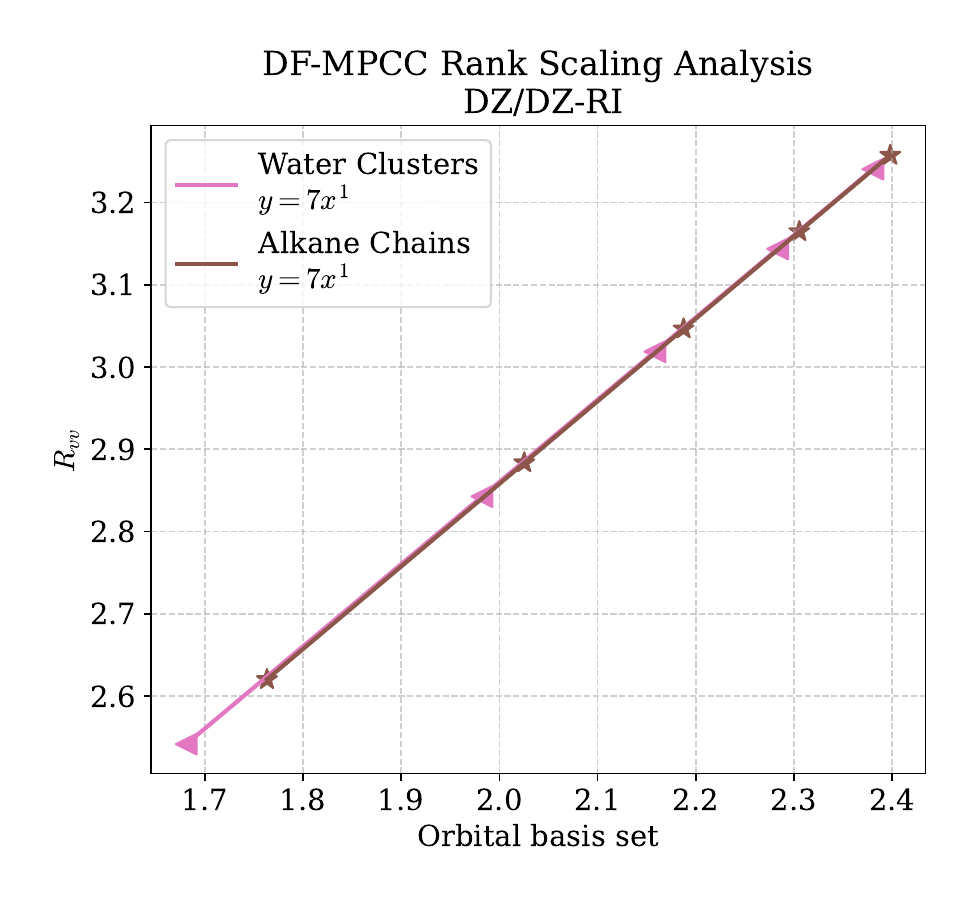}
\caption{Modeling the growth of the CP rank with system size for water molecule clusters and alkane chains in the DZ/DZ-RI basis using a threshold of 0.5mH per non-hydrogen atom.}
\label{fig:mpcc_dz_scaling_curve}
\end{figure}
\begin{figure}[htbp]
        \centering
        \begin{subfigure}{0.49\textwidth}
        \includegraphics[width=\linewidth]{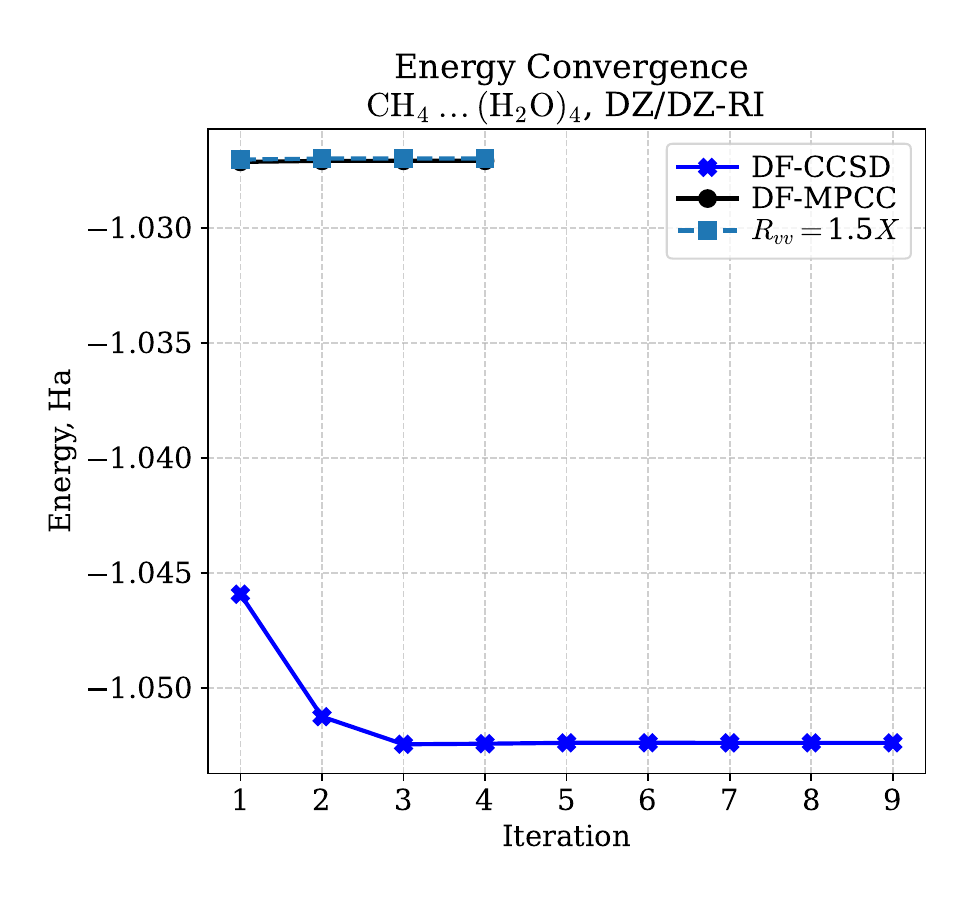}
        \end{subfigure}
        \begin{subfigure}{0.49\textwidth}
            \includegraphics[width=\linewidth]{plots_for_articles/dissociation_error_CH4_H2O4_MPCC_cc-pvdz_Lvv_fix_Loo_1_ccsd.pdf}
        \end{subfigure}
\caption{(a)Energy reported as function of DF-CCSD and MPCC iteration for a $\ce{CH_4}\dots\ce{(H_2O)_4}$ cluster in the DZ/DZ-RI basis.
(b) Error in MPCC dissociation energy compared to DF-CCSD for a $\ce{CH_4}\dots\ce{(H_2O)_4}$ cluster in the DZ/DZ-RI basis.}
\label{fig:mpcc_m4w_ccsd}
\end{figure}

\cref{fig:MPCC_err_alk} considers the MPCC convergence behavior for a $\ce{C_6H_14}$ molecule in the DZ/DZ-RI and TZ/TZ-RI basis.
\cref{fig:MPCC_err_h2o} considers the MPCC convergence behavior for a $\ce{({H}_2{O})_6}$ cluster in the DZ/DZ-RI basis and
These results closely match those found in \cref{fig:ll_err_h2o,fig:ll_err_alk,fig:hl_err_h2o,fig:hl_err_alk} illustrating that CPD approximation does not significantly influence the convergence of the MPCC solver as a whole.
In \cref{fig:mpcc_mols}, we consider the error in the converged MPCC energies for water clusters and alkane chains of increasing size.
These results mirror those in \cref{fig:cc2_err_scan,fig:ccsd_err_scan} including the marginally higher error in the $\ce{(H_2O)_4}$ molecule in the DZ/DZ-RI basis and the $\ce{C_3H_8}$ in the TZ/TZ-RI basis.
With these results, we demonstrate that the error in the LL solver introduced by the CPD does not compound over the MPCC optimization procedure.
In \cref{fig:mpcc_dz_ccsd}, we show the MPCC energy convergence compared the DF-CCSD energy convergence for a 6-water cluster in the DZ/DZ-RI basis.
With this figure, we show that the deviations in the MPCC energy introduced by the CPD are significantly smaller
than the overall error of the MPCC method compared to canonical DF-CCSD.
In \cref{fig:mpcc_dissociation_subplots} we show the error in water cluster disassociation energies in the DZ/DZ-RI basis.
We demonstrate that the error in the dissociation energy introduced by the CPD approximation is relatively small compared to the overall error of the MPCC method.
Next in \cref{fig:mpcc_dz_scaling_curve} we show the relationship between the CP rank and OBS dimension at a fixed absolute error of 0.5mH per non-hydrogen atom for water clusters and alkane chains in the DZ/DZ-RI basis.
This figure shows that the CP rank scales linearly with system size in the DZ/DZ-RI basis.
Finally in \cref{fig:mpcc_m4w_ccsd} we consider the MPCC convergence behavior and dissociation energy error for a $\ce{CH_4}...\ce{(H_2O)_4}$ cluster in the DZ/DZ-RI basis.
These results further demonstrate that the introduction of the adds a relatively small degree of error into the MPCC procedure and does not significantly influence the accuracy of relevant chemical energy differences.

\newpage
\newpage
\bibliography{kmp5refs}